\documentclass[journal]{IEEEtran}

\usepackage[T1]{fontenc}
\usepackage{amsmath,amssymb,amsthm}
\usepackage{mathtools}
\usepackage{graphicx}
\usepackage{booktabs}
\usepackage{array}
\usepackage{multirow}
\usepackage{xcolor}
\usepackage{cite}
\usepackage{pgfplots}
\pgfplotsset{compat=1.17}
\usepgfplotslibrary{groupplots}
\usetikzlibrary{calc,arrows.meta,positioning,fit,backgrounds}
\usepackage{hyperref}

\newcommand{\sysname}{BioBind}
\newcommand{\ES}{\mathsf{ES}}%
\newcommand{\AFE}{\mathsf{ACM}}%
\newcommand{\Bk}{\mathsf{B}}%

\newcommand{\vv}{\mathbf{v}}%
\newcommand{\vp}{\mathbf{v}'}%
\newcommand{\dist}{\mathsf{dist}}%
\newcommand{\Dup}{\mathsf{Dup}}
\newcommand{\tauauth}{\tau_{\mathsf{auth}}}
\newcommand{\taudup}{\tau_{\mathsf{dup}}}

\newcommand{\Cid}{C_{\mathsf{id}}}%
\newcommand{\Cen}{C_{\mathsf{en}}}%
\newcommand{\Cenr}{C_{\mathsf{enr}}}%
\newcommand{\Cprb}{C_{\mathsf{prb}}}%
\newcommand{\renr}{r_{\mathsf{enr}}}
\newcommand{\rprb}{r_{\mathsf{prb}}}
\newcommand{\ComH}{\mathsf{Com}}
\newcommand{\Enc}{\mathsf{Enc}}
\newcommand{\Pt}{\mathsf{Pt}}
\newcommand{\decl}{\mathsf{decl}}
\newcommand{\link}{\mathsf{link}}
\newcommand{\sid}{\mathsf{sid}}
\newcommand{\cred}{\mathsf{cred}}
\newcommand{\slot}{\mathsf{slot}}
\newcommand{\Mig}{\mathsf{Mig}}
\newcommand{\Gid}{\mathcal{G}}%

\newcommand{\hP}{h_{\mathcal{P}}}%
\newcommand{\hchi}{h_{\chi}}%
\newcommand{\HF}{H_{\mathbb{F}}}%
\newcommand{\idc}{\mathsf{idc}}
\newcommand{\nf}{\mathsf{nf}}
\newcommand{\scope}{\mathsf{scope}}
\newcommand{\Rcred}{R_{\mathsf{cred}}}
\newcommand{\Rdev}{R_{\mathsf{dev}}}

\newcommand{\Adv}{\mathsf{Adv}}
\newcommand{\Gunq}{\mathsf{G}^{\mathsf{unq}}}
\newcommand{\Gauth}{\mathsf{G}^{\mathsf{auth}}}
\newcommand{\Glink}{\mathsf{G}^{\mathsf{link}}}
\newcommand{\fnm}{f_{\mathsf{nm}}}
\newcommand{\fmax}{f_{\mathsf{max}}}

\definecolor{bbBlueA}{HTML}{86B6EF}%
\definecolor{bbBlueB}{HTML}{5598E7}%
\definecolor{bbBlueC}{HTML}{2A78D6}%
\definecolor{bbBlueD}{HTML}{1C5CAB}%
\definecolor{bbBlueE}{HTML}{0D366B}%
\definecolor{bbOrange}{HTML}{EB6834}%
\definecolor{bbGray}{HTML}{8F8D86}%
\definecolor{bbInk}{HTML}{52514E}%
\pgfplotsset{
  bbaxis/.style={
    font=\scriptsize,
    tick label style={font=\scriptsize, text=bbInk},
    label style={font=\scriptsize},
    axis line style={bbInk, line width=0.4pt},
    tick style={bbInk, line width=0.4pt},
    major grid style={black!10, line width=0.3pt},
    grid=major,
    legend style={font=\scriptsize, draw=none, fill=white, fill opacity=0.85,
                  text opacity=1, inner sep=1pt, row sep=-1pt},
    legend cell align=left,
    every axis plot/.append style={line join=round},
  },
}

\definecolor{bbRed}{HTML}{C0392B}
\tikzset{
  nohyph/.style={execute at begin node={\hyphenpenalty=10000\exhyphenpenalty=10000}},
  box/.style={nohyph, draw, rounded corners=3pt, line width=0.6pt, align=center,
    font=\scriptsize, inner sep=3pt, minimum height=0.85cm, text depth=0pt},
  trusted/.style={box, draw=bbOrange, fill=bbOrange!10},
  backend/.style={box, draw=bbGray, fill=bbGray!14},
  chain/.style={box, draw=bbBlueC, fill=bbBlueC!8},
  adversary/.style={box, draw=bbInk, dash pattern=on 2pt off 1.5pt, fill=white},
  person/.style={box, draw=bbInk!60, fill=white},
  flow/.style={-{Latex[round,length=4pt,width=3.5pt]}, line width=0.6pt, bbInk},
  lab/.style={nohyph, font=\fontsize{6}{7}\selectfont, text=bbInk, align=center, inner sep=1.5pt},
  lane/.style={font=\scriptsize\bfseries, text=bbInk, anchor=west},
}

\newtheorem{theorem}{Theorem}
\newtheorem{lemma}{Lemma}
\newtheorem{corollary}{Corollary}
\theoremstyle{definition}
\newtheorem{definition}{Definition}

\title{BioBind: Binding Biometric Deduplication to Zero-Knowledge Authentication}
\author{Taotao Wang, Xiaoxiao Wu, and Shengli Zhang%
\thanks{The authors are with the College of Electronics and Information Engineering, Shenzhen University, Shenzhen 518060, China (e-mail: ttwang@szu.edu.cn; xxwu.eesissi@szu.edu.cn; zsl@szu.edu.cn).}}

\begin{document}
\maketitle

\begin{abstract}
Biometric identity systems for decentralized applications need credentials that resist repeated enrollment and support private, publicly verifiable authentication.
Capture and large-scale private
deduplication run off chain, while smart contracts authorize credential issuance
and use. These stages span distinct trust domains, and a correct verdict
covers only the checked inputs and gallery snapshot. Without binding and
state checks, concurrent enrollments can issue duplicate credentials from
correct verdicts, while substituted inputs can authorize unrelated credentials.
We present BioBind, a protocol whose four backend interface conditions
bind the deduplication query, gallery insertion and issued commitment to one
captured vector, authenticate gallery transitions, and require a sound
uniqueness verdict valid at finalization. Our encrypted instance uses bound
requests, stale-session rejection and committee-confirmed gallery migration.
Capture-bound zero-knowledge authentication links credential use to
the enrolled template and a fresh device-originated probe bound to the
challenge. We give a
conditional multi-session composition argument under explicit capture,
backend and cryptographic assumptions. A prototype using
LFW face features rejects the tested composition attacks; its unlinkable
authentication circuit uses 31,445 constraints and approximately 305,000
gas. An equal-public-field baseline quantifies the cost of
keeping matching verifiable, with witness-plus-proving medians of 1.989 s
versus 1.119 s when the device is trusted with matching. Separate tests of encrypted enrollment complete three
histories of 16,384 registrations without decoding errors at the tested checkpoints.
Biometric accuracy bounds person-level uniqueness: on the Bosphorus
database, cooperative subjects who change their appearance evade
deduplication in 1.8\% of re-enrollment attempts at the operating threshold. Full backend integration
remains open.
\end{abstract}


\section{Introduction}
\label{sec:intro}

Decentralized applications use identity credentials to limit repeated
participation and allocate one-per-person benefits without exposing a
user's identity at every interaction~\cite{douceur2002sybil,worldwhitepaper}.
The one-person-one-account goal has also been proposed as a design
requirement for Web 3.0 accounts that bind each account to one person
while keeping its use anonymous~\cite{wang2023account,wang2025zkbid}.
Formal proof-of-personhood work separates uniqueness, authenticated
personhood and unlinkable use~\cite{choudhuri2026pop}.
Biometrics can connect this goal to physical people: 1:N deduplication
checks a newcomer against an enrollment gallery, and 1:1 authentication
checks a later probe against an enrolled template. World ID combines a
trusted capture device, private uniqueness checking and credential
use~\cite{worldwhitepaper}, while Kothari et al. demonstrate
privacy-preserving fingerprint authentication with on-chain
verification~\cite{kothari2026fingerprint}. These
settings motivate our question: how can a deduplication result become an
enrollment record that subsequent private authentication can rely on?

The question is concrete at the boundary between off-chain biometric
processing and on-chain credential management. Capture depends on physical
devices, while encrypted search over large galleries is computationally
intensive. These operations therefore typically take place off chain, with
holders supplying proofs and contracts verifying evidence to authorize
issuance and use. The stages span distinct trust domains, and an enrollee
seeking multiple credentials can control its own host, submitted data and
request timing. Without bindings among the capture record,
deduplication request and gallery insertion, a valid verdict for one input
may authorize a credential or gallery update derived from another.
For example, if the ciphertext used to query the gallery and the one
inserted afterwards are supplied separately without a consistency check, an
enrollee could query with its captured template, receive a correct unique
verdict, and insert an encrypted zero vector. The gallery then holds no
entry for this person, and a second enrollment passes deduplication again.
Even when query and insertion are consistent, concurrent
enrollments can change the
gallery after the verdict is computed, so uniqueness relative to an earlier
snapshot need not justify issuance at finalization.

Existing systems address different parts of this boundary.
World ID's published MPC protocol and BioZKFHE provide
private 1:N deduplication through secret sharing and verifiable encrypted
evaluation, respectively~\cite{bloemen2024ampc,xin2026biozkfhe}. Their
adversary models differ, but either route still needs an enrollment protocol
that makes a unique verdict justify the gallery update and credential that
follow it. World ID describes storing the fragments used in a successful
uniqueness check~\cite{worldwhitepaper}; BioZKFHE instead updates its
gallery with a separately encrypted insertion ciphertext, which the enrollment
protocol must bind to the queried one.

On the authentication side, prior work proves a match against a
committed enrollment template and verifies the proof on
chain~\cite{lin2025biozero,kothari2026fingerprint}.
These authentication proofs do not establish 1:N enrollment uniqueness or
the provenance of a fresh capture.
World ID's WIP-201/202 draft instead binds a trusted matching service's
result to a credential and challenge in zero knowledge~\cite{wip201,wip202}.
Taken together, these systems leave open what a contract can rely on when
it issues a credential: which captured vector was queried, which entry was
added to the gallery, which commitment was issued, and whether the unique
verdict still held when the entry was added. \sysname{} makes these checks
explicit under malicious enrollees and concurrent sessions, with capture and
backend trust stated separately.

\begin{table*}[!t]
	\centering
	\caption{Which link of the uniqueness-to-authentication chain each system makes checkable, and where the inputs to that link come from. The table characterizes the cited versions, not a deployment audit. NE means not established by the inspected sources; -- means outside the compared operation.}
	\label{tab:related}
	\footnotesize
	\setlength{\tabcolsep}{3pt}
	\begin{tabular}{@{}lcccccccc@{}}
		\toprule
		System & 1:1 match & Uniqueness & Probe bound & Enrollment & Gallery write & Verdict & Live state & Off-device \\
		& verifiable & & to capture & provenance & verifiable & linearized & at acceptance & material \\
		\midrule
		Kothari et al.~\cite{kothari2026fingerprint} & yes & no & no & -- & -- & -- & NE & ciphertext \\
		BioZero~\cite{lin2025biozero} & yes & no & no & -- & -- & -- & NE & commitments \\
		BioZKFHE~\cite{xin2026biozkfhe} & -- & yes & -- & honest client & NE & NE & -- & ciphertexts \\
		World ID iris-mpc~\cite{bloemen2024ampc} & -- & yes & -- & operator & NE & NE & -- & shares \\
		World ID WIP-201/202~\cite{wip201,wip202} & enclave & -- & engine & -- & -- & -- & NE & images to enclave \\
		\sysname{} (this paper) & yes & relative to backend & yes ($\AFE$) & yes (I1) & yes (I2) & yes (I3) & yes & plaintext at ES \\
		\bottomrule
	\end{tabular}
\end{table*}

Two failures illustrate why these checks matter. In a snapshot race, one
person opens two enrollment sessions on a snapshot that contains no entry of
theirs. Both queries correctly return unique. If both sessions insert into
fresh slots and issue credentials, that person enrolls twice even though each
query and write is correct. In a dropped merge, the failure lies with the
backend rather than the enrollee: the chain stores only a commitment to each
gallery block and cannot recompute a merge of megabyte-sized ciphertexts, so a
compute node can discard the new ciphertext and still report an updated
commitment. The person is then absent from the gallery and can enroll again.
We reproduce both on a plaintext model of BioZKFHE's gallery encoding, as
detailed in Section~\ref{sec:gaps}.
These cases show why a valid off-chain verdict must be connected to the
actual gallery transition before it can authorize on-chain issuance.

We present \sysname{}, a protocol and composition model that binds the
deduplication query, gallery insertion and issued commitment to one
captured enrollment vector. Later authentication proves a match between
that commitment and a probe bound to the current challenge. The protocol
requires the 1:N backend to meet four conditions under malicious enrollees
and concurrent sessions, listed below. Our instance builds on BioZKFHE, which
supplies the gallery encoding and the verifiable similarity evaluation. A
certified enrollment station plays the honest client that BioZKFHE assumes,
and its existing committee additionally confirms gallery integrity under the
same corruption threshold. At authentication, raw features remain on the
user side and the contract verifies the match without an online matching
service. We also
analyze how the conditions map onto World ID's secret-sharing protocol, where
verdict soundness rests on a different trust model.
We make three contributions.
\begin{itemize}
\item \textbf{Secure enrollment composition conditions.} We identify four
  conditions a 1:N backend must meet: input binding, so that the queried,
  inserted and committed templates are one captured vector (I1);
  authenticated gallery transitions, so that an update appends exactly the
  declared entry (I2); verdict linearization, so that a verdict is used only
  if the gallery has not changed since it was computed (I3); and verdict
  soundness, so that a unique verdict is correct for the queried template and
  snapshot (I4). Counterexamples show the role of these conditions in the
  stated composition model.
\item \textbf{Construction and conditional security.} For a
  BioZKFHE-based backend, requests are bound to one captured vector when they
  are made, a session whose gallery snapshot has gone stale is rejected, and
  the committee confirms each merge of a new ciphertext into the gallery.
  Capture-bound authentication then connects the issued commitment to later
  use. A multi-session argument sorts failures into cryptographic ones,
  trusted-device assumptions, state invariants and biometric errors, and a
  separate conditional argument covers hiding which credential and device
  authenticated from an on-chain observer.
\item \textbf{Implementation and evaluation.} We implement the contracts,
  circuits and enrollment bridge, test them against the composition
  attacks, each paired with an honest run, and run the pipeline end to end on
  real LFW face features. A baseline with the same public inputs, in which the
  capture device attests to the match instead of the circuit proving it,
  measures the proving cost of keeping matching verifiable. Separate component
  tests run real BGV encryption through three histories of 16{,}384
  registrations without decoding errors at the tested checkpoints, while the
  end-to-end pipeline uses a simulated committee. LFW measurements and
  evasive re-enrollment on the Bosphorus database show how biometric accuracy
  and limits on enrollment attempts bound person-level uniqueness: under
  controlled changes of appearance, 1.8\% of re-enrollment attempts evade
  deduplication, and the end-to-end run records one missed duplicate.
\end{itemize}

\section{Background and Related Work}
\label{sec:background}

\begin{figure*}[t]
	\centering
	\input{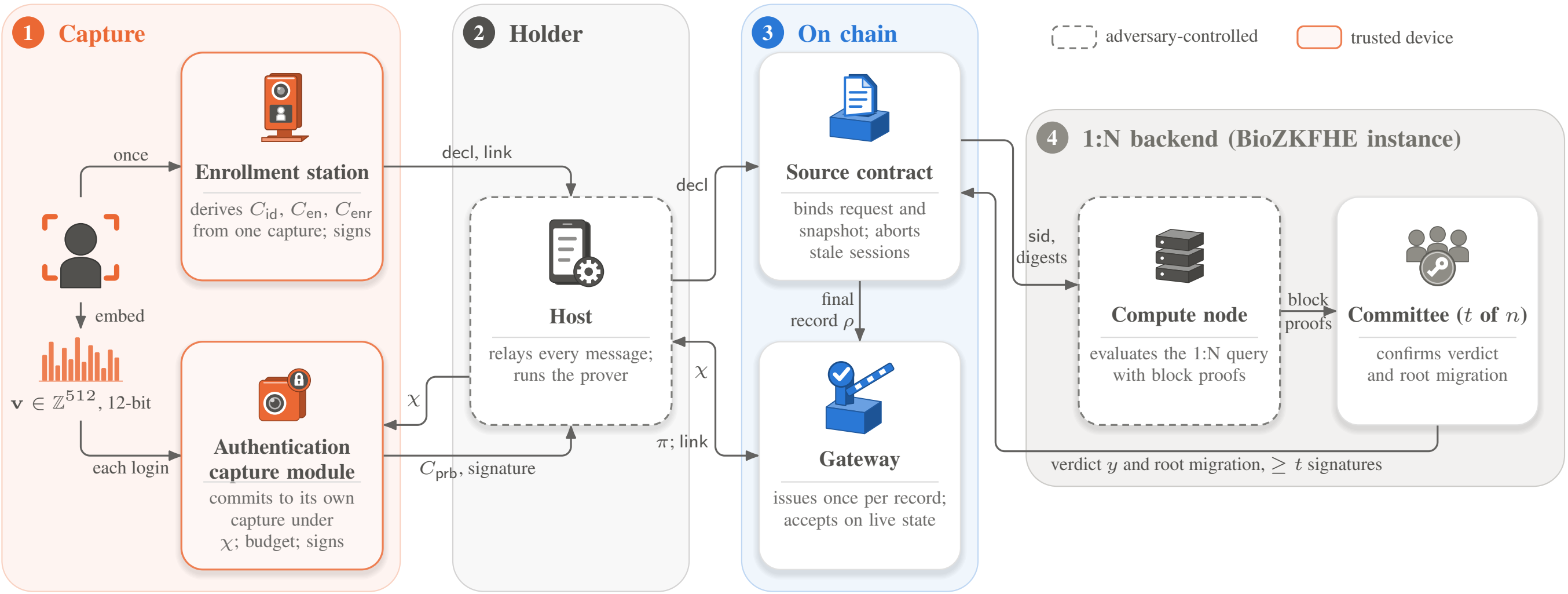}
\begin{tikzpicture}[x=1cm, y=1cm,
  card/.style={draw=black!12, rounded corners=7pt, fill=white,
    minimum width=2.3cm, minimum height=2.6cm, inner sep=0pt},
  keycard/.style={card, draw=bbOrange!85, line width=0.6pt, fill=bbOrange!5},
  advcard/.style={card, draw=bbInk!75, line width=0.6pt, dash pattern=on 2.2pt off 1.6pt},
  ctitle/.style={font=\bfseries\scriptsize, text=bbInk, align=center,
    text width=2.2cm, anchor=center, inner sep=0pt, nohyph},
  crole/.style={font=\fontsize{6}{7}\selectfont, text=bbInk!85, align=center,
    text width=2.1cm, anchor=north, inner sep=0pt, nohyph},
  phase/.style={font=\bfseries\footnotesize, anchor=west, text height=1.6ex, text depth=0.3ex},
  num/.style={circle, text=white, font=\bfseries\scriptsize, inner sep=1.6pt},
  flow/.style={-{Latex[round, length=1.8mm]}, draw=bbInk!90, line width=0.6pt, rounded corners=3pt},
  alab/.style={lab},
]
\def\ra{0}\def\rb{-3.3}\def\rh{-1.65}\def\xf{-2.3}
\newcommand{\shadow}[2]{%
  \foreach \k in {0.4,0.8,1.2,1.6,2.0}
    \fill[black, opacity=0.022, rounded corners=7pt+\k pt]
      ([xshift=-\k pt, yshift=-1.3pt-\k pt]#1-1.15,#2-1.3)
      rectangle ([xshift=\k pt, yshift=-1.3pt+\k pt]#1+1.15,#2+1.3);}
\newcommand{\content}[6]{%
  \node[inner sep=0pt] at (#1,#2+0.66) {\BioIcon{#3}{#4cm}};
  \node[ctitle] at (#1,#2-0.06) {#5};
  \draw[black!14, line width=0.4pt] (#1-0.9,#2-0.3) -- (#1+0.9,#2-0.3);
  \node[crole] at (#1,#2-0.4) {#6};}

\begin{scope}[on background layer]
  \draw[draw=bbOrange!30, fill=bbOrange!7, rounded corners=10pt] (-3.2,-4.85) rectangle (1.35,1.85);
  \draw[draw=bbInk!25, fill=bbInk!4, rounded corners=10pt] (1.95,-4.85) rectangle (4.65,1.85);
  \draw[draw=bbBlueC!30, fill=bbBlueC!7, rounded corners=10pt] (5.25,-4.85) rectangle (7.95,1.85);
  \draw[draw=bbGray!45, fill=bbGray!12, rounded corners=10pt] (8.5,-3.65) rectangle (14.65,0.65);
  \shadow{0}{\ra}\shadow{0}{\rb}\shadow{3.3}{\rh}\shadow{6.6}{\ra}\shadow{6.6}{\rb}\shadow{10.25}{\rh}\shadow{13.2}{\rh}
\end{scope}
\node[num, fill=bbOrange] at (-2.9,1.53) {1};
\node[phase, text=bbOrange] at (-2.68,1.53) {Capture};
\node[num, fill=bbInk] at (2.25,1.53) {2};
\node[phase, text=bbInk] at (2.47,1.53) {Holder};
\node[num, fill=bbBlueC] at (5.55,1.53) {3};
\node[phase, text=bbBlueC] at (5.77,1.53) {On chain};
\node[num, fill=bbGray] at (8.8,0.33) {4};
\node[phase, text=bbGray!80!black] at (9.02,0.33) {1:N backend (BioZKFHE instance)};

\node[inner sep=0pt] (face) at (\xf,-0.95) {\BioIcon{facecapture}{1.08cm}};
\draw[flow] (\xf,-1.55) -- node[alab, right, xshift=3pt] {embed} (\xf,-1.85);
\foreach \i/\h in {0/0.30,1/0.45,2/0.18,3/0.38,4/0.25,5/0.50,6/0.22,7/0.41,8/0.33,9/0.15,10/0.36,11/0.27}
  \fill[bbOrange!85, rounded corners=0.4pt] (\xf-0.44+0.075*\i,-2.45) rectangle ++(0.055,\h);
\draw[bbOrange!60, line width=0.4pt] (\xf-0.48,-2.45) -- (\xf+0.48,-2.45);
\node[alab, anchor=north] at (\xf,-2.5) {$\vv\in\mathbb{Z}^{512}$, 12-bit};
\draw[flow] (\xf,-0.35) |- node[alab, pos=0.75, above] {once} (-1.15,\ra);
\draw[flow] (\xf,-2.9) |- node[alab, pos=0.75, below] {each login} (-1.15,\rb);

\node[keycard] (es)   at (0,\ra)    {};
\node[keycard] (afe)  at (0,\rb)    {};
\node[advcard] (host) at (3.3,\rh)  {};
\node[card]    (src)  at (6.6,\ra)  {};
\node[card]    (gw)   at (6.6,\rb)  {};
\node[advcard] (cn)   at (10.25,\rh) {};
\node[card]    (com)  at (13.2,\rh) {};
\content{0}{\ra}{es}{0.89}{Enrollment station}{derives $\Cid$, $\Cen$, $\Cenr$ from one capture; signs}
\content{0}{\rb}{afe}{0.69}{Authentication\\capture module}{commits to its own capture under $\chi$; budget; signs}
\content{3.3}{\rh}{host}{0.81}{Host}{relays every message; runs the prover}
\content{6.6}{\ra}{source}{0.81}{Source contract}{binds request and snapshot; aborts stale sessions}
\content{6.6}{\rb}{gateway}{0.89}{Gateway}{issues once per record; accepts on live state}
\content{10.25}{\rh}{compute}{0.70}{Compute node}{evaluates the 1:N query with block proofs}
\content{13.2}{\rh}{committee}{0.80}{Committee ($t$ of $n$)}{confirms verdict and root migration}

\draw[flow] (es.east) -| node[alab, pos=0.25, above] {$\decl$, $\link$} (host.north);
\draw[flow] (afe.east) -| node[alab, pos=0.25, below] {$\Cprb$, signature} (host.south);
\draw[flow] ([yshift=-0.75cm]host.west) -- (1.65,\rh-0.75) -- node[alab, left] {$\chi$} (1.65,\rb+0.35) -- ([yshift=0.35cm]afe.east);
\draw[flow] ([yshift=0.35cm]host.east) -| (4.95,\ra) -- (src.west);
\node[alab, anchor=east] at (4.93,-0.2) {$\decl$};
\draw[flow, arrows={{Latex[round, length=1.8mm]}-{Latex[round, length=1.8mm]}}] ([yshift=-0.35cm]host.east) -| (4.95,\rb) -- (gw.west);
\node[alab, anchor=east] at (4.93,-3.15) {$\pi$; $\link$};
\node[alab, anchor=east] at (4.93,-2.35) {$\chi$};
\draw[flow] (src.south) -- node[alab, left, align=right] {final\\record $\rho$} (gw.north);

\draw[flow] ([yshift=0.3cm]src.east) -| (8.35,\rh+0.3) -- ([yshift=0.3cm]cn.west);
\node[alab, align=center] at (8.68,\rh+0.72) {$\sid$,\\digests};
\draw[flow] (cn.east) -- node[alab, above] {block} node[alab, below] {proofs} (com.west);
\draw[flow] (com.south) |- node[alab, pos=0.75, below] {verdict $y$ and root migration, $\ge t$ signatures} (8.15,\rh-1.6) -- (8.15,-0.3) -- ([yshift=-0.3cm]src.east);

\draw[bbInk!75, line width=0.6pt, dash pattern=on 2.2pt off 1.6pt, rounded corners=2pt] (8.8,1.35) rectangle (9.3,1.6);
\node[lab, anchor=west] at (9.35,1.475) {adversary-controlled};
\draw[bbOrange!85, line width=0.6pt, fill=bbOrange!5, rounded corners=2pt] (11.6,1.35) rectangle (12.1,1.6);
\node[lab, anchor=west] at (12.15,1.475) {trusted device};
\end{tikzpicture}
	\caption{Enrollment and authentication in the evaluated BGV instance, linked by the enrollment commitment. The upper path binds capture, private deduplication, gallery migration and issuance; the lower path binds a fresh probe to credential use. Panels group components by trust domain; dashed cards are adversary-controlled. Features are processed inside the attended $\ES$ at enrollment and remain on the user side at authentication, under the trust assumptions detailed in Section~\ref{sec:model}.}
	\label{fig:overview}
\end{figure*}

\subsection{Biometric Matching and Template Protection}

A 1:1 check compares one probe and template, whereas 1:N deduplication
searches a gallery. Matching errors therefore have different consequences
as the gallery grows. ISO/IEC 24745 specifies protection goals for stored
biometric references~\cite{iso24745}. We focus on whether protected query,
inserted template and credential originate from one capture; additional
source comparisons appear in Supplementary
Appendix~\ref{app:related-literature}.

\subsection{Private Deduplication and Enrollment}

BioZKFHE provides verifiable encrypted similarity evaluation for a
gallery snapshot under an honest-client assumption~\cite{xin2026biozkfhe}.
It encrypts the gallery under the BGV homomorphic encryption scheme and
packs templates into ciphertext blocks. A compute node evaluates each 1:N
query over these blocks and attaches a publicly verifiable proof for each
block, and a threshold committee decrypts only the verdict. We call our
instantiation on BioZKFHE the \emph{BGV instance}. World
ID's cited MPC protocol evaluates iris uniqueness under semi-honest,
honest-majority assumptions~\cite{bloemen2024ampc}. In humanitarian aid,
Wang et al. assume unique registration before private token
use~\cite{wang2023nyadid}; Janus studies the biometric registration step
with MPC, somewhat homomorphic encryption and trusted
hardware~\cite{edalatnejad2024janus}. xDup addresses private deduplication
across organizations using fuzzy private set
intersection~\cite{rausch2026xdup}. Our interface conditions concern the
link from a bound private verdict to an authorized gallery
transition and credential.

\subsection{Biometric Authentication and Credentials}

BAZKP, zkSABER, BioZero and Kothari et al. demonstrate different
zero-knowledge 1:1 biometric checks~\cite{mao2023bazkp,chinen2025zksaber,
lin2025biozero,kothari2026fingerprint}. Krul et al. systematize trust in
self-sovereign identity~\cite{krul2024sok}; CanDID combines decentralized
credentials with Sybil resistance~\cite{maram2021candid}; and a formal
personhood framework separates uniqueness from unlinkable
use~\cite{choudhuri2026pop}. For anonymous credential use, zk-creds proves
claims over existing identity sources and Semaphore uses membership proofs
and scoped nullifiers~\cite{rosenberg2023zkcreds,semaphore}. \sysname{}
links private enrollment deduplication, capture-bound authentication
and live credential state.

\subsection{World ID Context}

World ID combines device-based enrollment, private uniqueness checking
and credential use~\cite{worldwhitepaper}. Its documented phone-local Face
Auth~\cite{worldfaq2026} is distinct from the cloud-enclave
re-authentication design in the WIP-201/202 drafts~\cite{wip201,wip202}.
The latter binds a matching result to a credential and challenge; our
comparison concerns the registration-state interface and trust allocation.
Table~\ref{tab:related} summarizes the cited operations and input
assumptions, without auditing current deployments.

\section{System and Trust Model}
\label{sec:model}

\begin{table}[t]
	\centering
	\caption{Parties and trust.}
	\label{tab:parties}
	\footnotesize
	\begin{tabular}{@{}>{\raggedright\arraybackslash}p{2.1cm}>{\raggedright\arraybackslash}p{5.9cm}@{}}
		\toprule
		Party & Role and trust \\
		\midrule
		Holder & Enrolls, holds credentials, authenticates. \textbf{Malicious}: wants two credentials, someone else's credential, or more attempts. \\
		Holder's host &Phone or PC that relays messages and runs the prover. \textbf{Malicious}: may alter, replay or substitute anything it relays. \\
		Enrollment station ($\ES$) & Enrollment capture, extraction and derivation. \textbf{Trusted}, certified and versioned; attended by default; erases the plaintext after sealing the session. \\
		Authentication capture module ($\AFE$) & Authentication capture, probe commitment, budget, signature. \textbf{Trusted}, certified; does not evaluate the policy. \\
		1:N backend & Private deduplication and authenticated gallery updates under I1--I4. Inherits its own model: malicious compute node and threshold committee in the BGV instance; passive, honest-majority computation in the cited MPC design. \\
		Source contract & On-chain contract that binds each enrollment request to its gallery snapshot and aborts stale sessions. Public and deterministic. \\
		Gateway & On-chain contract that checks signatures, digests, proofs and state, and advances state atomically. Public and deterministic. \\
		Relying party & Sets policy, reads results. Curious; may collude to link users. \\
		Observer & Reads all public data. Curious; links across applications. \\
		Device authority & Registers, rotates and revokes device versions. Trust root. \\
		\bottomrule
	\end{tabular}
\end{table}

\subsection{Parties}

Figure~\ref{fig:overview} shows the off-chain capture and private
deduplication paths connected to the on-chain gateway.
Table~\ref{tab:parties} lists the parties and their trust responsibilities.
Two certified devices form the capture trusted computing base. The
\emph{enrollment station} ($\ES$) derives the protected query, insertion
input and enrollment commitment from one captured, quantized vector; the
protected inputs are ciphertexts in the BGV instance. Its default deployment
is attended: staff supervise liveness and limit attempts before a person
has a unique identifier. The \emph{authentication capture module} ($\AFE$)
captures and quantizes each probe, commits to it under the current
challenge, enforces an attempt budget and signs the result. Neither device
decides whether a probe matches. That decision is split between two other
parties. The holder's host, an ordinary phone or PC that we do not trust,
produces a zero-knowledge proof that the committed probe lies within the
registered threshold of the enrolled template; the on-chain gateway
verifies this proof before it accepts. The devices therefore only need to
capture honestly and sign, and a malicious host gains nothing from running
the prover, because soundness prevents it from proving a match that does
not exist.

\subsection{Division of Trust}

World ID's MPC protocol and verifiable encrypted evaluation protect templates
and establish verdict integrity under different adversary
models~\cite{bloemen2024ampc,xin2026biozkfhe}. I1--I4 specify their enrollment
interface conditions. In the BGV instance, enrollment confidentiality relies on BGV and fewer than $t$
corrupted committee members, who threshold-decrypt the verdict. The $\ES$
instantiates the honest client assumed by that backend. \sysname{} assigns
the same committee an additional integrity responsibility: members confirm
the gallery-root migration in the finalization message, as detailed
in Section~\ref{sec:rootmig}. Thus the party set and corruption threshold are
unchanged, but the committee's duty extends beyond confidentiality.
Everyday authentication relies on the user-side $\AFE$, the proof-system
setup and the gateway contract, with no online committee or matching server,
as detailed in Section~\ref{sec:defs}.

\subsection{Privacy Model}

At authentication, raw images and embeddings stay in the $\AFE$ and
the holder's host. At enrollment, the certified $\ES$ processes them, derives
the protected inputs and commitment, seals the session and erases the
plaintext without exposing it to its operator. This is the explicit
$\mathsf{NoRetain}$ assumption of Section~\ref{sec:defs}; only the
self-enrollment variant also keeps enrollment on the user side.
The $\ES$ receives the holder's identity commitment $\idc$, a hash of a
secret $s$ chosen by the holder, but never $s$ itself.
Because the holder's host runs the prover of the matching proof, it
receives the opening of the probe commitment, namely the quantized probe
vector $\vp$ and its commitment randomness $\rprb$. Our privacy claim
therefore does not hide the probe from a malicious host; hiding it would
require an attested prover reached over an encrypted channel.

\sysname{} offers two authentication profiles that differ in what
appears on chain. In the \emph{public profile}, every authentication
exposes the holder address, the credential handle, the enrollment
commitment, the device key and a per-device counter, so an observer can
link all authentications of one credential. The \emph{unlinkable profile},
as detailed in Section~\ref{sec:unlink}, keeps these handles in the
zero-knowledge witness. In their place, the holder derives from $s$ a
\emph{nullifier} for each application, a pseudonym that stays fixed within
that application and differs across applications. Because $s$ never
reaches the $\ES$, the station cannot compute these nullifiers to link the
holder's authentications. Issuance calldata, the nullifiers themselves and
funding edges remain observable. Section~\ref{sec:eval} reports the
trace audit, and Supplementary Appendix~\ref{app:privacy} states the
conditional observer model.

\section{Security Definitions}
\label{sec:defs}

\subsection{Syntax}

The backend interface binds the protected query and insertion inputs,
the queried snapshot and the finalization record to one enrollment
commitment. Below we give the syntax of this interface for the BGV
instance, in which the protected inputs are ciphertexts. Supplementary
Appendix~\ref{app:world-mpc} maps the same conditions to an MPC
backend, in which the protected inputs are secret shares.
A feature vector $\vv \in \mathbb{Z}^L$ has coordinates bounded by
$|v_i| \le \Gamma$. The 1:1 relation uses squared Euclidean distance
$\dist(\cdot,\cdot)$ and a policy threshold $\tauauth$; the backend uses its
own similarity and threshold $\taudup$, and the duplicate predicate
$\Dup(\vv,G)$ is true when the similarity between $\vv$ and some entry of
the gallery $G$ reaches $\taudup$.

\emph{Enrollment.} The source contract opens a session $\sid$ whose context
binds the current gallery snapshot $s$ and the number $N$ of templates
enrolled so far. The $\ES$ captures $\vv$ from a person $p$ and derives
three outputs from this single vector, so that the queried, inserted and
committed templates coincide. The \emph{query ciphertext}
$\Cid=\Enc(\vv)$ is what the backend compares against the snapshot to
decide whether $p$ is already enrolled. The \emph{write ciphertext} $\Cen$
encodes $\vv$ at slot $N$; the gallery stores templates in slots
numbered in enrollment order, so slot $N$ is the position reserved for the
next template. After a unique verdict, $\Cen$ is added to
the gallery, so that later enrollments are checked against $p$ as well. The
\emph{enrollment commitment}
\begin{equation}
\Cenr=\ComH(\texttt{enroll},\hP,0,\vv;\renr)
\label{eq:cenr}
\end{equation}
is a hiding commitment to $\vv$ under the policy digest $\hP$, with
randomness $\renr$. Its third field is a context field, which binds a
probe commitment to its authentication challenge; it is zero here because
enrollment answers no challenge. The issued credential carries
$\Cenr$, and every later authentication proves a match against its
opening, which ties credential use to the enrolled template without
revealing it. The $\ES$ seals a pending record before releasing any output and signs the
declaration
\begin{equation}
\decl=\mathsf{Sig}_{\ES}\bigl(\sid,H(\mathsf{ctx}),N,H(\Cid),H(\Cen),\slot,\Cenr\bigr).
\label{eq:decl}
\end{equation}
Here $\mathsf{ctx}$ is the session context, $H$ is a collision-resistant
hash that replaces each ciphertext by a short digest, because BGV
ciphertexts are too large to sign or store on chain, and $\slot$ is the
write position declared for $\Cen$. The declaration is the $\ES$'s
statement, made before the backend runs, that the three outputs come from
one capture and belong to session $\sid$ at count $N$. The source contract
accepts the session's inputs only with a valid declaration from a
registered $\ES$, so a host that substitutes any output after signing is
rejected.
The backend identifies $\Cid$ against $s$ and finalizes the session with a
record $\rho=(\sid,\mathsf{ctx},s,y,H(\Cid),H(\Cen),\slot)$, where
$y\in\{\texttt{unique},\texttt{dup}\}$, and updates the gallery. The gateway
issues a credential by consuming $\rho$ atomically together with the
$\ES$ link signature
\begin{equation}
\link=\mathsf{Sig}_{\ES}\bigl(H(\rho),\Cenr,\hP,\text{holder},\mathsf{ver}\bigr),
\label{eq:link}
\end{equation}
where $\mathsf{ver}$ collects the model, quantization and template
versions. The two $\ES$ signatures guard the two ends of the backend.
Whereas $\decl$ binds the inputs to the session before evaluation, $\link$
binds the verdict to the credential after it. The record $\rho$ carries
only ciphertext digests and not $\Cenr$, so without $\link$ a valid unique
record could be presented together with an unrelated commitment, for
instance one to another person's vector. The $\ES$ therefore signs $\link$
only when $\rho$ matches the pending record it sealed, and takes $\Cenr$
from that record rather than from the host. The credential thus carries
the commitment derived from the deduplicated vector and is bound to its
holder.

\emph{Authentication.} The gateway issues a challenge $\chi$. The
$\AFE$ captures $\vp$ and commits to it under the challenge digest
$\hchi=\HF(\chi)$, placed in the context field:
\begin{equation}
\Cprb=\ComH(\texttt{probe},\hP,\hchi,\vp;\rprb).
\label{eq:cprb}
\end{equation}
The probe commitment fixes the probe before any proof is produced and ties
it to this challenge, so a probe captured for one challenge cannot be
reused for another. The $\AFE$ then consumes one unit of its attempt budget
and signs the capture statement
\begin{equation}
\sigma_{\AFE}=\mathsf{Sig}_{\AFE}\bigl(\Cprb,\hchi,\hP,\mathsf{dev},\mathsf{ctr},\mathsf{exp}\bigr),
\label{eq:capsig}
\end{equation}
where $\mathsf{dev}$ is the model and firmware measurement, $\mathsf{ctr}$
a per-device counter and $\mathsf{exp}$ an expiry time. The signature
attests that $\Cprb$ commits to a fresh capture made by a certified device
for this challenge and policy; the counter lets the gateway reject a
replayed statement, and the expiry bounds how long it remains valid.
Without this signature, the host could commit to a vector of its own
choice, such as the enrolled template itself, and prove a distance of
zero. With the probe thus fixed by a trusted device, the holder's host
produces a zero-knowledge proof $\pi$ that it knows openings
of~\eqref{eq:cenr} and~\eqref{eq:cprb}, namely the two vectors and their
randomness, that satisfy the match relation
\begin{equation}
\dist(\vv,\vp)=\sum_{i=1}^{L}(v_i-v'_i)^2\le\tauauth .
\label{eq:dist}
\end{equation}
The commitments, $\hP$, $\hchi$ and $\tauauth$ are public inputs, while
the openings stay in the witness, so the gateway learns that the probe
matches without seeing either vector. The gateway verifies $\pi$ and
accepts after reading the current state of the credential, the policy and
the device version, as detailed in Section~\ref{sec:auth}.

\emph{Design choice.} Three routes can check a fresh probe against an
enrolled template. First, BioZKFHE treats 1:1 verification as a special
case of its encrypted evaluation, comparing the probe against a single
gallery slot; each authentication then needs the compute node and an
online committee to decrypt the result. Second, World ID's draft
re-authentication design~\cite{wip201,wip202} runs a closed matching
engine inside a cloud enclave and uses a zero-knowledge proof to bind the
enclave's signed result to a credential and challenge; the raw face leaves
the user side, and matching correctness rests on the enclave and its
provider. Third, zero-knowledge 1:1 checks such as
BioZero~\cite{lin2025biozero,kothari2026fingerprint} let the holder prove a distance
bound against a committed template. We take the third route, so that
everyday authentication involves only the $\AFE$, the holder's host and
the gateway, keeps raw features on the user side and makes the comparison
itself publicly verifiable. What \sysname{} adds is the binding around the
proof: the commitment that $\pi$ opens is the one that $\link$ ties to the
deduplicated vector, and the probe it opens is fixed by a certified
capture under the current challenge. Supplementary
Appendix~\ref{app:comparison} compares the trust placed in
\sysname{} and in the World ID draft in more detail.

\emph{Lifecycle.} After issuance, the state that authentication reads
keeps changing. The holder or the relying party can revoke a credential,
the device authority can rotate or revoke a device version, and a relying
party can migrate its policy or model. Only the legitimate owner of each
object issues these operations, but the adversary chooses when each one
takes effect relative to pending challenges and authentications. The
gateway must therefore decide on the state current at acceptance.

\subsection{Adversary Model}

The adversary controls all holders and their hosts, which run the
provers, as well as the network, relayers and transaction ordering. It may
open any number of concurrent enrollment and authentication sessions,
including requests built directly against the source contract rather than
through an $\ES$, and it receives every opening handed to a holder's host.
It may have the $\ES$ and the $\AFE$ capture any person in
$\mathcal{P}_{\mathsf{adv}}$, the set of persons it can present to a
device, and it chooses when each lifecycle operation takes effect. Inside
the backend it corrupts whatever the backend's own model allows, which in
the BGV instance is the compute node and up to $t-1$ committee members.

The adversary does not control the keys of unrevoked devices, the
derivation inside a device, or the captures of honest users, who form the
set $\mathcal{P}_{\mathsf{hon}}$, and it cannot break the assumptions on
the cryptographic primitives. Corrupted devices and a malicious owner of a
credential, device or policy form separate threat profiles that this paper
does not cover. Retention is a separate assumption that concerns privacy
only. Under $\mathsf{NoRetain}$, the $\ES$ erases the plaintext image and
vector after sealing the session, and its operator never receives them. If
retention fails, that user's template leaks, but neither game below is
affected. A failure of the derivation inside the $\ES$ is more serious,
because it can produce a Sybil identity.

To state the winning conditions, the challenger keeps records of what
actually happened. These records define the games only; no party reads
them. The enrollment log $L_{\mathsf{enroll}}$ holds each $\ES$ derivation
$(\sid,p,\vv,\Cid,\Cen,\Cenr,\renr)$, and the capture log
$L_{\mathsf{capture}}$ holds each $\AFE$ capture. The \emph{linearization
point} of an operation is the transaction in which it takes effect on
chain; for an enrollment session, it is the transaction in which the
session finalizes and writes the gallery. The term comes from the theory
of concurrent objects~\cite{herlihy1990linearizability}. An operation such
as an enrollment session spans several transactions and off-chain
computation, but its effect is ordered as if it happened at this single
point, and the adversary, who orders transactions, chooses how these
points interleave. The ideal gallery $\Gid$ is the
map $\slot\mapsto\vv$ that the gallery should hold: at the linearization
point of each session $\rho$ that finalizes as unique, it gains the entry
$(\slot_\rho,\vv_\rho)$. Finally, the challenger holds the FHE secret key,
so the plaintext $\Pt(C)$ of any ciphertext $C$ is well defined.

\subsection{Multi-Session Uniqueness Game}

The uniqueness game asks whether the adversary can obtain a credential
that honest enrollment would not have produced. It may run many sessions
concurrently and wins in one of three ways.

\begin{definition}[Game $\Gunq$]
\label{def:gunq}
The adversary wins if, at the end of the experiment, one of the following
holds.
\begin{itemize}
\item[W1] \emph{Commitment provenance}: some issued credential with final record $\rho$
  has no record in $L_{\mathsf{enroll}}$ whose
  $(\sid,H(\Cid),H(\Cen),\slot)$ equals that of $\rho$ and whose $\Cenr$
  equals the credential's commitment.
\item[W2] \emph{Gallery integrity}: at some time, the decrypted and decoded gallery
  differs from $\Gid$ in some slot $j<N$, or some slot $j\ge N$ is non-empty.
\item[W3] \emph{Verdict linearization}: some issued credential with session
  $\rho$ and vector $\vv_\rho$ has, immediately before $\rho$'s atomic insertion, an entry
  $g\in\Gid$ with $\Dup(\vv_\rho,\{g\})$.
\end{itemize}
The \emph{uniqueness advantage} of the adversary is its probability of
winning, $\Adv^{\mathsf{unq}}=\Pr[\mathrm{W1}\vee\mathrm{W2}\vee\mathrm{W3}]$.
\end{definition}

Each condition describes a different way in which a credential can
escape deduplication. W1 is a credential whose commitment does not come
from the vector that was deduplicated, which is the gap that $\link$
closes. W2 is a gallery that no longer equals the sequence of unique
enrollments, so that later queries compare against the wrong templates.
W3 is a credential issued to a person who was already in the gallery when
the session took effect.

W3 is evaluated against the gallery at finalization, not against the
snapshot the backend queried. The difference between the two is exactly the
concurrency that the snapshot race exploits, as detailed in
Section~\ref{sec:gaps}. It is also why the game must range over many
sessions: a game with a single acceptance cannot express it.

\subsection{Capture-Bound Authentication Game}

The authentication game asks whether the adversary can make the
gateway accept a submission that is not a genuine match between an issued
credential and a fresh capture. It wins if an accepted submission violates
any of four conditions.

\begin{definition}[Game $\Gauth$]
\label{def:gauth}
The adversary wins if the gateway accepts a submission while one of the
following fails.
\begin{itemize}
\item[U1] \emph{Credential origin}: the credential is in the issuance log,
  its $\Cenr$ matches an $L_{\mathsf{enroll}}$ record, and its record $\rho$
  was consumed exactly once.
\item[U2] \emph{Capture origin}: $(\Cprb,\chi)$ appears in
  $L_{\mathsf{capture}}$.
\item[U3] \emph{Whole match}: bounded openings of $\Cenr$ and $\Cprb$ exist,
  agree with the logged vectors and satisfy $\dist(\vv,\vp)\le\tauauth$.
\item[U4] \emph{Live state}: credential, policy, device version, holder and
  freshness are valid and consistent at the linearization point, and success
  advances state atomically.
\end{itemize}
The \emph{authentication advantage} of the adversary is its probability of
winning, $\Adv^{\mathsf{auth}}=\Pr[\neg\mathrm{U1}\vee\neg\mathrm{U2}\vee\neg\mathrm{U3}\vee\neg\mathrm{U4}]$,
where $\neg\mathrm{U}i$ denotes that some accepted submission violates
U$i$.
\end{definition}

Each condition corresponds to one of the bindings in the syntax. U1
requires that the credential come from a consumed enrollment record,
through $\link$. U2 requires that the probe come from a certified capture
for this challenge, through $\sigma_{\AFE}$. U3 requires that the two
committed vectors actually match, through $\pi$. U4 requires that the
gateway decide on the state current at acceptance, as the lifecycle
operations demand.

\subsection{Person-Level Sybil and Impersonation}

The two games speak about vectors, but the goals of the system speak
about people. Two further events capture these goals.
$\mathsf{SYB}$ is the event that some person in
$\mathcal{P}_{\mathsf{adv}}$ holds two issued credentials, and
$\mathsf{Imp}$ is the event that the credential $\cred_h$ of a fixed honest
target user is accepted on a capture that is not that user's. Cryptography
alone cannot exclude either event, because both also depend on how well
the biometric matcher separates people.

Two biometric quantities capture this dependence. The miss rate $\fnm$
bounds the probability, over any reachable history, that a new enrollment
capture of a person fails to match that person's earlier gallery entry;
$q_E$ bounds the number of enrollment captures summed over all adversarial
persons, or the capture count of the target person when one person is
fixed. The false-match bound $\fmax$ bounds the probability that one
capture of an adversarial person falls within $\tauauth$ of an honest
user's template, and $B$ bounds the captures the adversary can expose
within one budget scope. Both $\fnm$ and $\fmax$ are taken under
adversarial presentation; the FNMR and FAR measured on a dataset are
non-adversarial reference values for them, not substitutes.

Theorem~\ref{thm:main} in Section~\ref{sec:security} bounds
$\mathsf{SYB}$ by $\Adv^{\mathsf{unq}}+q_E\,\fnm$ and, when the $\AFE$
enforces its capture budget, bounds $\mathsf{Imp}$ by
$\Adv^{\mathsf{auth}}+B\,\fmax$. The first bound shows that enrollment
also needs an attempt budget. Because a person cannot be counted before
deduplication, we treat $q_E$ as a deployment parameter, which an attended
station can enforce.

\subsection{Unlinkability Game}

The unlinkability game asks whether an observer of the public chain can
tell which of two honest holders authenticated in an application. It
concerns the unlinkable profile, and the observer only reads the chain
instead of controlling holders' hosts.

\begin{definition}[Game $\Glink$]
\label{def:glink}
The observer receives the entire public chain transcript, including
issuance calldata, all nine public inputs of each authentication, proof
bytes, transaction senders and funding edges. Two honest candidates use the
same policy and valid tree roots, and each can authenticate successfully;
their hosts, devices and identity secrets remain private. The observer
chooses a fresh scope in which neither candidate has authenticated, a
candidate $b\in\{0,1\}$ is chosen uniformly at random and authenticates
once in that scope, and the observer outputs a guess $b'$. The observer may
query either candidate in other scopes, but not in the challenge scope,
before or after the challenge. The \emph{unlinkability advantage} of the
observer is $\Adv^{\mathsf{link}}=|\Pr[b'=b]-1/2|$.
\end{definition}

The game excludes corrupting a candidate and revoking selectively so that
the two candidates differ in eligibility. The accounts, funding, timing,
ordering and outcome of the challenge transaction must also have the same
distribution whichever candidate is chosen. This equal-metadata condition
does not require equal nullifiers, probe commitments or proof bytes, since
their indistinguishability is exactly what the profile must achieve. Under
the zero-knowledge, commitment-hiding and pseudorandomness assumptions
stated in Supplementary Appendix~\ref{app:privacy},
$\Adv^{\mathsf{link}}$ is negligible. The claim is conditional and holds
at the protocol level. It excludes identification through the network,
compromise of a holder's host or device, and metadata that differs between
the candidates, and it does not promise anonymity from fresh addresses
alone.

\section{Conditions for Secure Enrollment Composition}
\label{sec:gaps}

\subsection{The Four Conditions}

The enrollment contract must connect an off-chain verdict to the vector
written into the gallery and the record consumed at issuance. A bound
snapshot identifies what was queried; additional checks establish what
was committed and whether the verdict was still applicable. We call a
backend, together with its source contract and enrollment station,
\emph{composable under malicious enrollees} if it meets the following four
conditions. They concern logical templates and gallery state, and their
realization depends on the backend representation and adversary model;
Section~\ref{sec:construction} realizes them for the BGV instance.
Authentication is governed separately by the conditions
U1--U4, as discussed in Section~\ref{sec:defs}.
\begin{itemize}
\item[I1] \emph{Input binding.} (a) At request time, the source binds
  the session and policy to query and insertion inputs derived by a
  registered $\ES$ from one captured template. (b) Authenticated input
  references identify the actual protected objects consumed by the backend;
  their template consistency is established by trusted derivation or a
  proof, not by hashes alone. (c) Issuance requires an $\ES$ link from the
  matching pending record to the final record and credential commitment.
\item[I2] \emph{Authenticated gallery transition.} (a) The source
  assigns the next slot from the authenticated count at write time.
  (b) An accepted transition appends the declared insertion template at that slot,
  preserves earlier entries and leaves later slots empty. (c) The write
  and unique finalization share one linearization point: each occurs
  exactly once with the other, and all other sessions leave the gallery
  unchanged.
\item[I3] \emph{Verdict linearization.} A session may finalize as unique only
  if no write occurred after the snapshot it queried; otherwise it aborts and
  the person retries with a new session and a new capture. Under I2(a), ``no
  write'' is equivalent to ``the count is unchanged'', which gives a stale-slot
  abort.
\item[I4] \emph{Verdict soundness.} Under the backend's stated
  corruption model, a unique verdict is sound for the bound query template
  and logical gallery snapshot, except with negligible probability over
  the permitted adaptive session history.
\end{itemize}

\subsection{Counterexamples}

We test each premise by removing it and exhibiting an adversary that
then wins, and Table~\ref{tab:counterexamples} summarizes the resulting
counterexamples, each with the removed premise, the adversary's action and
the violated winning condition. The table groups them by the premise they
remove. Besides the enrollment premises, it lists the two authentication
premises that the same chain evaluation tests, so that the evidence for
all executed counterexamples appears in one place. The E-series runs on a
local chain against our gateway with synthetic vectors. The A-series and
the dropped merge run on a plaintext model of BioZKFHE's packed gallery
with blocks of twelve templates; they do not execute its block proofs,
threshold decryption or committee protocol.

\begin{table*}[t]
\centering
\caption{Counterexamples, grouped by the premise they remove: removed premise, adversary action, violated condition and evidence.}
\label{tab:counterexamples}
\footnotesize
\begin{tabular}{@{}l>{\raggedright\arraybackslash}p{3.2cm}>{\raggedright\arraybackslash}p{6.6cm}l>{\raggedright\arraybackslash}p{3.0cm}@{}}
\toprule
ID & Removed premise & Adversary action & Violates & Status (run) \\
\midrule
\multicolumn{5}{@{}l}{\emph{Enrollment premises (I1--I4)}} \\
E1 & Issuance checks $\ES$ link (I1c) & Present a valid unique receipt with an unrelated commitment, e.g.\ of another person's vector & W1 & Executed, chain$^{\dagger}$ \\
A1 & Request-time input binding (I1a) & Query $\Enc(\vv_A)$ but write $\Enc(\mathbf{0})$; enroll again later & W1 & Executed, model (R-BRIDGE-20261025) \\
A2 & Slot from authenticated count (I2a) & Declare $N=1$ while $N=10$; template is added onto user 1's slot & W2 & Executed, model; misidentification, not Sybil \\
A3 & Write-time slot, serialization & Two honest sessions prepared at the same $N$ both write slot $N{+}1$ & W2 & Executed, model \\
A3$'$ & Verdict linearization (I3) & Same person opens two sessions on one snapshot; both judged unique & W3 & Executed, model (R-A3P-20261027) \\
Drop & Verifiable root migration (I2b) & Compute node drops $\Cen$ at a merge; caller supplies the next root & W2 & Executed, model (R-ROOTMIG-20261027) \\
\midrule
\multicolumn{5}{@{}l}{\emph{Authentication premises (U1--U4)}} \\
E2 & $\AFE$ derives its own probe & Signing API accepts a host-built $\Cprb$; host sets $\vp=\vv$ and proves distance 0 & U2 & Executed, chain$^{\dagger}$ \\
E3 & Live state at acceptance & Gateway caches authorization at challenge time; owner revokes before submission & U4 & Executed, chain$^{\dagger}$; our gateway rejects \\
\bottomrule
\end{tabular}\\[2pt]
\raggedright\footnotesize $^{\dagger}$\,\texttt{build/provenance-evaluation/result.json}; R-PROV-20261028 (local synthetic evaluation).
\end{table*}

The enrollment rows also show that each condition is independent of the
other three: for each I$i$ there is a backend that meets the other three,
violates I$i$, and admits an adversary that wins $\Gunq$ with probability
close to one. E1 and A1 remove parts of I1; A2, A3 and the dropped merge
remove parts of I2; and A3$'$ removes I3 while I1, I2 and I4 hold. For I4
the witness is trivial: a backend that answers unique without comparing
satisfies I1--I3 and lets the same person enroll twice. Independence is a
statement about this set of conditions only; it is not a lower bound over
all designs.

Three cautions apply. First, the counterexamples show that a premise is
needed for the composition; they do not show that a strong signature
baseline fails, since a trusted matcher that signs correct results also
excludes E1--E3. Second, A1--A3 need a malicious enrollment client, which
BioZKFHE's model excludes; they are conditions the composition must add, not
vulnerabilities of BioZKFHE. Third, A2 does not let the adversary escape
deduplication, because inner products are linear; its effect is
misidentification and a wrong count.

\subsection{The Snapshot Race}

A3$'$ deserves separate attention because every component is correct in it.
Assume I1 and I2: every write takes a fresh slot at write time, and its
content is bound to the declared digest. BioZKFHE binds the gallery snapshot
into the session context, and its block proofs authenticate each block against the
snapshot resolved from that context. A person opens two sessions
$\rho_1,\rho_2$ on the same snapshot $s$, which contains no vector of theirs.
Both verdicts are unique; $\rho_1$ finalizes and writes; $\rho_2$ finalizes
and writes another fresh slot. At $\rho_2$'s finalization, $\Gid$ already
contains $\vv_1$ and $\Dup(\vv_2,\{\vv_1\})$ holds, so W3 occurs while I1, I2
and I4 hold.
This is an abstract backend variant that permits trusted re-encoding of the
same captured vector into a fresh write slot. It is not obtained by merely
removing a check from our fixed-slot ciphertext implementation: that
implementation additionally binds the declaration to the snapshot count.

\begin{figure}[t]
\centering
\begin{tikzpicture}[x=0.9cm,y=1cm,
  sess/.style={line width=3.2pt, bbBlueC!35},
  ev/.style={circle, fill=bbBlueC, inner sep=1.3pt},
  evlab/.style={lab, anchor=south},
]
\def\tA{4.1}\def\tB{6.4}
\node[lab, anchor=east] at (0.15,2.55) {gallery\\count};
\node[lab, anchor=east] at (0.15,1.6) {session $\rho_1$};
\node[lab, anchor=east] at (0.15,0.75) {session $\rho_2$};

\draw[bbInk, line width=0.8pt] (0.3,2.4) -- (\tA,2.4) -- (\tA,2.7) -- (8.2,2.7);
\draw[bbRed, line width=0.8pt, dash pattern=on 2pt off 1.5pt] (\tB,2.7) -- (\tB,3.0) -- (8.2,3.0);
\node[lab, anchor=south west] at (0.3,2.4) {$N$};
\node[lab, anchor=south west] at (\tA,2.7) {$N{+}1$};
\node[lab, anchor=south west, text=bbRed] at (\tB,3.0) {$N{+}2$ (no I3)};

\draw[sess] (0.8,1.6) -- (\tA,1.6);
\node[ev] (o1) at (0.8,1.6) {}; \node[evlab] at (o1.north) {open on $s$ ($N$)};
\node[ev] (i1) at (2.75,1.6) {}; \node[evlab] at (i1.north) {$y$ on $s$: unique};
\node[ev] (f1) at (\tA,1.6) {}; \node[lab, anchor=north] at (f1.south) {finalize,\\write slot $N$};

\draw[sess] (1.3,0.75) -- (\tB,0.75);
\node[ev] (o2) at (1.3,0.75) {}; \node[lab, anchor=north] at (o2.south) {open on $s$ ($N$)};
\node[ev] (i2) at (3.3,0.75) {}; \node[lab, anchor=north] at (i2.south) {$y$ on $s$: unique};
\node[ev] (f2) at (\tB,0.75) {};

\node[lab, anchor=south, text=bbRed] at (\tB,0.95) {snapshot semantics:\\write slot $N{+}1$,\\second credential (W3)};
\node[lab, anchor=north] at (\tB,0.55) {with I3: count $\ne N$\\$\Rightarrow$ AbortedStaleSlot};

\draw[bbInk!50, line width=0.3pt, dash pattern=on 1pt off 1pt] (\tA,0.2) -- (\tA,2.4);
\draw[bbInk!50, line width=0.3pt, dash pattern=on 1pt off 1pt] (\tB,1.75) -- (\tB,2.7);
\draw[flow] (0.3,-0.35) -- (8.3,-0.35) node[lab, anchor=north east] {time};
\node[lab, anchor=north] at (\tA,-0.35) {$t_1$};
\node[lab, anchor=north] at (\tB,-0.35) {$t_2$};
\end{tikzpicture}
\caption{Snapshot race A3$'$. Two sessions of the same person open on snapshot $s$ with count $N$, and both verdicts on $s$ are unique. Under snapshot semantics the second session also writes, yielding two credentials despite correct queries and writes; with I3 the changed count aborts the second session.}
\label{fig:a3p}
\end{figure}
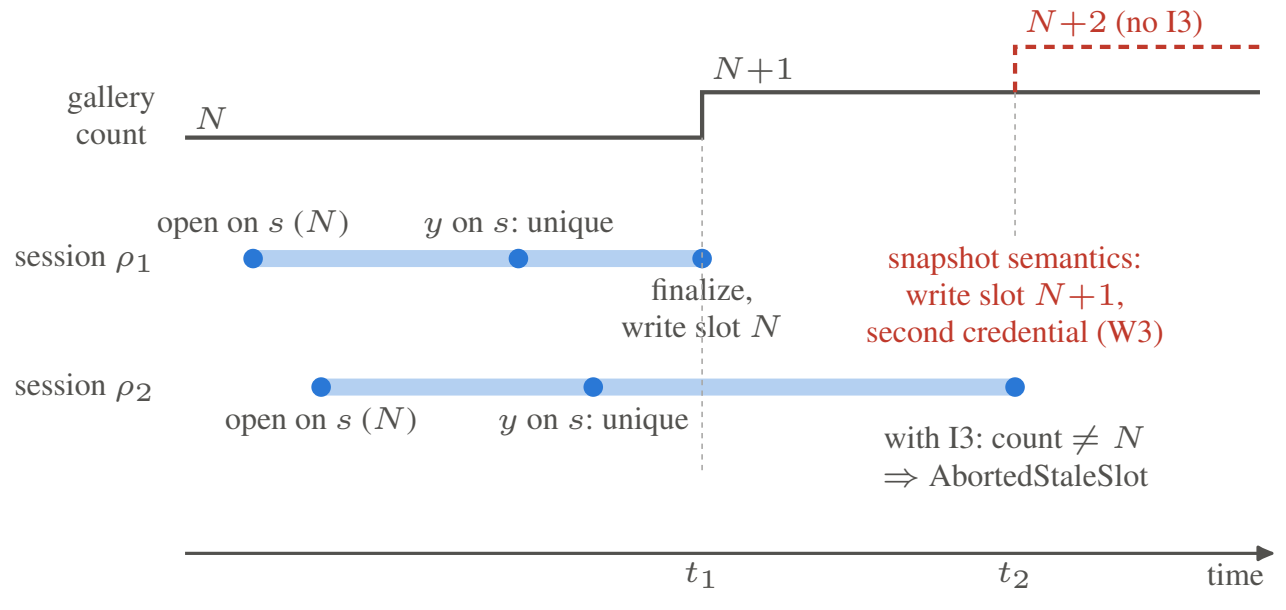

We execute A3$'$ after ten honest enrollments. Table~\ref{tab:a3p} reports
the outcome under three verdict semantics. With the gallery at finalization,
the second session is a duplicate. With BioZKFHE's snapshot semantics, the person is accepted twice.
With snapshot semantics and a stale abort, the second session aborts and its
retry is a duplicate. The abort also stops one of two different people who
enroll concurrently; that person retries successfully, so the cost is
availability, not security. Figure~\ref{fig:a3p} illustrates the timeline.

\begin{table}[t]
\centering
\caption{A3$'$ under three verdict semantics (plaintext gallery model with blocks of twelve templates, after ten honest enrollments).}
\label{tab:a3p}
\footnotesize
\setlength{\tabcolsep}{4pt}
\begin{tabular}{@{}>{\raggedright\arraybackslash}p{2.3cm}l>{\raggedright\arraybackslash}p{1.5cm}>{\raggedright\arraybackslash}p{2.0cm}@{}}
\toprule
Verdict semantics & A, 1st & A, 2nd & Two people \\
\midrule
Gallery at finalization & unique & dup & both pass \\
Session snapshot & unique & \textbf{unique} & both pass \\
Snapshot + stale abort (I3) & unique & abort; retry dup & 2nd aborts; retry passes \\
\bottomrule
\end{tabular}
\end{table}

\subsection{Existing Backends}

Table~\ref{tab:conditions} assesses three configurations against I1--I4.
BioZKFHE's published protocol binds $h_{\mathsf{id}}=H(\mathsf{ctx}\,\|\,\Cid)$
but not $\Cen$, and its count, slot and gallery root are supplied by the
client or caller; its public repository is an experiment reproduction package
without an enrollment path or an implemented gallery update, so I3 is
unspecified. We implement the bound extension ourselves, as detailed in
Section~\ref{sec:construction}. World ID's published MPC protocol studies
private iris deduplication under semi-honest, honest-majority
assumptions~\cite{bloemen2024ampc}. Supplementary
Appendix~\ref{app:world-mpc} maps I1--I4 to a proposed World ID-inspired
MPC instance: input binding and serialized issuance remain necessary,
while verdict integrity depends on the backend's adversary model.
This interface comparison does not establish properties of the deployed
World ID service. A ``no'' denotes an unmet composition condition for malicious enrollees
outside BioZKFHE's model, not a backend flaw.

\begin{table}[t]
\centering
\caption{Interface conditions for three backend configurations. NE marks an interface condition not established by the sources inspected, rather than an absent deployment property. The World ID comparison uses the cited MPC protocol, not a full service audit.}
\label{tab:conditions}
\footnotesize
\setlength{\tabcolsep}{4pt}
\begin{tabular}{@{}>{\raggedright\arraybackslash}p{0.5cm}>{\raggedright\arraybackslash}p{2.3cm}>{\raggedright\arraybackslash}p{2.3cm}>{\raggedright\arraybackslash}p{2.4cm}@{}}
\toprule
 & BioZKFHE~\cite{xin2026biozkfhe} & \sysname{} on BioZKFHE & World ID MPC~\cite{bloemen2024ampc} \\
\midrule
I1 & no: $\Cen$ unbound & yes, trusting $\ES$ derivation & NE \\
I2 & no: client-supplied $N$, slot, root & yes (Section~\ref{sec:rootmig}) & NE \\
I3 & unspecified & yes: stale abort & NE \\
I4 & yes, plus two implementation requirements & yes & semi-honest, honest majority \\
\bottomrule
\end{tabular}
\end{table}

\section{Concrete Construction of \sysname{}}
\label{sec:construction}

This section makes each step of the syntax in Section~\ref{sec:defs}
concrete, in the same order: enrollment in the first two subsections,
issuance and lifecycle operations in the third, authentication in the
fourth and the unlinkable profile in the last. It realizes the interface
conditions I1--I4 for the BGV instance and connects off-chain biometric
processing to two on-chain decisions: issuance from a finalized enrollment
record and authentication against the commitment that the record carries.
On the enrollment side, we extend BioZKFHE's session handling with a source
contract that binds the inputs, checks the gallery state and migrates the
gallery, and with a finalization adapter that exposes each finalized record
to the gateway. On the authentication side, the capture-bound proof links
the issued commitment to the current challenge.

The BGV instance realizes the four conditions as follows. I1 relies on
digests of canonically serialized ciphertexts, the
declaration~\eqref{eq:decl} and the link signature~\eqref{eq:link}. I2
relies on the committee-confirmed gallery update, as detailed in
Section~\ref{sec:rootmig}, under the correctness conditions C1--C4 stated
in Section~\ref{sec:security}. For I4, a unique verdict means that the
decrypted query matches no entry of the decrypted snapshot gallery $G_s$,
that is, $\neg\Dup(\Pt(\Cid),\Pt(G_s))$. This follows from BioZKFHE's
result integrity, provided that the gateway checks the committee roster
and accepts only block-proof openings verified against their quadratic
arithmetic programs. I3 is the stale-session check of the finalization
step below. Section~\ref{sec:security} separates the state argument
common to all backends from these instance-specific bounds.

\subsection{Bound Enrollment}

\emph{Session creation.} When a session opens, the source contract
binds into its context a digest of the on-chain gallery state, which is the
snapshot the backend will query. The chain stores only a commitment to each
ciphertext block of the gallery, so this state consists of a hash chain
over the closed blocks, the commitment $u_{M-1}$ of the last block, the
number $M$ of blocks and the template count $N$.

\emph{Request binding.} The $\ES$ performs the single-capture derivation
and signs the declaration~\eqref{eq:decl}, as discussed in
Section~\ref{sec:defs}; the source contract then checks the declaration
against the on-chain state. Each gallery block holds $\ell d$ templates,
arranged in $d$ columns and $\ell$ packed layers, so the slot of the next
template lies in column $h=a\bmod d$ and layer $k=\lfloor a/d\rfloor$ of
the last block, where $a=N\bmod \ell d$. The source accepts the request only if
the signer is a registered $\ES$, the declared $N$ equals both the snapshot
count and the current count, and the declared column and layer equal
$(h,k)$; these checks realize I1(a) and I2(a). We index from zero;
BioZKFHE indexes the same slot from one. A holder's host that swaps
$\Cen$ after the $\ES$ has signed, or omits the declaration, is rejected
because the signature no longer verifies for a registered station.

\begin{figure}[t]
\centering
\begin{tikzpicture}[x=1cm,y=1cm,
  st/.style={chain, minimum width=1.9cm, minimum height=0.6cm},
  chk/.style={lab, align=left, anchor=west},
]
\node[st] (cr) at (1.3,4.6) {Created};
\node[st] (bd) at (1.3,3.0) {Bound};
\node[st] (du) at (1.3,0) {Duplicate};
\node[st] (ab) at (3.9,0) {AbortedStaleSlot};
\node[st, line width=1.2pt] (fi) at (6.6,0) {Finalized};

\draw[flow] (1.3,5.45) -- node[chk, right=2pt] {\texttt{createSession}: bind snapshot $s$\\(block commitments, $M$, $N$) into $\mathsf{ctx}$} (cr);
\draw[flow] (cr) -- node[chk, right=2pt] {\texttt{bindRequest}: registered $\ES$ signed $\decl$;\\declared $N$ = snapshot $N$ = current $N$;\\$(h,k)$ = slot of $N$} (bd);

\coordinate (sp) at (1.3,1.55);
\draw[line width=0.6pt, bbInk] (bd) -- (sp);
\node[chk] at (1.45,2.2) {\texttt{finalize}: $\ge t$ distinct roster signatures on\\one confirmation ($y$, outputs, migration)};
\draw[flow] (sp) -- node[lab, left] {$y=\texttt{dup}$} (du);
\draw[flow] (sp) -| node[lab, pos=0.75, right, align=left] {$y=\texttt{unique}$,\\$N$ changed (I3)} (ab);
\draw[flow] (sp) -| node[lab, pos=0.75, right, align=left] {$y=\texttt{unique}$,\\$N$ unchanged} (fi);

\node[lab, align=left, anchor=north] at (6.6,-0.4) {append: $u_{\mathsf{new}}=h_{\mathsf{en}}$\\merge: $t$ signatures on $\Mig$\\$N\gets N{+}1$, same transaction};
\node[lab, align=left, anchor=west, text=bbInk!80] at (2.55,3.0) {(any failed check reverts the call;\\the session keeps its state)};
\end{tikzpicture}
\caption{On-chain state machine of an enrollment session, with the check on each transition. Only Finalized leads to issuance; a stale or duplicate session cannot be issued a credential.}
\label{fig:fsm}
\end{figure}
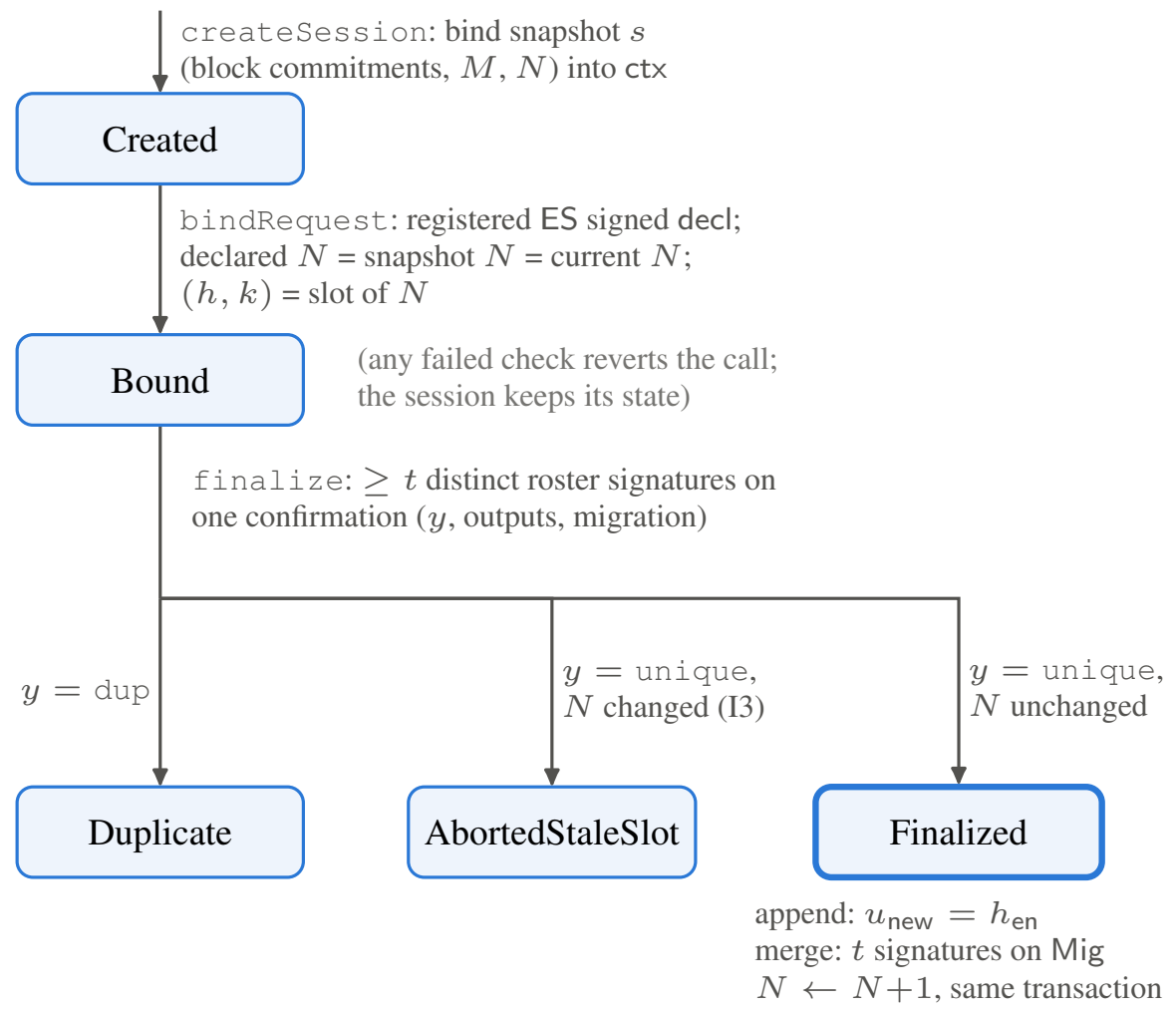

\emph{Finalization.} At least $t$ members of the registered committee,
called the roster, sign one confirmation that covers the verdict $y$, the
digest of the backend's outputs and the gallery migration. If
$y=\texttt{unique}$ but $N$ has changed since the snapshot, the session
ends as \texttt{AbortedStaleSlot}, as I3 requires. Otherwise the contract
checks the migration, as detailed in Section~\ref{sec:rootmig}, updates the
block commitments and $N$, and marks the session finalized in the same
transaction, as I2(c) requires. Signers must appear in strictly increasing
address order, so a repeated signature cannot be counted twice, and
addresses outside the roster do not count. Figure~\ref{fig:fsm}
illustrates the resulting state machine.

\subsection{Gallery-Root Migration}
\label{sec:rootmig}

After a unique verdict, the gallery update either appends or merges. When the
last block is full ($N \bmod \ell d = 0$), $\Cen$ becomes a new block and its
commitment is $H(\Cen)$. Otherwise the last block becomes
$C_M\boxplus\Cen$ under homomorphic addition, and its new commitment
$H(C_M\boxplus\Cen)$ is not a function of $H(\Cen)$. The chain can therefore
check an append directly, but cannot check a merge: BGV ciphertexts are
megabytes in size, so neither $C_M$ nor the addition fits on chain. If the
next root is supplied by the caller, as in BioZKFHE's public
implementation, a compute node can drop $\Cen$ at a merge and still report
success.

Instead of recomputing a merge, the chain accepts it when the
committee that the backend already has confirms it. A merge migration is
the statement $\Mig=(\sid,m,u_{\mathsf{old}},h_{\mathsf{en}},\slot,N_{\mathsf{old}},N_{\mathsf{new}},u_{\mathsf{new}})$,
signed as $H(\texttt{BioBind/gallery-migration/v1},\Mig)$, where
$m$ is the index of the merged block, $u_{\mathsf{old}}$ and
$u_{\mathsf{new}}$ are its commitments before and after the merge, $N_{\mathsf{old}}$
and $N_{\mathsf{new}}$ are the template counts, and
$h_{\mathsf{en}}=H(\Cen)$ is the digest bound at request time. An honest
member signs only if it holds canonical, uncompressed serializations of
$C_M$ and $\Cen$ with $H(C_M)=u_{\mathsf{old}}$ and $H(\Cen)=h_{\mathsf{en}}$
and its own $H(C_M\boxplus\Cen)$ equals $u_{\mathsf{new}}$. Members use the same
deterministic BGV addition algorithm, including correction-factor handling;
with matching levels and correction factors it reduces to coefficient-wise
modular addition. The contract requires
$\slot=N$ and $N_{\mathsf{new}}=N+1$; for an append it requires
$u_{\mathsf{new}}=h_{\mathsf{en}}$ with no signature; for a merge it requires
$u_{\mathsf{old}}=u_{M-1}$ and $t$ distinct roster signatures. The finalization confirmation binds the session, inputs, verdict and
migration; Supplementary Appendix~\ref{app:bgv-confirmation} specifies
its roster, data-availability and byte-level requirements.

The rule changes what the committee is responsible for. Corrupting $t$
members now breaks gallery integrity as well as confidentiality. A compute
node that withholds the ciphertexts only prevents finalization, which is an
availability failure: the session aborts and the person retries. Liveness
needs at least $t$ honest members online.

\subsection{Issuance and Lifecycle Operations}

Issuance consumes the final record atomically and requires the link
signature~\eqref{eq:link}, whose role is discussed in
Section~\ref{sec:defs}; each record therefore yields at most one
credential. The lifecycle operations of Section~\ref{sec:defs} change
on-chain records that the gateway reads in the transaction that accepts an
authentication: the credential's status, the policy and the registered
device versions. Revoking any of them therefore takes effect at the next
acceptance, without notifying holders or invalidating pending challenges.
Template replacement is limited. Because the gallery only grows, a person
cannot pass deduplication again under the same model, so same-model
template refresh would need a new backend operation. Model or quantization
migration requires a new gallery, a continuity check and a frozen old
gallery.

\subsection{Authentication}
\label{sec:auth}

In the public profile the circuit has five public signals
$(\Cenr,\Cprb,\hP,\hchi,\tauauth)$. It constrains the openings of
the two commitments~\eqref{eq:cenr} and~\eqref{eq:cprb}, checks every
coordinate's range, and checks the match relation~\eqref{eq:dist}. In one transaction
the gateway checks that the sender is the challenge's beneficiary,
the address for which the challenge was issued, and that the
challenge is pending and unexpired; that the credential, policy and device
version are active in the current state; that the public signals equal the
credential's $\Cenr$, the policy digest, the registered threshold and the
reduced challenge digest; that the $\AFE$ statement~\eqref{eq:capsig} carries the same $\Cprb$
and context, matching device metadata and a counter above the last one; that
the statement's signature recovers the device's registered key; and that the
Groth16 verifier accepts. On success it deletes the challenge and advances the
counter in the same state transition.

\subsection{Unlinkable Profile}
\label{sec:unlink}

As discussed in Section~\ref{sec:model}, the public profile exposes the
holder's credential handles at every authentication. The unlinkable profile
keeps enrollment public and makes use private, following the Semaphore pattern~\cite{semaphore}. Before issuance
the holder chooses an identity secret $s$ and submits the identity
commitment $\idc=\mathsf{Poseidon}(\texttt{id},s)$~\cite{grassi2021poseidon}, which the $\ES$ link signature
covers. Issuance inserts the leaf
$\mathsf{Poseidon}(\texttt{leaf},\Cenr,\idc,\hP,\text{version})$ into an
incremental Poseidon Merkle tree of depth 20 in the same transaction that
consumes the record. Devices use EdDSA keys on BabyJubJub. Each person carries one capture
module, so active device versions are leaves in an equally deep device tree.

The circuit of the unlinkable profile takes nine public inputs: the
two tree roots, $\Cprb$, the policy digest and threshold, the challenge
digest, the capture expiry, a scope and a nullifier.
It proves that a credential leaf built from $\Cenr$ and $\mathsf{Poseidon}(\texttt{id},s)$
is in the credential tree; that $\Cenr$ and $\Cprb$ open to vectors
satisfying~\eqref{eq:dist}; that a device leaf is in the device tree and its
key signed $(\Cprb,\hchi,\hP,\text{model},\text{quantization},\text{firmware},\text{expiry})$,
a variant of the capture statement~\eqref{eq:capsig} without the
per-device counter and without any credential handle; and that
$\nf=\mathsf{Poseidon}(\texttt{nf},s,\scope)$. With
$\scope=H_{\mathbb{F}}(\texttt{app-domain},\text{appId})$, where
$H_{\mathbb{F}}$ hashes into the field of the proof system, the nullifier is a stable
pseudonym within an application, and with an action identifier added it is a
one-time token with a distinct action domain. Both scopes use Keccak ABI
encoding reduced modulo the field order in the implementation. Conditional
on person-level uniqueness and preservation of $\idc$ across any future
migration, one person has at most one nullifier per application. A missed
duplicate can violate that premise.

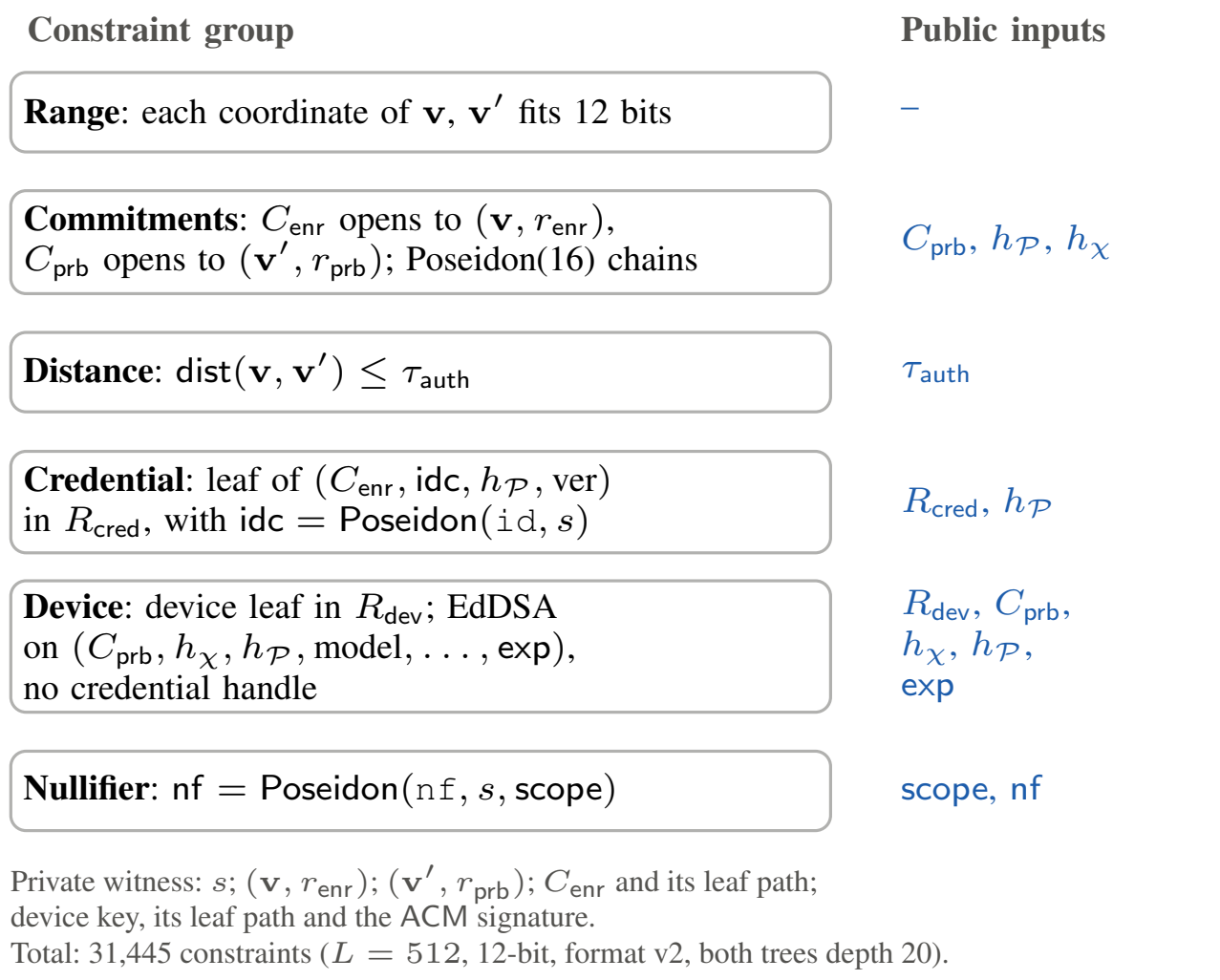
\begin{figure}[t]
\centering
\begin{tikzpicture}[x=1cm,y=1cm,
  grp/.style={box, draw=bbInk!45, fill=white, minimum height=0.55cm,
    align=left, text width=5.5cm, anchor=west, inner sep=2.5pt,
    text depth=},%
  pub/.style={anchor=west, text=bbBlueD, font=\scriptsize, nohyph, align=left},
]
\node[anchor=west, font=\scriptsize\bfseries, text=bbInk] at (0,6.3) {Constraint group};
\node[anchor=west, font=\scriptsize\bfseries, text=bbInk] at (6.05,6.3) {Public inputs};

\node[grp] (g1) at (0,5.75) {\textbf{Range}: each coordinate of $\vv$, $\vp$ fits 12 bits};
\node[grp] (g2) at (0,4.85) {\textbf{Commitments}: $\Cenr$ opens to $(\vv,\renr)$,\\$\Cprb$ opens to $(\vp,\rprb)$; Poseidon(16) chains};
\node[grp] (g3) at (0,3.95) {\textbf{Distance}: $\dist(\vv,\vp)\le\tauauth$};
\node[grp] (g4) at (0,3.05) {\textbf{Credential}: leaf of $(\Cenr,\idc,\hP,\text{ver})$\\in $\Rcred$, with $\idc=\mathsf{Poseidon}(\texttt{id},s)$};
\node[grp] (g5) at (0,2.05) {\textbf{Device}: device leaf in $\Rdev$; EdDSA\\on $(\Cprb,\hchi,\hP,\text{model},\dots,\mathsf{exp})$,\\no credential handle};
\node[grp] (g6) at (0,1.05) {\textbf{Nullifier}: $\nf=\mathsf{Poseidon}(\texttt{nf},s,\scope)$};

\node[pub] at (6.05,5.75) {--};
\node[pub] at (6.05,4.85) {$\Cprb$, $\hP$, $\hchi$};
\node[pub] at (6.05,3.95) {$\tauauth$};
\node[pub] at (6.05,3.05) {$\Rcred$, $\hP$};
\node[pub] at (6.05,2.05) {$\Rdev$, $\Cprb$,\\$\hchi$, $\hP$,\\$\mathsf{exp}$};
\node[pub] at (6.05,1.05) {$\scope$, $\nf$};

\node[lab, anchor=north west, align=left] at (-0.05,0.6) {Private witness: $s$; $(\vv,\renr)$; $(\vp,\rprb)$; $\Cenr$ and its leaf path;\\device key, its leaf path and the $\AFE$ signature.\\Total: 31{,}445 constraints ($L=512$, 12-bit, format v2, both trees depth 20).};
\end{tikzpicture}
\caption{Constraint groups of the authentication circuit of the unlinkable profile and the public inputs each group binds. The credential handle, the enrollment commitment and the device key stay in the witness.}
\label{fig:circuit}
\end{figure}

The gateway accepts if the challenge is pending and the sender is its
beneficiary; both roots are in their
root histories at or after the minimum valid epoch; the policy and threshold
match; a one-time nullifier is unused; and the proof verifies. The
current contract has no interface for delegated relayers, so the holder
submits the transaction itself. Revocation
zeroes the leaf and raises the minimum valid epoch to the current one, so
proofs against older roots are rejected and revocation takes effect at the
next acceptance. The price is that every holder must re-prove against the new
root after any revocation. The per-device counter is dropped, because it
would identify the device; the attempt budget then rests on the $\AFE$
alone. Figure~\ref{fig:circuit} shows the circuit's constraint groups and
the public inputs each group binds.

\section{Security Analysis}
\label{sec:security}

The security argument follows the enrollment record across the off-chain
and on-chain boundary. The gallery invariant ties decoded entries to the
ideal registration history; finalization and single-use issuance tie each
credential to that history. Authentication then ties its template to the
current device-originated probe and live authorization state. We state the
multi-query bounds below; Supplementary Appendix~\ref{app:proofs}
gives the full conditional arguments.

\subsection{Backend-Independent State Argument}

The first part of the argument does not depend on how the backend
represents templates. Starting from an empty gallery and with single-use
issuance, I1 binds each credential to its captured template. Induction
over the I2 transitions shows that the logical gallery equals the history
of finalized enrollments. I3 makes the snapshot that a session queried
equal to that history at the session's finalization, and I4 then excludes
an earlier matching entry. Supplementary Lemma~\ref{lem:interface}
bounds the probability that any of these steps fails over the entire
adaptive execution. Each backend must establish the premises I1--I4 under
its own adversary model.

\subsection{BGV Instance and Concrete Bounds}

For the BGV instance, the correctness conditions are explicit: (C1) the packed similarity
scores decode exactly, without wrapping around the plaintext modulus.
Coordinate bounds alone do not ensure this, so the $\ES$ normalizes each
embedding to unit length before scaling by $\Gamma$ and rounding, which
bounds every query--template inner product; (C2) every
block decrypts correctly after its allowed enrollment additions and full
query evaluation; (C3) addition is deterministic under fixed parameters,
representation and canonical serialization, including BGV correction
factors; and (C4) an enrollment encoding changes only its designated empty
slot. Supplementary Appendix~\ref{app:proofs} gives the
precise domain and capacity requirements. These are correctness premises,
not consequences of the signature or proof assumptions.

Let $q_S$ be the number of enrollment sessions, $k_{\ES}$ and $k_{\AFE}$ the
numbers of registered device versions, $q_{\ES}$ and $q_{\mathsf{cap}}$ the
signing queries per device key, $n$ and $t$ the committee size and threshold,
$q_M\le q_S$ the migration signatures per member, and $q_{\mathsf{acc}}$ the
accepted authentications. $\Adv^{\mathsf{euf}}_{\mathsf{Sig}}(q)$ is the
single-key EUF-CMA advantage with $q$ queries, $\Adv^{\mathsf{cr}}_H$ the
collision advantage against $H$ with domain separation, and
$\Adv^{\mathsf{res\text{-}int}}_{\Bk}$ the advantage against BioZKFHE's
result-integrity experiment for one session. Likewise,
$\Adv^{\mathsf{cr}}_{\HF}$ is the collision advantage against the hash
$\HF$ into the circuit field, $\Adv^{\mathsf{bind}}_{\ComH}$ the binding
advantage against $\ComH$, and $\Adv^{\mathsf{ks}}_{\Pi}$ the
knowledge-soundness advantage against the proof system $\Pi$. Three
events are not cryptographic: $\mathsf{DerivFail}_{\ES}$ and
$\mathsf{DerivFail}_{\AFE}$ are the events that a device's internal
derivation deviates from its specification, and $\mathsf{BadState}$ is
the event that the deployed contracts violate their state invariant.

\begin{theorem}[Lifecycle binding for the BGV instance]
\label{thm:main}
Suppose the backend meets I1--I4 and the correctness conditions
(C1)--(C4) of Supplementary Appendix~\ref{app:proofs} hold, at most $t-1$ committee members
are corrupted, the $\ES$ and $\AFE$ signatures are EUF-CMA, $H$ is collision
resistant with unambiguous, domain-separated encodings, $\ComH$ is
computationally binding, and the proof system is knowledge sound for the
matching relation of Section~\ref{sec:defs}, given by Eqs.~\eqref{eq:cenr}, \eqref{eq:cprb} and~\eqref{eq:dist}, in the presence of honest-proof auxiliary
input, as specified in Supplementary Appendix~\ref{app:privateauth}. Then for every PPT adversary
\begin{align}
\Adv^{\mathsf{unq}} \le{}& k_{\ES}\Adv^{\mathsf{euf}}_{\mathsf{Sig}}(q_{\ES}) + n\,\Adv^{\mathsf{euf}}_{\mathsf{Sig}}(q_M) + \Adv^{\mathsf{cr}}_H \notag\\
&+ \Pr[\mathsf{DerivFail}_{\ES}] + q_S\,\Adv^{\mathsf{res\text{-}int}}_{\Bk}, \label{eq:unq}\\
\Adv^{\mathsf{auth}} \le{}& k_{\ES}\Adv^{\mathsf{euf}}_{\mathsf{Sig}}(q_{\ES}) + \Pr[\mathsf{DerivFail}_{\ES}] \notag\\
&+ k_{\AFE}\Adv^{\mathsf{euf}}_{\mathsf{Sig}}(q_{\mathsf{cap}}) + \Pr[\mathsf{DerivFail}_{\AFE}] \notag\\
&+ \Adv^{\mathsf{cr}}_H + \Adv^{\mathsf{cr}}_{\HF} + q_{\mathsf{acc}}\Adv^{\mathsf{ks}}_{\Pi} \notag\\
&+ \Adv^{\mathsf{bind}}_{\ComH} + \Pr[\mathsf{BadState}], \label{eq:auth}
\end{align}
and at the level of people,
\begin{align}
\Pr[\mathsf{SYB}] &\le \Adv^{\mathsf{unq}} + q_E\,\fnm, \label{eq:syb}\\
\Pr[\mathsf{Imp}] &\le \Adv^{\mathsf{auth}} + q_h\,\fmax, \label{eq:imp}
\end{align}
where $q_h$ is the number of captures presented against $\cred_h$, and $q_h\le B$ when the $\AFE$ budget holds.
\end{theorem}

For this instance, the backend term uses BioZKFHE's single-session
result integrity. Its extension to concurrent sessions requires local
soundness for adaptively chosen statements under shared parameters;
$q_S$ counts all initiated sessions, including aborts and duplicates.
Supplementary Appendix~\ref{app:bgv-integrity-bound} gives the concrete
bound. Enrollment history and malicious insertion inputs are handled by
W1--W3, separately from that backend result.

The authentication bound of the public profile extends to the
unlinkable profile with one additional term $\epsilon_{\mathsf{mem}}$,
which bounds a membership failure anywhere in the execution, including an
attempt to open an empty leaf as a valid credential or device.
Supplementary Appendix~\ref{app:privateauth} gives the argument and
its assumptions.

\subsection{Proof Outline}

Supplementary Appendix~\ref{app:proofs} proves the theorem through
three lemmas, whose ideas we outline here. Lemma~U shows that if none of
W1--W3 occurs, no two issued credentials carry duplicate vectors: the chain
orders their finalizations, the earlier vector enters $\Gid$ and stays
there because the gallery only grows, and $\neg$W3 at the later
finalization excludes a duplicate. BioZKFHE's per-slot inner-product
threshold is entry-wise and symmetric, as this argument requires. Lemma~R
bounds the acceptance of a merge with a wrong new commitment by
$n\,\Adv^{\mathsf{euf}}_{\mathsf{Sig}}(q_M)+\Adv^{\mathsf{cr}}_H$. Any $t$
accepted roster signatures include one from an honest member, which is
either a forgery or was computed from ciphertexts that collide with the
true ones; the lemma therefore reuses the threshold assumption of
BioZKFHE's result integrity instead of adding a new one. Lemma~S then shows
that, in the absence of a forgery, a collision, a derivation failure, a bad
migration and a bad verdict, the link signature excludes W1, an induction
over gallery changes excludes W2, and I3 and I4 together exclude W3.
Guessing the session reduces a bad verdict to BioZKFHE's single-session
experiment, at the cost of the factor $q_S$.

At the level of people, I3 also excludes the capture order that W3 alone
would miss: a session that opened before a person's first vector entered
the gallery but finalized after it sees a changed count and aborts. A
conditional union bound over the remaining captures gives $q_E\fnm$
without assuming independence, and Eq.~\eqref{eq:imp} follows in the same
way for authentication captures, because an honest user's own capture is
bound to that user's challenge. The authentication bound replaces each
check of the gateway by the corresponding assumption. Because the
adversary sees honest proofs, extraction must succeed in their presence;
we state this as an explicit assumption, since ordinary knowledge
soundness and re-randomizability do not establish it, and whether the
deployed Groth16 instantiation meets it still requires independent
review. This part of the proof is a standard hybrid. The uniqueness part
is not: the induction over gallery transitions, verdict linearization,
whose necessity the snapshot race shows, and the root migration of
Lemma~R cannot be argued one acceptance at a time.

\subsection{Nature of the Error Terms}

Table~\ref{tab:errors} classifies the terms. Only the first row consists of
standard assumptions that can be cited as such. $\Adv^{\mathsf{cr}}_{\HF}$ is
heuristic: the context and policy digests are Keccak outputs reduced modulo
the field order $r\approx2^{254}$, which is not implied by Keccak's collision
resistance, and in the random-oracle model it is at most
$(q_\chi+q_{\mathcal{P}})^2/r$ for $q_\chi$ challenges and $q_{\mathcal{P}}$
policies. Placing the full 256-bit digest in the circuit as two field
elements would remove it. Knowledge soundness requires a trusted setup, which
our local single-party benchmark setup does not provide. The derivation
failures are trusted-computing-base assumptions and are not reduced. The
state term is a deterministic program invariant: it is zero if the deployed
bytecode meets its specification, and establishing that needs an audit and
adversarial ordering tests. The biometric terms cannot be replaced by dataset
metrics.

\begin{table}[t]
\centering
\caption{Error terms of Theorem~\ref{thm:main}.}
\label{tab:errors}
\footnotesize
\begin{tabular}{@{}>{\raggedright\arraybackslash}p{2.5cm}>{\raggedright\arraybackslash}p{2.3cm}>{\raggedright\arraybackslash}p{2.9cm}@{}}
\toprule
Term & Type & Current status \\
\midrule
$\Adv^{\mathsf{euf}}_{\mathsf{Sig}}$, $\Adv^{\mathsf{cr}}_H$, $\Adv^{\mathsf{bind}}_{\ComH}$ & Standard assumptions & Citable \\
$\Adv^{\mathsf{cr}}_{\HF}$ & Heuristic (ROM birthday bound) & Removable by a 256-bit in-circuit context \\
$\Adv^{\mathsf{ks}}_{\Pi}$ & Extraction with honest-proof auxiliary input & Needs trusted setup; prototype setup does not qualify \\
$\Adv^{\mathsf{res\text{-}int}}_{\Bk}$ & BioZKFHE result integrity & Concrete per session (its Cor.~1); multi-session union is ours \\
$\Pr[\mathsf{DerivFail}_{\ES}]$, $\Pr[\mathsf{DerivFail}_{\AFE}]$ & TCB assumptions & Not reduced; software devices only \\
$\Pr[\mathsf{BadState}]$ & Implementation invariant & Needs audit and ordering tests \\
$\fnm$, $\fmax$ & Biometric, adversarial presentation & Assumptions; dataset rates are non-adversarial references \\
\bottomrule
\end{tabular}
\end{table}

\section{Prototype Implementation}
\label{sec:impl}

The enrollment source has reference and compact versions with the same
checks. The compact version stores record digests and authenticates records
supplied as calldata; its adapter re-encodes each final record and checks
the digest before issuance. The public gateway implements the live-state
and lifecycle checks; the gateway of the unlinkable profile uses two
incremental Poseidon trees with epoch-tagged root histories. The evaluated
circuit of the unlinkable profile uses
$L=512$, 12-bit coordinates ($\Gamma=2047$), commitment format v2 and
depth-20 trees. Format v2 absorbs fifteen packed chunks per Poseidon call
instead of two, reducing each commitment by about 2{,}175 constraints.
Supplementary Appendix~\ref{app:implementation} gives the record
encoding, commitment formats and hash-library optimization.

A software enrollment bridge and a mock $\AFE$ implement the $\ES$ and $\AFE$
roles with software keys. The bridge derives transparent stand-ins for the
query and insertion ciphertexts together with the enrollment commitment
from one vector, seals a pending record, and signs declarations and links.
The simulated committee evaluates the 1:N query on plaintext snapshots and merges
the stand-ins; block proofs, the check that every block is proved, and
threshold decryption are not executed. Separate C++ experiments, detailed in Section~\ref{sec:bgv}, execute
real BGV components that are not integrated into this pipeline.

We align faces with MTCNN~\cite{zhang2016mtcnn} and extract
$L_2$-normalized, 512-dimensional InceptionResnetV1 embeddings pretrained
on VGGFace2~\cite{cao2018vggface2}. Quantization scales by $\Gamma$ and
rounds half away from zero. Contracts run on local Ganache; circuits use
Circom and snarkjs. Groth16 uses local single-party benchmark setups whose
keys are not deployable. The contracts, circuits, evaluation scripts and the
outputs behind every reported number are available at
\url{https://github.com/plan-lab-szu/biobind}; its README lists the toolchain versions and maps each result to
its scripts and run identifier, and Supplementary
Appendix~\ref{app:implementation} gives further build details.

\section{Experimental Evaluation}
\label{sec:eval}

We evaluate whether the implementation rejects the composition attacks,
what verifiable matching adds relative to trusted matching, and what the
selected circuits and contracts cost. We then examine the separate BGV
component requirements and the biometric accuracy that limits person-level
uniqueness. Local experiments use software keys and benchmark-only Groth16
setups; the enrollment pipeline uses a simulated committee. Gas is measured
per transaction, proving times follow the reported repeated-run protocols,
and the cost composition explicitly separates published BioZKFHE data from
our measurements. Run identifiers link the results to the reproduction
records.

\subsection{Negative Controls and End-to-End Runs}

The implementation rejects the tested input substitutions, slot and
snapshot races, invalid migration confirmations, repeated issuance,
unauthorized authentication and stale-root proofs. Each attack has a
successful honest control; all 45 tests pass, and fourteen honest
enrollments across a block boundary preserve the expected last-block
commitment. Supplementary Appendix~\ref{app:evaluation-details}
lists the controls and run identifiers.

An end-to-end run with real LFW features passes all eight checks.
Two people enroll through the software bridge and simulated committee,
authenticate with other images of themselves and obtain stable scope
pseudonyms; an impostor's probe fails witness generation.
However, the calibrated deduplication threshold misses a re-enrollment
using another image, and that person obtains a second credential.
This illustrates the biometric-error contribution $q_E\fnm$ in
Eq.~\eqref{eq:syb}: the protocol executes correctly, but the tested
operating point has a non-adversarial missed-duplicate rate of about 17\%.
Authentication costs 305{,}060--305{,}072 gas in this run; the matched
proving-time comparison follows below.

\subsection{Comparison with a Strong Signature Baseline}
\label{sec:signature-baseline}

A trusted matcher with the same capture and live-state checks also
excludes E1--E3. We compare retaining verifiable opening and matching
constraints with additionally trusting those operations to the device.
The main comparison uses equal public fields in the unlinkable profile;
Supplementary Appendix~\ref{app:public-baselines} reports the
public-profile signature variants.

For an equal-public-field comparison in the unlinkable profile, we implemented a second
relation that retains both hidden memberships, the identity secret and
scope nullifier. The device checks the enrollment opening and match, then
signs the enrollment commitment and threshold together with the existing
capture fields under a separate domain. The proof
verifies that signature without exposing which device or credential was
used. Both relations have the same nine public inputs and depth-20 trees;
matching is the changed trust assumption. This baseline still needs ZK
for membership privacy, so public-profile signature costs cannot serve as
its gas estimate.

With the same real 512-dimensional LFW inputs, the attested relation
has 15{,}534 constraints versus 31{,}445 for ZK matching, a 50.6\% reduction.
Its authentication costs 305{,}048--305{,}060 gas, essentially unchanged
because the public input count and gateway checks are unchanged
(R-PRIVATT-20261030). The first challenge costs 140{,}577 gas and subsequent
ones 123{,}477. All eight end-to-end controls, seven statement-binding
controls and rejection of a capture-only signature pass. A deliberately
false match signed by the authorized software device is accepted, making
the added trust assumption explicit. In five alternating-order rounds
after one warmup, witness plus proving medians are 1.989\,s for ZK
matching and 1.119\,s for attested matching (R-PRIVTIME-20261030); these
exclude capture, TCB matching/opening checks, signing and network time.
The baseline removes verifiable
matching from the circuit; it does not remove trust in capture from either
profile.

\subsection{Circuit Cost}

Table~\ref{tab:circuit} reports circuit costs.
Format v2 reduces the 12-bit, 512-dimensional circuit below $2^{15}$
constraints, halving the Groth16 evaluation domain. In five alternating
v1/v2 rounds after one warmup, proving time falls by 26\% and total time by
15\%, despite slower witness generation with $\mathsf{Poseidon}(16)$.
The depth-20 configuration takes 1{,}801\,ms in this benchmark, which is
separate from the matched signature comparison above. Real LFW vectors
pass for the genuine user and fail witness generation for the impostor in
all three tested v2 configurations. Supplementary
Appendix~\ref{app:circuit-details} records the implementation tradeoffs
and run identifiers.

\begin{table}[t]
\centering
\caption{Circuit of the unlinkable profile ($L=512$ unless noted): constraints and proving time (median of five alternating runs, one local CPU).}
\label{tab:circuit}
\footnotesize
\setlength{\tabcolsep}{4pt}
\begin{tabular}{@{}lrrrr@{}}
\toprule
Circuit & Constraints & Witness & Prove & Total \\
\midrule
v1, 12-bit, device tree 10 & 33{,}463 & 297\,ms & 1{,}761\,ms & 2{,}040\,ms \\
v2, 12-bit, device tree 10 & 29{,}025 & 414\,ms & 1{,}310\,ms & 1{,}724\,ms \\
\textbf{v2, 12-bit, device tree 20} & \textbf{31{,}445} & 423\,ms & 1{,}354\,ms & 1{,}801\,ms \\
v2, 8-bit, device tree 10 & 24{,}863 & \multicolumn{3}{c}{not timed} \\
v2, 16-bit, device tree 10 & 34{,}331 & \multicolumn{3}{c}{not timed} \\
v2, $L=128$, 16-bit & 19{,}247 & \multicolumn{3}{c}{not timed} \\
\bottomrule
\end{tabular}
\end{table}

\subsection{On-Chain Cost}

Table~\ref{tab:gas} reports per-operation gas with $t=3$ and $n=5$.
Compact enrollment uses 228{,}746 gas across three transactions, a 36.6\%
reduction from the reference source; each additional committee signature
adds about 16{,}000 gas. Duplicate rejection and stale abort also incur
transaction costs. Authentication in the unlinkable profile costs less than in the public
profile because it removes on-chain signature recovery, the device counter
and several state reads. Tree updates shift cost to issuance and
revocation: issuance in the unlinkable profile costs about twice as much as public issuance,
and credential revocation costs about one million gas to recompute both
paths.

\begin{table*}[t]
\centering
\caption{Gas per operation (local Ganache, single runs, $t=3$, $n=5$). Neither profile includes BioZKFHE's block proofs. Public challenge and authentication come from the current 128-dimensional baseline run; on-chain verification cost does not depend on $L$. Other runs report authentication in the unlinkable profile at 305{,}048--305{,}072 gas, because gas varies slightly with calldata contents.}
\label{tab:gas}
\footnotesize
\begin{tabular}{@{}lrrl@{}}
\toprule
Operation & Public & Unlinkable & Run \\
\midrule
\texttt{createSession} & \multicolumn{2}{c}{61{,}285} & R-ENRSRC2 \\
\texttt{bindRequest} & \multicolumn{2}{c}{48{,}075} & R-ENRSRC2 \\
\texttt{finalize}, merge & \multicolumn{2}{c}{119{,}386} & R-CALLDATA \\
\texttt{finalize}, append & \multicolumn{2}{c}{142{,}307} & R-CALLDATA \\
\texttt{finalize}, stale abort / dup & \multicolumn{2}{c}{78{,}889 / 76{,}180} & R-ENRSRC2 \\
Issuance (merge) & 438{,}059 & 856{,}135 & R-CALLDATA / R-PRIVGW2 \\
Enrollment + issuance (merge) & 666{,}841 & 1{,}085{,}317 & R-CALLDATA / R-PRIVE2E \\
Challenge & 108{,}121 & 140{,}314 & R-BASE-20261030 / R-PRIVGW-20261030 \\
Authentication, app scope & 336{,}239 & 304{,}736 & R-BASE-20261030 / R-PRIVGW2 \\
Authentication, one-time scope & -- & 327{,}140 & R-PRIVGW2 \\
Revoke credential (depth 20) & not measured & 1{,}042{,}325 & R-PRIVGW2 \\
Register / revoke device (depth 10)$^{*}$ & -- & 388{,}095 / 554{,}550 & R-PRIVGW2 \\
\bottomrule
\end{tabular}\\[2pt]
{\raggedright\footnotesize$^{*}$Measured with a depth-10 device tree. The evaluated circuit uses depth 20, whose device updates were not measured; they use the same tree contract as the depth-20 credential-tree updates above.\par}
\end{table*}

\subsection{Composition with BioZKFHE}

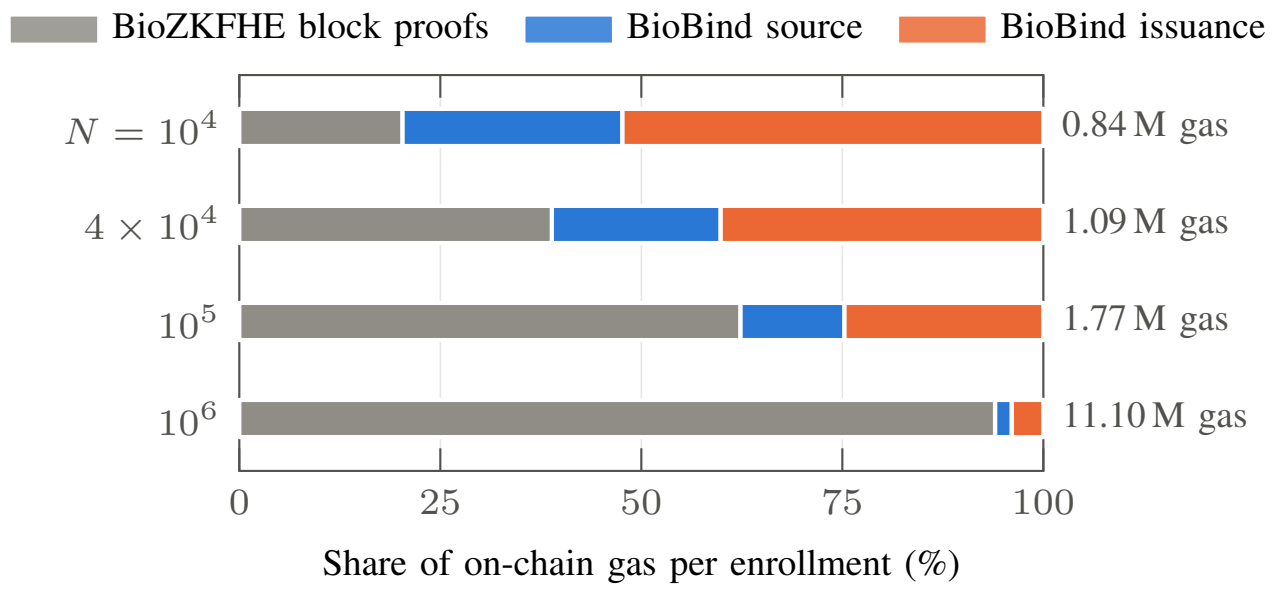
\begin{figure}[t]
\centering
\begin{tikzpicture}
\begin{axis}[bbaxis,
  width=6.9cm, height=4.2cm,
  xbar stacked, bar width=7pt,
  xmin=0, xmax=100, xtick={0,25,50,75,100},
  xlabel={Share of on-chain gas per enrollment (\%)},
  ytick=data, y dir=reverse,
  symbolic y coords={10000,40000,100000,1000000},
  yticklabels={{$N=10^4$},{$4\times10^4$},{$10^5$},{$10^6$}},
  ymajorgrids=false, enlarge y limits=0.18,
  clip=false,
  legend style={at={(0.5,1.03)}, anchor=south, legend columns=3,
    /tikz/every even column/.append style={column sep=4pt}},
  area legend,
  every axis plot/.append style={draw=white, line width=0.8pt},
]
\addplot[fill=bbGray] table[x=pvscPct, y=N, col sep=comma]{figures/data/cost_facenet_n1.csv};
\addlegendentry{BioZKFHE block proofs}
\addplot[fill=bbBlueC] table[x=sourcePct, y=N, col sep=comma]{figures/data/cost_facenet_n1.csv};
\addlegendentry{BioBind source}
\addplot[fill=bbOrange] table[x=issuePct, y=N, col sep=comma]{figures/data/cost_facenet_n1.csv};
\addlegendentry{BioBind issuance}
\node[font=\scriptsize, text=bbInk, anchor=west] at (axis cs:100,10000) {0.84\,M gas};
\node[font=\scriptsize, text=bbInk, anchor=west] at (axis cs:100,40000) {1.09\,M gas};
\node[font=\scriptsize, text=bbInk, anchor=west] at (axis cs:100,100000) {1.77\,M gas};
\node[font=\scriptsize, text=bbInk, anchor=west] at (axis cs:100,1000000) {11.10\,M gas};
\end{axis}
\end{tikzpicture}
\caption{Composition of on-chain gas per enrollment when \sysname{} replaces BioZKFHE's session control (FaceNet, $\ell=1$, $t=3$; BioZKFHE gas from its published data), with totals at right. Source and issuance are the transactions that \sysname{} adds or replaces.}
\label{fig:cost}
\end{figure}

BioZKFHE reports the gas cost of each function of its enrollment contract,
which lets us split one of its enrollment sessions exactly into session creation, $M$ block-proof submissions, $t$
committee confirmations and finalization (R-COST-20261027). In the composed
deployment the block proofs are kept unchanged; session creation,
confirmations and finalization are replaced by \sysname{}'s source, which
verifies the $t$ signatures in one transaction; issuance is new. The replaced
part falls from 329{,}563 to 228{,}746 gas, mostly because $t$ confirmation
transactions become one, so the bound request, the stale abort and the root
migration make BioZKFHE's session no more expensive. Figure~\ref{fig:cost}
presents the breakdown for FaceNet with one packed layer, $\ell=1$, in which
\emph{\sysname{} source} denotes the three source transactions above and
\emph{\sysname{} issuance} the public-profile credential issuance of
438{,}059 gas, and Supplementary Table~\ref{tab:compose} lists the totals for further
profiles. Block proofs grow with the number of blocks, while \sysname{}'s
source and issuance are constant. At $10^4$ people, source and issuance
therefore account for 80\% of the total, issuance alone for about half; at
$10^6$ almost all gas goes to BioZKFHE's block proofs, and the share of
source and issuance falls to 6\%, with issuance alone at 4--10\% across the
listed profiles. BioZKFHE's session totals are its own per-block
extrapolations, which we adopt.

\subsection{Real BGV Enrollment and Merge Components}
\label{sec:bgv}

We evaluate a unified BGV profile, denoted Unified512-2, with
$L=512$ and $\Gamma=2047$ matching the authentication configuration.
It uses degree $d=8192$, two packed layers (16{,}384 registrations),
$T=8{,}473{,}312$, a 48-bit plaintext prime and coefficient-modulus limb
sizes $(60,50,50,50)$. Microsoft SEAL 4.1.2 retains its 128-bit parameter
check. These components execute separately from the software bridge;
the Supplementary Material records the exact primes and the controls
with BioZKFHE's original parameter sets.

For serial timing, one warmup precedes five measured rounds. Each round
encrypts an enrollment into the final empty slot of a freshly encrypted
tail block, merges it and evaluates an encrypted query. Every decoded slot
and packed layer agrees with the plaintext result in all six rounds.
A second context recomputes identical merge bytes within the same process.
Table~\ref{tab:bgv} reports component medians; enrollment and query preparation
include construction, encoding and encryption. Key generation, capture,
signatures, network transfer and chain execution are excluded.

\begin{table}[t]
\centering
\caption{Unified512-2 serial component measurements: one warmup and
five measured rounds. The timing experiment appends one enrollment to a
freshly encrypted gallery; full-history correctness is tested separately.}
\label{tab:bgv}
\footnotesize
{
\begin{tabular}{@{}lr@{}}
\toprule
Operation / representation & Median \\
\midrule
Enrollment preparation & 2312.12 ms \\
Query preparation & 2204.09 ms \\
Load both input frames & 268.76 ms \\
Homomorphic addition & 31.88 ms \\
Serialize merged frame (uncompressed) & 186.05 ms \\
Encrypted query evaluation & 266.20 ms \\
Enrollment frame (uncompressed / zstd) & 192.06 / 179.89 MiB \\
\bottomrule
\end{tabular}}
\end{table}

To test accumulated noise, we separately execute three complete
histories, each with a newly generated key and a different plaintext seed.
Every registration independently encrypts all 512 rows before addition;
a fixed encrypted query is evaluated at 16 checkpoints per history,
including the 8192/8193 layer boundary and capacity 16{,}384. At all 48
checkpoints, every accumulated row, query slot and unpacked score is correct.
Fresh encryption of the same gallery provides a separate control and never
replaces the accumulated ciphertexts. The final minimum row noise budget is
90 bits in every run; query budgets are 17, 17 and 15 bits.

This validates the three tested full-capacity histories for the aligned
profile. The Supplementary Material gives the admissible input domain,
checkpoint results and a source-dependent sufficient noise bound; finite
synthetic tests do not establish a failure probability for all inputs and
keys. BioZKFHE's published parameter sets, from its experiment E3, were
evaluated on galleries encrypted as whole blocks; under the accumulated
noise of merge-based enrollment, two of them fail within 32 registrations
because the query's noise budget runs out. This motivates the larger
modulus chain, and Supplementary Appendix~\ref{app:bgv-details}
retains these controls. Unified512-2 aligns the vector domain with authentication, while integration
with block proofs, threshold decryption and the enrollment bridge remains separate
work.

\subsection{Biometric Accuracy under Quantization}

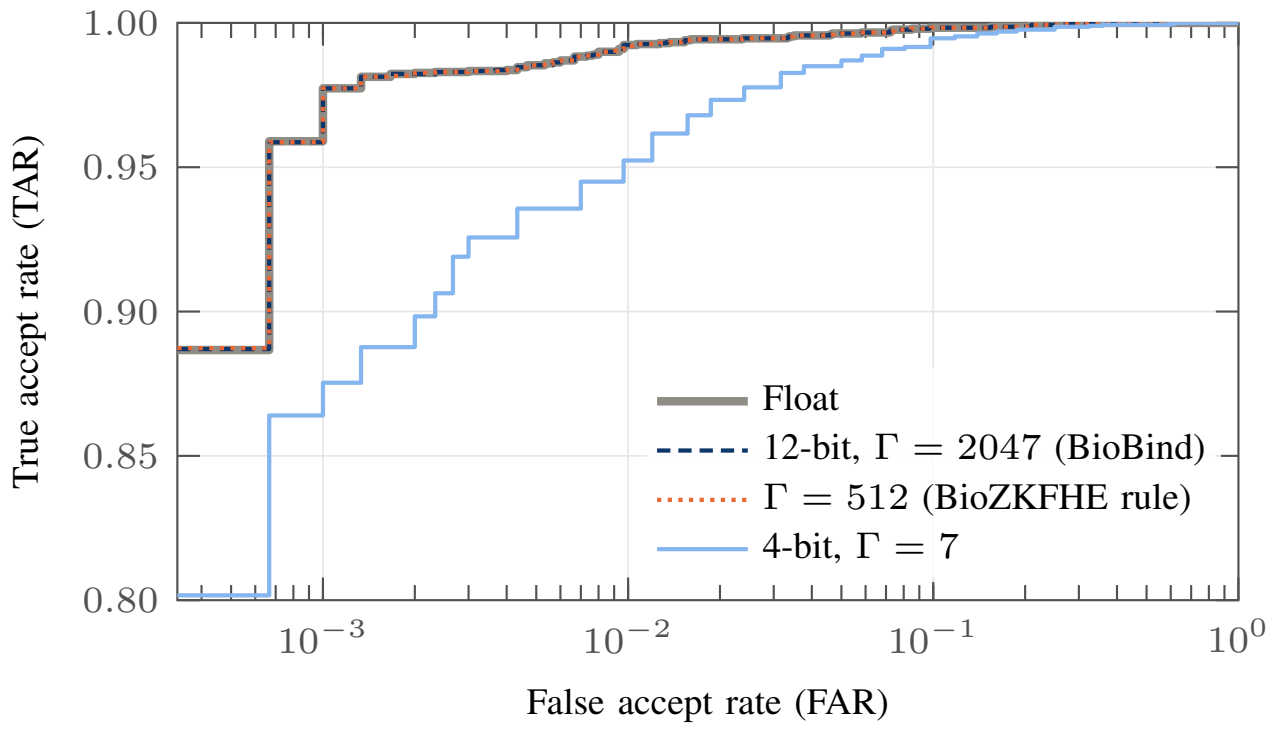
\begin{figure}[t]
\centering
\begin{tikzpicture}
\begin{axis}[bbaxis,
  width=8.6cm, height=5.4cm,
  xmode=log, xmin=3.3333e-4, xmax=1, ymin=0.80, ymax=1.0,
  xlabel={False accept rate (FAR)}, ylabel={True accept rate (TAR)},
  ytick={0.80,0.85,0.90,0.95,1.00},
  yticklabel style={/pgf/number format/fixed, /pgf/number format/precision=2,
                    /pgf/number format/fixed zerofill},
  legend style={at={(0.985,0.03)}, anchor=south east},
]
\addplot[const plot, bbGray, line width=1.6pt] table[x=far,y=tar,col sep=comma]{figures/data/roc_float.csv};
\addlegendentry{Float}
\addplot[const plot, bbBlueE, line width=0.8pt, dash pattern=on 2.5pt off 1.5pt] table[x=far,y=tar,col sep=comma]{figures/data/roc_int12.csv};
\addlegendentry{12-bit, $\Gamma=2047$ (BioBind)}
\addplot[const plot, bbOrange, line width=0.8pt, dash pattern=on 0.8pt off 1.2pt] table[x=far,y=tar,col sep=comma]{figures/data/roc_biozkfhe_gamma512.csv};
\addlegendentry{$\Gamma=512$ (BioZKFHE rule)}
\addplot[const plot, bbBlueA, line width=0.8pt] table[x=far,y=tar,col sep=comma]{figures/data/roc_int4.csv};
\addlegendentry{4-bit, $\Gamma=7$}
\end{axis}
\end{tikzpicture}
\caption{ROC on LFW View 2 (6{,}000 pairs, 3{,}000 impostor) for the float embedding and quantized encodings, scored by squared Euclidean distance. The 12-bit encoding used by \sysname{} and the $\Gamma=512$ rule of BioZKFHE lie on the float curve; so do the 6-, 8-, 10- and 16-bit encodings (not drawn; AUC within $2\times10^{-6}$ of float). Only the 4-bit encoding departs visibly from the float curve. The left edge is one impostor pair ($\mathrm{FAR}=1/3000$); View 2 cannot resolve lower FARs.}
\label{fig:roc}
\end{figure}

We use LFW~\cite{huang2007lfw} View 2 with its ten fixed folds of 300 genuine
and 300 impostor pairs, and fix the protocol before scoring: each fold's
threshold is chosen on the other nine by balanced accuracy, and errors are
counted on the held-out fold only. MTCNN aligned all 7{,}701 images without a
fallback. Table~\ref{tab:lfw} reports the results (R-LFW-20261027). From 12
bits up, quantized vectors make exactly the same errors as float vectors.
BioZKFHE's rule $\Gamma=512$ loses about 0.2 percentage points, as do 8 to 10
bits, because the coordinates of a 512-dimensional unit vector are small and
$\Gamma=512$ uses only about 8 bits of range. Figure~\ref{fig:roc} presents
the ROC curves. With only 3{,}000 impostor pairs, LFW supports no low-FAR claim.
Our untuned accuracy is 99.12\%, below the model documentation's 99.6\%;
we attribute the gap to alignment settings and our held-out threshold protocol.

\begin{table}[t]
\centering
\caption{LFW View 2 with InceptionResnetV1 (VGGFace2): held-out accuracy and errors over 3{,}000 genuine and 3{,}000 impostor pairs. AUC and EER are descriptive over all 6{,}000 pairs.}
\label{tab:lfw}
\footnotesize
\begin{tabular}{@{}lrrrrr@{}}
\toprule
Encoding & $\Gamma$ & Accuracy (\%) & FA & FR & EER (\%) \\
\midrule
Float & -- & 99.12 $\pm$ 0.43 & 29 & 24 & 0.93 \\
4-bit & 7 & 97.73 $\pm$ 0.80 & 56 & 80 & 2.32 \\
6-bit & 31 & 99.02 $\pm$ 0.48 & 30 & 29 & 0.98 \\
8-bit & 127 & 98.92 $\pm$ 0.44 & 27 & 38 & 0.97 \\
10-bit & 511 & 98.93 $\pm$ 0.45 & 27 & 37 & 0.93 \\
\textbf{12-bit} & 2047 & \textbf{99.12} $\pm$ 0.43 & 29 & 24 & 0.93 \\
16-bit & 32767 & 99.12 $\pm$ 0.43 & 29 & 24 & 0.93 \\
BioZKFHE rule & 512 & 98.93 $\pm$ 0.45 & 26 & 38 & 0.93 \\
\bottomrule
\end{tabular}
\end{table}

\subsection{Deduplication at Scale}

A 1:N verdict compounds per-comparison errors over the gallery, so we
calibrate it separately (R-DEDUP-20261028). Scores are the 12-bit inner
products that the encrypted 1:N query evaluates. We split identities by a hash of the name into
a calibration half (2{,}112 identities) and a test half (2{,}169), and choose
thresholds on the calibration half only, as the smallest $\taudup$ whose
per-pair false match rate is at most $10^{-3}$ to $10^{-6}$; we also report
the model's original threshold $\Gamma^2/2$ (cosine 0.5). On the test half,
with 7{,}478{,}879 impostor and 3{,}767 genuine pairs, we report per-pair FMR
and FNMR with Clopper--Pearson intervals~\cite{clopper1934}. We then split the
test half again into a gallery pool of 1{,}645 identities with one image each
and 524 newcomers with 978 images, and count false duplicates, in which a newcomer
is flagged as enrolled, and missed duplicates, in which an enrolled person's
second image is not flagged, over 592 probes. Beyond the measured range we extrapolate false
duplicates by $1-(1-\mathrm{FMR})^N$, which assumes independent comparisons.

\begin{figure}[t]
\centering
\begin{tikzpicture}
\begin{groupplot}[bbaxis,
  group style={group size=2 by 1, horizontal sep=0.9cm},
  width=4.55cm, height=4.6cm, ymin=0, ymax=1,
  ytick={0,0.25,0.5,0.75,1}, yticklabels={0,25,50,75,100},
  every axis plot/.append style={mark size=1.4pt},
]
\nextgroupplot[width=4.3cm, xmode=log, xmin=100, xmax=1e6, xlabel={Gallery size $N$},
  ylabel={Rate (\%)},
  title={(a) False duplicate}, title style={font=\scriptsize, yshift=4pt},
  legend to name=dedupleg, legend columns=3,
  legend style={/tikz/every even column/.append style={column sep=3pt}}]
\draw[bbInk, line width=0.3pt, dash pattern=on 1pt off 1pt] (axis cs:1645,0) -- (axis cs:1645,1);
\foreach \t/\c/\m/\lw in {cur/bbBlueA/o/0.7pt, e3/bbBlueB/square/0.7pt, e4/bbBlueC/triangle/1.2pt, e5/bbBlueD/diamond/0.7pt, e6/bbBlueE/pentagon/0.7pt}{
  \edef\tmp{\noexpand\addplot[\c, line width=\lw, dash pattern=on 2pt off 1.5pt, forget plot] table[x=N,y=fd,col sep=comma]{figures/data/dedup_extrap_\t.csv};}\tmp
  \edef\tmp{\noexpand\addplot[\c, line width=\lw, mark=\m, mark options={solid, fill=white}] table[x=N,y=fd,col sep=comma]{figures/data/dedup_meas_\t.csv};}\tmp
}
\legend{$\tau=\Gamma^2/2$ (cos 0.5), FMR $\le 10^{-3}$, FMR $\le 10^{-4}$ (demo), FMR $\le 10^{-5}$, FMR $\le 10^{-6}$}
\nextgroupplot[width=4.85cm, xmin=-0.55, xmax=4.55, xtick={0,1,2,3,4},
  xticklabels={$\Gamma^2\!/2$, $10^{-3}$, $10^{-4}$, $10^{-5}$, $10^{-6}$},
  xlabel={Threshold}, xmajorgrids=false,
  title={(b) Missed duplicate, $N=1{,}645$}, title style={font=\scriptsize, yshift=4pt},
  every axis plot/.append style={ybar, bar width=9pt, bar shift=0pt, draw=none,
    error bars/y dir=both, error bars/y explicit,
    error bars/error bar style={bbInk, line width=0.5pt},
    error bars/error mark options={bbInk, mark size=1.5pt, rotate=90}}]
\foreach \i/\c in {0/bbBlueA, 1/bbBlueB, 2/bbBlueC, 3/bbBlueD, 4/bbBlueE}{
  \edef\tmp{\noexpand\addplot[fill=\c] table[x=x, y=md, y error plus=errPlus, y error minus=errMinus, col sep=comma,
    restrict expr to domain={\noexpand\thisrow{x}}{\i:\i}]{figures/data/dedup_missed_last.csv};}\tmp
}
\end{groupplot}
\begin{scope}[font=\scriptsize, bbInk]
\coordinate (k) at ($(group c1r1.north east)+(0,0.13cm)$);
\node[anchor=east, inner sep=0pt] (kx) at (k) {extrap.};
\draw[bbGray, line width=0.7pt, dash pattern=on 2pt off 1.5pt] ($(kx.west)+(-0.38cm,0)$) -- ++(0.32cm,0);
\node[anchor=east, inner sep=0pt] (km) at ($(kx.west)+(-0.48cm,0)$) {meas.};
\draw[bbGray, line width=0.7pt] ($(km.west)+(-0.38cm,0)$) -- ++(0.32cm,0);
\fill[white] ($(km.west)+(-0.22cm,0)$) circle (1.4pt);
\draw[bbGray, line width=0.5pt] ($(km.west)+(-0.22cm,0)$) circle (1.4pt);
\end{scope}
\node[anchor=south] at ($(group c1r1.north)!0.5!(group c2r1.north)+(0,0.72cm)$) {\pgfplotslegendfromname{dedupleg}};
\end{tikzpicture}
\caption{Open-set deduplication on the identity-disjoint test half of LFW (12-bit inner product; thresholds chosen on the calibration half). (a) Share of newcomers wrongly flagged as duplicates; markers are measured up to $N=1{,}645$ (dotted line), and dashed curves extrapolate $1-(1-\mathrm{FMR})^N$. (b) Share of re-enrollments with a second image that go undetected at $N=1{,}645$, with 95\% Clopper--Pearson intervals over 592 probes. ``Demo'' marks the end-to-end threshold.}
\label{fig:dedup}
\end{figure}
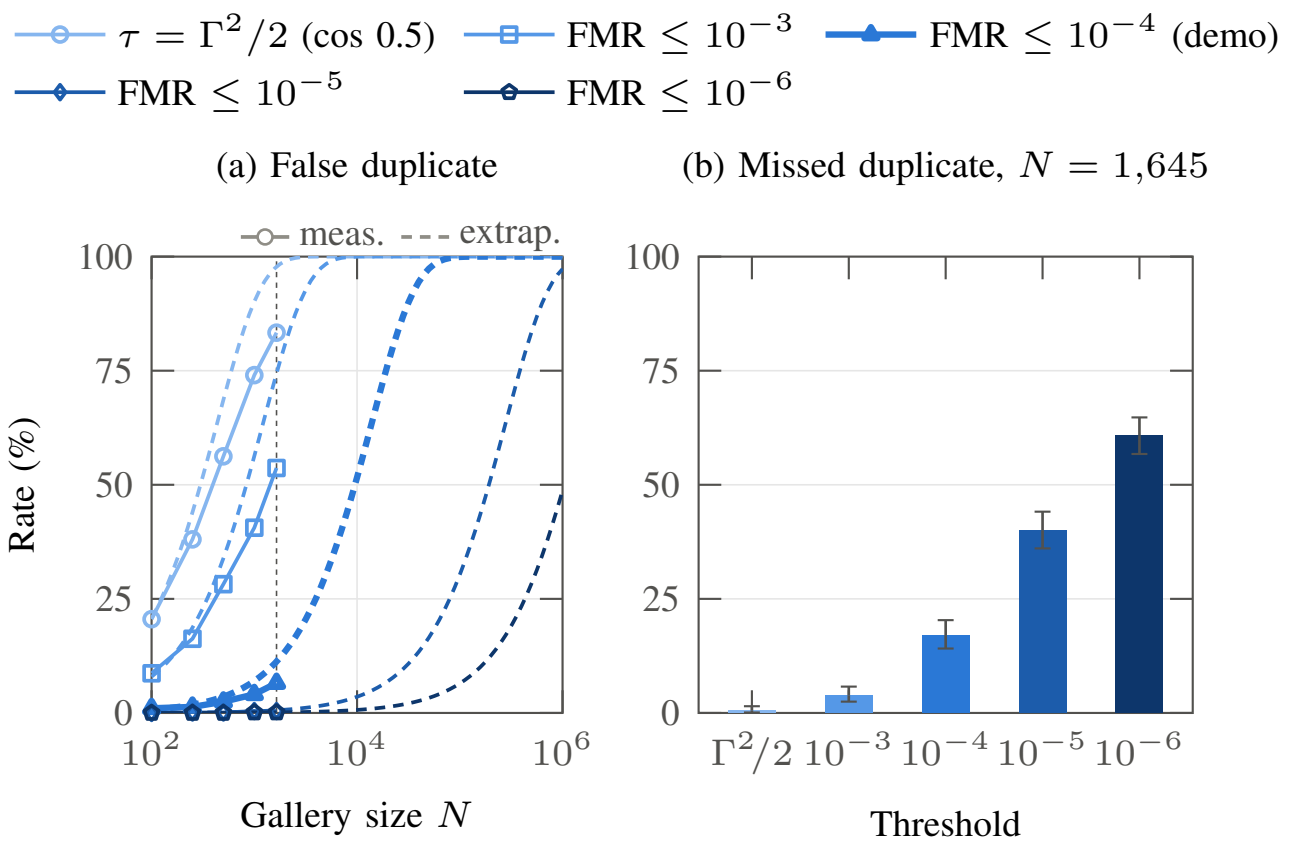

Figure~\ref{fig:dedup} presents the results, and Supplementary
Table~\ref{tab:dedup} lists the thresholds and error rates. The
original threshold flags 83\% of newcomers as duplicates at only 1{,}645
enrolled people. No threshold balances the two errors. To
keep false duplicates acceptable at $10^6$ people the per-pair FMR must be far
below $10^{-6}$, where more than 60\% of re-enrollments already go
undetected; to keep missed duplicates in single digits, more than half of
newcomers are rejected at a thousand people. Missed duplicates change by at
most 8 percentage points across the measured $N$, as expected, since they
depend on the threshold. These numbers are pessimistic for deployment: LFW
contains unconstrained web photographs of celebrities, we use one public
2018 model with one image per enrolled person, and the extrapolation
ignores dependence among comparisons. Within the measured range it
overestimates the false-duplicate rate, for example 98\% against the
measured 83\% at $\Gamma^2/2$, so it indicates the trend only. They are also not the
adversarial $\fnm$ of Theorem~\ref{thm:main}, and do not establish a bound on it.
The tested thresholds expose a tradeoff for this model and dataset. The
uniqueness guarantee of \sysname{} remains relative to the backend's
biometric accuracy at the intended gallery size.

The rate $\fnm$ of Theorem~\ref{thm:main} is taken under adversarial
presentation, and the simplest such presentation is a change of appearance
before re-enrolling. The Bosphorus database~\cite{savran2008bosphorus}
offers controlled examples of such changes, since it holds a neutral image
of each subject together with images under occlusion, expression and pose
changes. We enroll the neutral image of each of 98 subjects and
present every other image of that subject as a re-enrollment attempt against
the thresholds calibrated on LFW, which we do not retune. A failed face
detection counts as a miss, as detailed in Supplementary
Appendix~\ref{app:bosphorus}, which also separates the conditions that an
attended station can accept from those that it would reject. Table~\ref{tab:bosphorus-main} lists the
missed duplicates by condition. For the accepted
conditions, which are occlusion, expression and rotations up to $30^\circ$,
1.8\% of 3{,}273 attempts are missed at the threshold with a per-pair false
match rate of $10^{-4}$ (95\% interval 1.4--2.3\%) and 4.6\% at $10^{-5}$;
occluding the eyes or hair is the most effective change, at about 6\% for
$10^{-4}$. A turn of $90^\circ$, which an attended station would reject, is
missed in 46\% of attempts at $10^{-4}$. The subjects are cooperative and
perform prescribed poses, so these rates are reference values for $\fnm$ under
controlled changes of appearance. They lie below the 17\% of second images
that LFW, with its unconstrained photographs, leaves undetected at the same
threshold. An adversary who selects the most effective disguise may exceed
them, and the attempt budget $q_E$ limits how often it can try.

\begin{table}[t]
\centering
\caption{Missed duplicates (\%, 95\% intervals) when 98 Bosphorus subjects re-enroll under changed appearance, at the thresholds calibrated on LFW for the per-pair false match rate in the header. A failed face detection counts as a miss. Supplementary Table~\ref{tab:bosphorus} lists all conditions and thresholds.}
\label{tab:bosphorus-main}
\footnotesize
\setlength{\tabcolsep}{3pt}
\begin{tabular}{@{}>{\raggedright\arraybackslash}p{2.7cm}rrr@{}}
\toprule
Condition & $n$ & $10^{-4}$ & $10^{-5}$ \\
\midrule
Eyes occluded & 98 & 6.1\,[2.3, 12.8] & 6.1\,[2.3, 12.8] \\
Mouth occluded & 98 & 1.0\,[0.0, 5.5] & 7.1\,[2.9, 14.2] \\
Glasses & 97 & 4.1\,[1.1, 10.2] & 8.2\,[3.6, 15.6] \\
Hair occluded & 65 & 6.2\,[1.7, 15.0] & 16.9\,[8.8, 28.3] \\
Six expressions & 421 & 1.9\,[0.8, 3.7] & 8.6\,[6.1, 11.6] \\
Lower-face action units & 1{,}443 & 1.7\,[1.1, 2.5] & 4.5\,[3.5, 5.7] \\
Yaw up to $30^\circ$ & 294 & 0.3\,[0.0, 1.9] & 0.7\,[0.1, 2.4] \\
\textbf{All attended conditions} & 3{,}273 & 1.8\,[1.4, 2.3] & 4.6\,[3.9, 5.4] \\
\addlinespace[2pt]
Yaw $90^\circ$ & 196 & 45.9\,[38.8, 53.2] & 83.2\,[77.2, 88.1] \\
Same person, neutral & 187 & 0.0\,[0.0, 1.9] & 0.0\,[0.0, 1.9] \\
\bottomrule
\end{tabular}
\end{table}

The dependence of person-level uniqueness on the matcher shapes three
design choices. First, every missed duplicate
is a chance to obtain a second credential, so enrollment needs an attempt
budget; because a person cannot be counted before deduplication,
\sysname{} treats $q_E$ as a deployment parameter. Second, the default
attended enrollment enforces this budget through supervision and liveness
checks, which self-enrollment across devices cannot provide. Third, the
matcher and modality must suit the intended gallery size. World ID, for
instance, uses the iris for uniqueness and the face only for
re-authentication, and any change of modality or capture conditions
requires a fresh evaluation.

\subsection{Public Observer Audit}

We audit calldata, events and metadata from five authentications in the unlinkable profile.
Same-scope nullifiers link repeated use, while nullifiers differ across
scopes; a common funding source remains observable but does not identify
the holder by itself. All six field and equality-pattern checks pass. These observations test
the exposed interface, not statistical anonymity or the cryptographic
assumptions of Supplementary Appendix~\ref{app:privacy}; the full
trace protocol is in Supplementary Appendix~\ref{app:observer-trace}.

\section{Limitations}
\label{sec:limitations}

The evaluation uses software capture devices and a simulated committee;
sensor provenance, presentation-attack detection and tamper-resistant
attempt budgets are unvalidated.
Real BGV components are measured separately; block proofs, the check that
every block is proved, and threshold decryption remain unintegrated.
Measured gas excludes block proofs; the composed estimate uses published
BioZKFHE costs. Ciphertext availability is assumed. Three full-capacity
BGV histories do not establish a general failure probability. The
single-model LFW evaluation cannot establish low-FAR performance or
adversarial bounds on $\fnm$ and $\fmax$.

Security and privacy depend on the stated extraction, membership and
related-output hiding assumptions; the proofs and noise bound await
independent review. Groth16 setups are local and benchmark-only, and
performance measurements carry no deployment-wide guarantee.
Stale-session rejection preserves safety at the cost of retries and does
not establish concurrent throughput. The count-based stale check assumes an
append-only gallery; template refresh and deletion require revised state
semantics. The gateway of the unlinkable profile implements credential and
device revocation but lacks policy revocation and model migration, and no
gateway supports same-model template refresh. Supplementary
Appendix~\ref{app:extensions} discusses the design alternatives,
hardware requirements and trust-reducing extensions, and
Supplementary Appendix~\ref{app:deployment-limitations} details the
remaining deployment requirements.

\section{Conclusion}
\label{sec:conclusion}

\sysname{} connects off-chain biometric enrollment to on-chain credential
issuance and capture-bound authentication for decentralized applications.
Its four enrollment interface conditions bind the queried vector, gallery
transition and issued commitment; authentication relates that commitment
to a fresh authorized probe. The snapshot race shows why correct queries
and writes still require a state check at finalization. The conditional
multi-session argument separates these common state conditions from
backend-specific correctness and trust. The BGV instance, built on BioZKFHE, realizes
them with bound requests, committee-confirmed root migration and
stale-session rejection; the World ID-inspired MPC mapping exposes the
alternative trust requirements. The software prototype rejects the tested
composition attacks on real face features. The signature baseline of
the unlinkable profile
quantifies the proving cost of retaining verifiable matching, and an aligned
BGV profile passes three full-capacity registration histories. With
biometric errors and trusted acquisition accounted for separately, the
protocol makes explicit which evidence a contract needs to rely on a
registration record at issuance and use.

\section*{Supplementary Material}
The appendices below contain the full conditional proofs, the extended trust
comparison, implementation details, evaluation controls and the unified BGV
capacity results and parameter-repair record. The BGV record includes three
completed 16{,}384-registration histories, serial component measurements and
their evidence boundaries. References prefixed by ``main paper'' point to the
numbered sections above, and the equations, tables, figures and lemmas of the
appendices carry an S prefix.

\renewcommand{\theequation}{S\arabic{equation}}
\renewcommand{\thetable}{S\arabic{table}}
\renewcommand{\thefigure}{S\arabic{figure}}
\renewcommand{\thelemma}{S\arabic{lemma}}
\setcounter{equation}{0}
\setcounter{table}{0}
\setcounter{figure}{0}
\setcounter{lemma}{0}
\def\theHequation{S\arabic{equation}}
\def\theHtable{S\arabic{table}}
\def\theHfigure{S\arabic{figure}}
\def\theHlemma{S\arabic{lemma}}

\appendices
\def\theHsection{app.\Alph{section}}
\section{Proofs}
\label{app:proofs}

\subsection{Abstract Enrollment Interface}
\label{app:interface-proof}

We view protected storage through its logical gallery, initially
empty. An abstract enrollment failure is an issued commitment without the
declared capture provenance, a gallery inconsistent with finalized
insertions, or issuance of a template that matched an earlier entry at
finalization. These specialize to W1--W3 in the BGV instance.

\begin{lemma}[Interface composition]
\label{lem:interface}
Suppose I3 is enforced, issuance consumes only unique final records
and consumes each at most once, and the gallery initially is empty.
Let $\epsilon_1$, $\epsilon_2$ and $\epsilon_4$ bound the probabilities
that I1, I2 and I4, respectively, fail anywhere in the permitted adaptive
execution. The probability of any abstract enrollment failure is at most
$\epsilon_1+\epsilon_2+\epsilon_4$. If $\Dup$ is entry-wise and symmetric,
the same bound applies to issuing two credentials for duplicate templates.
\end{lemma}

\begin{proof}
Condition on the absence of all three interface-failure events.
I1 associates every issued commitment with its session's captured template
and the actual query and insertion inputs. Initially the logical gallery
and finalized insertion history are empty. At each unique finalization,
I2 appends the declared insertion template at the next slot and preserves
all other entries. I1 identifies it with the captured query template;
non-finalizing sessions do not change the gallery. Induction therefore
preserves equality with the finalized history. I3 ensures that this history
has not changed since the queried snapshot, and I4 excludes a match at
finalization. Distinct issued credentials consume distinct final records;
the later finalization is checked against a gallery containing the earlier
template. Entry-wise matching and symmetry exclude a duplicate pair.
A union bound over the three excluded events proves the claim, without
independence assumptions.
\end{proof}

The bounds here concern entire adaptive executions, not marginal
single-session experiments. A concrete backend must justify them,
including representation correctness, authenticated state and output
delivery, under its specified corruption model. The following BGV analysis
derives concrete terms under C1--C4; the MPC mapping in
Appendix~\ref{app:world-mpc} identifies conditions but supplies no second
implementation or instantiated cryptographic bound.

\subsection{BGV Instance: Setting}

\emph{Correctness conditions.} These are parameter choices, not
cryptographic assumptions. (C1) Packed-score decoding is correct on the backend's
admissible domain. For unit-normalized vectors scaled by $\Gamma$ and rounded,
$\|\vv\|_2\le\Gamma+\sqrt{L}/2$, so any query--template inner product has
absolute value at most $D=(\Gamma+\sqrt{L}/2)^2$. Require $T>2D$ and plaintext
modulus $p>2D\sum_{k=0}^{\ell-1}T^k$. Centered base-$T$ decoding then recovers
every slot's score without modular wraparound. Coordinate bounds alone do
not establish this claim. The trusted station must enforce this domain.
(C2) Under the chosen BGV parameters, a block still decrypts
correctly after at most $\ell d-1$ enrollment additions per row and the full
1:N query evaluation: $L$ ciphertext products followed by $L-1$ additions. (C3) The specified addition algorithm is deterministic under
fixed parameters, ciphertext representation and canonical serialization.
Coefficient-wise addition is sufficient only when the BGV correction
factors and other required states agree. (C4) The
enrollment encoding of $\vv$ at slot $N$ is non-zero only in the column and
layer of slot $N$, so adding it to a block whose slot $N$ is empty writes
$\vv$ into slot $N$ and changes no other slot.
Main-paper Section~\ref{sec:bgv} reports three complete
registration histories for Unified512-2; Appendix~\ref{app:bgv-details}
gives a source-dependent sufficient bound and the failure controls with
BioZKFHE's original parameter sets. The theorem remains conditional on (C2); finite tests do not
replace its general correctness premise.

\emph{$\ES$ signing rule.} An honest $\ES$ keeps, per derivation, a pending
record $\mathsf{rec}=(\sid,\mathsf{ctx},N,\slot,\Cid,\Cen,\Cenr,\vv)$. It signs
$\decl$ once with $\slot=N$; it signs a $\link$ only if $\rho$ is a unique
final record equal to $\mathsf{rec}$ on $(\sid,H(\Cid),H(\Cen),\slot)$, taking
$\Cenr$ from $\mathsf{rec}$; and it signs at most one link per $\sid$.
$\mathsf{DerivFail}_{\ES}$ is the event that the $\ES$ violates this rule or
that $\Cid,\Cen,\Cenr$ are not honest encryptions and a commitment of $\vv$.

\emph{Real and ideal gallery.} The chain stores block commitments
$u_0,\dots,u_{M-1}$ and $N$. The challenger sees every ciphertext in the
experiment, so for each $u_m$ it can name the ciphertext with that hash; two
different such ciphertexts form a collision, event $\mathsf{Coll}$. Under
$\neg\mathsf{Coll}$ the decoded gallery $\Pt(G)(j)$ is well defined for
$j<N$.

\emph{Multi-key signatures.} For $K$ keys, guessing the index gives
$\Pr[\text{some key is forged}]\le K\cdot\Adv^{\mathsf{euf}}_{\mathsf{Sig}}(q)$.

\subsection{Lemma U}

\begin{lemma}
\label{lem:u}
Let $\Dup$ be entry-wise ($\Dup(\vv,G)\Leftrightarrow\exists g\in G:\Dup(\vv,\{g\})$)
and symmetric. If W1, W2 and W3 do not occur in an execution, then any two
issued credentials $\cred_j\ne\cred_k$ satisfy $\neg\Dup(\vv_j,\{\vv_k\})$.
\end{lemma}

\begin{proof}
By $\neg$W1 each credential has an $L_{\mathsf{enroll}}$ record; let its
vector be the credential's vector and $\rho(\cred)$ its unique final session.
Issuance consumes $\rho$ atomically, so distinct credentials have distinct
sessions; their finalizations are totally ordered on chain, and we let
$\rho_j$ come first. At $\rho_j$'s linearization point $(\slot_j,\vv_j)$
enters $\Gid$, and since $\Gid$ only grows it stays there. At $\rho_k$'s
later finalization $\Gid$ therefore contains $\vv_j$, and $\neg$W3 gives
$\neg\Dup(\vv_k,\{\vv_j\})$. Symmetry gives the converse.
\end{proof}

The proof uses W1 and W3 only. W2 is used in Lemma~\ref{lem:s}, where it
makes the gallery the backend queries equal to $\Gid$, so that I4's
statement about the real snapshot becomes $\neg$W3.

\subsection{Lemma R}

A merge migration $\Mig=(\sid,m,u_{\mathsf{old}},h_{\mathsf{en}},\slot,N_{\mathsf{old}},N_{\mathsf{new}},u_{\mathsf{new}})$
is signed as $D(\Mig)=H(\texttt{BioBind/gallery-migration/v1},\Mig)$. Event
$\mathsf{BadMig}$: the contract accepts some merge whose $u_{\mathsf{new}}$
differs from $H(C_M\boxplus\Cen)$, where $C_M,\Cen$ are the experiment's
ciphertexts hashing to $u_{\mathsf{old}},h_{\mathsf{en}}$.

\begin{lemma}
\label{lem:r}
If the set $\mathit{Cor}$ of corrupted members has $|\mathit{Cor}|\le t-1$,
there are adversaries $\mathcal{B}_1,\mathcal{B}_2$ with
\begin{equation}
\Pr[\mathsf{BadMig}]\le(n-|\mathit{Cor}|)\,\Adv^{\mathsf{euf}}_{\mathsf{Sig}}(\mathcal{B}_1,q_M)+\Adv^{\mathsf{cr}}_H(\mathcal{B}_2).
\label{eq:lemr}
\end{equation}
\end{lemma}

\begin{proof}
An accepted merge carries $t$ signatures from distinct roster addresses, so
at least one belongs to an honest member $i$. \emph{Case A}: $i$ never signed
$D(\Mig)$. Then the signature is a forgery. $\mathcal{B}_1$ guesses an honest
index, embeds its EUF-CMA challenge key there, answers that member's signing
queries through its oracle, simulates the rest and outputs the forgery; it
guesses right with probability at least $1/(n-|\mathit{Cor}|)$.
\emph{Case B}: $i$ signed $D(\Mig)$. If it signed some $\Mig'\ne\Mig$ with
$D(\Mig')=D(\Mig)$, the unambiguous encoding gives a collision. Otherwise $i$
signed $\Mig$ itself and, by its rule, held $C^{(i)}_M,C^{(i)}_{\mathsf{en}}$
with $H(C^{(i)}_M)=u_{\mathsf{old}}$, $H(C^{(i)}_{\mathsf{en}})=h_{\mathsf{en}}$ and
$H(C^{(i)}_M\boxplus C^{(i)}_{\mathsf{en}})=u_{\mathsf{new}}$. Under
$\mathsf{BadMig}$ this pair differs from $(C_M,\Cen)$ while having the same
hashes, so one of the two components is a collision; (C3) rules out two
results for the same pair. $\mathcal{B}_2$ runs the whole experiment and
outputs any collision it observes. An append is checked on chain as
$u_{\mathsf{new}}=h_{\mathsf{en}}$ and contributes no term.
\end{proof}

\subsection{Lemma S}

\begin{lemma}
\label{lem:s}
If the backend meets I1--I4, (C1)--(C4) hold and $|\mathit{Cor}|\le t-1$,
then $\Pr[\mathrm{W1}\vee\mathrm{W2}\vee\mathrm{W3}]$ is bounded by the
right-hand side of main-paper Eq.~\eqref{eq:unq}.
\end{lemma}

\begin{proof}
Let $\mathsf{Bad}$ be the union of $\mathsf{ForgeES}$ (an accepted $\ES$
signature on an unsigned message), $\mathsf{Coll}$, $\mathsf{DerivFail}_{\ES}$,
$\mathsf{BadMig}$ and $\mathsf{BadVerdict}$ (some session finalizes as unique
although $\Dup(\Pt(\Cid),\Pt(G_s))$ holds for its bound snapshot $G_s$).

\emph{$\neg\mathsf{Bad}\Rightarrow\neg$W1.} Issuance needs a valid $\ES$ link,
as I1(c) requires; without forgery an $\ES$ produced it, and without derivation failure
it did so only for a pending record equal to $\rho$ on the session fields,
with $\Cenr$ taken from that record.

\emph{$\neg\mathsf{Bad}\Rightarrow\neg$W2.} We induct over gallery changes
with the invariant that $\Pt(G)(j)=\Gid(j)$ for $j<N$ and every slot $j\ge N$
is empty. Initially $N=0$. By I2(c) the commitments and $N$ change only in
the transaction that finalizes a unique session $\rho$, and by I2(a) each
change increments $N$ once, while $\Gid$ appends $(\slot_\rho,\vv_\rho)$ at
the same point. By I1(a), $h_{\mathsf{en}}$ comes from an $\ES$-signed
declaration for a record $\mathsf{rec}$ with $\slot=N$, so
$h_{\mathsf{en}}=H(\mathsf{rec}.\Cen)$ and, under $\neg\mathsf{Coll}$, the
ciphertext is $\mathsf{rec}.\Cen$. In an append, the new block is
$\mathsf{rec}.\Cen$, which by (C4) decodes to $\mathsf{rec}.\vv$ at slot $N$.
In a merge, $\neg\mathsf{BadMig}$ makes the new last block
$C_M\boxplus\mathsf{rec}.\Cen$ with $C_M$ the previous last block; by
(C2)--(C4), (C1) and the empty slot $N$, it decodes to $\mathsf{rec}.\vv$ at
slot $N$ and leaves other slots unchanged. The record is the one $\rho$'s
digests come from, so $\mathsf{rec}.\vv=\vv_\rho$ and the invariant holds
after the change. Sessions that do not finalize cannot change the gallery,
and unique finalization always writes.

\emph{$\neg\mathsf{Bad}\Rightarrow\neg$W3.} Let $\rho$ have snapshot $G_s$
with count $N_0$. By I3 the count at finalization is $N_0$, and since every
change increments $N$, the gallery did not change: it equals $G_s$, and by
the invariant $\Pt(G_s)=\Gid$. By $\neg\mathsf{BadVerdict}$ and
$\Pt(\Cid)=\vv_\rho$, $\neg\Dup(\vv_\rho,\Gid)$, and entry-wise $\Dup$ gives
$\neg$W3.

\emph{Bounds.} $\Pr[\mathsf{ForgeES}]\le k_{\ES}\Adv^{\mathsf{euf}}_{\mathsf{Sig}}(q_{\ES})$;
$\Pr[\mathsf{Coll}]\le\Adv^{\mathsf{cr}}_H$, counted once because the same
$\mathcal{B}_2$ outputs the collisions of Lemma~\ref{lem:r};
$\Pr[\mathsf{BadMig}\wedge\neg\mathsf{Coll}]\le n\Adv^{\mathsf{euf}}_{\mathsf{Sig}}(q_M)$;
and $\Pr[\mathsf{BadVerdict}]\le q_S\Adv^{\mathsf{res\text{-}int}}_{\Bk}$, where
$\mathcal{B}_4$ guesses the session and plays BioZKFHE's single-session
experiment while simulating the $\ES$, the gateway, other sessions and the
corrupted members. A union bound completes the proof.
\end{proof}

\subsection{Concrete BioZKFHE Integrity Bound}
\label{app:bgv-integrity-bound}

For BioZKFHE, the backend term is concrete. Its result-integrity experiment
covers a single session, and its Corollary~1 bounds a session with $M$
blocks by $M(K_m+1)\epsilon_{\mathsf{loc}}+\epsilon_H$. In BioZKFHE's
notation, $\epsilon_{\mathsf{loc}}$ is the soundness error of one local
proof instance and $\epsilon_H$ the hash term;
$K_m=n_q n_{\mathsf{chunk}}(2L-1)$ counts the local proof instances of a
block, and the extra instance per block is its binding check. Taking the
union over sessions gives
$q_S\,\Adv^{\mathsf{res\text{-}int}}_{\Bk}\le
q_S M_{\max}(K_{\max}+1)\epsilon_{\mathsf{loc}}+\epsilon_{H,\mathsf{total}}$,
with authentication and threshold decryption treated as ideal functionalities.
This multi-session step is ours, not BioZKFHE's. It requires local soundness
to hold for adaptively chosen statements over shared parameters, and $q_S$
counts every session the adversary initiates, including aborted and duplicate
ones, not only those that finalize. The enrollment history, concurrent
sessions and malicious $\Cen$ are covered by W1--W3, not by this term.

\subsection{Person-Level Corollary for Uniqueness}

Order person $p$'s enrollment captures as $X^p_1,\dots,X^p_{q_E(p)}$. When
the $i$th capture happens and $p$ already has a vector in $\Gid$, let $g_p$ be
the earliest and $F^p_i$ the event $\neg\Dup(X^p_i,\{g_p\})$; $\fnm$ is the
supremum of $\Pr[F^p_i\mid\text{history}]$.

\begin{corollary}
$\Pr[\mathsf{SYB}]\le\Pr[\mathrm{W1}\vee\mathrm{W2}\vee\mathrm{W3}]+q_E\fnm$
with $q_E=\sum_{p\in\mathcal{P}_{\mathsf{adv}}}q_E(p)$ for the global
existential event defined in main-paper Section~\ref{sec:defs}. For a fixed target
person $p$, replace $q_E$ by $q_E(p)$.
\end{corollary}

\begin{proof}
Assume $\mathsf{SYB}$ and $\neg(\mathrm{W1}\vee\mathrm{W2}\vee\mathrm{W3})$. Some $p$ holds two credentials with
vectors $\vv_j,\vv_k$ that are $p$'s captures, and $\rho_j$ finalizes first,
so $g_p$ entered $\Gid$ no later than $\vv_j$. As in Lemma~\ref{lem:u},
$\neg\Dup(\vv_k,\{g_p\})$. If the capture of $\vv_k$ happened after $g_p$
entered $\Gid$, then $F^p_i$ occurred. Otherwise $\rho_k$ opened before the
write of $g_p$, since the $\ES$ captures only after a session opens, and
finalized after it, so its snapshot count differs from its final count and
I3 forbids a unique finalization, a contradiction. Hence some $F^p_i$
occurs, and a conditional union bound over $p$'s captures gives $q_E(p)\fnm$.
\end{proof}

\subsection{Authentication}

The accepted predicate is that of the public gateway in
main-paper Section~\ref{sec:auth}. Additional events are $\mathsf{ForgeACM}$,
$\mathsf{DerivFail}_{\AFE}$ (the $\AFE$ signs something other than its own
capture and derivation under the challenge supplied by the holder's host),
$\mathsf{CollF}$ (two different challenge or policy encodings with equal
reduced digests), $\mathsf{ExtFail}$ (no witness extracted for an accepted
proof), $\mathsf{BindFail}$ (one commitment opened two ways) and
$\mathsf{BadState}$ (the gateway deviates from its specification). Fix an
accepted submission. U1 follows from the issuance path as in the W1 case of
Lemma~\ref{lem:s}, with single consumption from $\neg\mathsf{BadState}$. For
U2, the device signature is genuine by $\neg\mathsf{ForgeACM}$ and refers to a
logged capture by $\neg\mathsf{DerivFail}_{\AFE}$; that capture's challenge
has the same reduced digest as the submitted challenge, so they are equal by
$\neg\mathsf{CollF}$. For U3, extraction gives openings that satisfy the
relation; the enrollment opening equals the $\ES$'s vector by
$\neg\mathsf{BindFail}$, provided the $\ES$ used the same policy field, which
the link signature covers; the
probe opening equals the $\AFE$'s vector likewise. U4 follows from the
same-transaction checks and $\neg\mathsf{BadState}$. Summing the events gives
main-paper Eq.~\eqref{eq:auth}. For main-paper Eq.~\eqref{eq:imp}, an honest user's own capture is
bound to that user's challenge, whose beneficiary is that user, so an
impersonating acceptance must use a capture of an adversarial person within
$\tauauth$ of the honest template, and a conditional union bound over the
captures for that credential gives $q_h\fmax$.
This bound excludes template leakage with a presentation attack.

\subsection{Authentication in the Unlinkable Profile}
\label{app:privateauth}

For the gateway of the unlinkable profile, interpret U1--U4 on the credential and device
extracted from an accepted proof. Holder validity means knowledge of the
identity secret and authorization by the challenge beneficiary; the issuance
and authentication addresses need not coincide. Let $\epsilon_{\mathsf{mem}}$
bound a failure of authenticated membership for either tree over the entire
execution. This includes opening a root to an unregistered leaf and opening
the empty leaf zero to a valid credential or device tuple. The latter needs
zero-preimage resistance of the domain-separated leaf hashes, in addition to
collision resistance. The device authority registers well-formed keys and
leaves, and the verifier is pinned to the configured relation and tree depths.

Under these conditions and the extraction assumption below,
main-paper Eq.~\eqref{eq:auth} applies with the additional term
$\epsilon_{\mathsf{mem}}$ and with the EdDSA signature scheme that the
unlinkable profile uses for the $\AFE$. This is a conditional extension, not a reduction of
Poseidon's zero-preimage resistance to collision resistance.

\begin{proof}
Extract a witness from an accepted proof. Membership soundness identifies
issued credential and registered device leaves. The credential leaf binds
the enrollment commitment, identity commitment, policy and version; atomic
issuance, the station's link signature and its derivation log then give U1.
The in-circuit device signature binds the probe commitment, context, policy,
model, quantization, firmware and expiry. Signature authenticity and correct
device derivation identify a capture log entry, while reduced-context
collision resistance identifies the challenge. This gives capture origin.
The two commitment openings and the bounded distance relation follow from
extraction; commitment binding identifies them with the station and capture
logs, and the gateway pins the threshold, giving U3.

An accepted historical root can precede insertions but cannot precede a
revocation: zeroing a leaf invalidates every earlier root. Insertions preserve
previous leaves. Thus membership under an accepted root implies that both
leaves are still active, assuming correct tree updates and no empty-leaf
opening. Sender, policy, deadline, scope and nullifier checks, followed by
atomic challenge consumption, give U4, including the live device. A
union bound over the original bad events and membership failure completes
the argument. The gateway of the unlinkable profile has no per-device counter, so this proof
does not establish the attempt budget.
\end{proof}

\emph{Extraction model.} We explicitly assume extraction for new accepted
statements in the presence of previously observed honest proofs. For an
already issued honest statement the reduction keeps its original witness.
We do not infer this property merely from Groth16's ordinary knowledge
soundness or from its re-randomizability. Instantiating the assumption for
the precise setup and adversary model remains an independent review
obligation. The term $q_{\mathsf{acc}}\Adv^{\mathsf{ks}}_\Pi$ denotes a
conservative bound on extraction failure over the execution.

\subsection{Conditional Protocol-Level Unlinkability}
\label{app:privacy}

The public observer receives all transaction calldata, events, public state
and metadata, including issuance commitments and identity commitments,
device registration fields, challenge callers, funding edges, proof bytes
and all nine public inputs. It does not receive the secrets or openings held by honest holders' hosts
and devices. This observer differs from the $\Gauth$ adversary, which
controls the holders' hosts.

In Game $\Glink$ of main-paper Definition~\ref{def:glink}, both candidates can successfully
authenticate under the same policy, root pair, scope and expiry. Transaction
accounts, funding, timing, ordering and subsequent public behavior are
chosen independently of the challenge bit. Neither candidate nor the
devices they require can be corrupted or selectively revoked to make only
one candidate eligible. Other-scope queries are allowed; neither candidate
may be queried in the challenge scope before or after the challenge.

We require three distributional assumptions: adaptive multi-statement zero
knowledge under an honest setup; hiding of the v2 probe commitment with
fresh high-entropy salts even given public enrollment records and known
candidate vectors; and pseudorandomness of
the nullifier $\mathsf{Poseidon}(\texttt{nf},s,\scope^*)$ at the fresh
challenge scope $\scope^*$, jointly with the public identity commitment
$\mathsf{Poseidon}(\texttt{id},s)$, credential leaf and other-scope outputs. The latter
two are explicit heuristic assumptions about Poseidon with domain
separation, not consequences of collision resistance alone. Their auxiliary
information includes the simulated setup and preceding public transcript,
but excludes the candidates' secrets.

\begin{proof}[Conditional hybrid argument]
For each candidate world, first replace honest proofs by simulated proofs.
Next replace the challenge probe commitment by a common reference
distribution using commitment hiding, and replace the fresh-scope nullifier
by an independent field element using the joint pseudorandomness assumption.
The simulator is an efficient function of the public statement under the
simulated setup. After replacing statement components it may run on a
statement without a witness; this step asserts indistinguishability of its
inputs and outputs, not real-setup provability of a false statement.
The query restrictions prevent a later response in the challenge scope from
revealing the substituted nullifier. Both candidate worlds now have the same
distribution because their external metadata and eligibility agree. The
triangle inequality bounds distinguishing advantage by the sum of the two
worlds' zero-knowledge, commitment-hiding and joint-pseudorandomness errors.
\end{proof}

This argument hides the selected member conditional on the stated metadata
and eligibility restrictions. It does not hide a funding graph, guarantee a
large anonymity set, or protect a corrupted holder's host. The five-session public
trace audit tests observable fields and intended equality patterns; it does
not establish these cryptographic assumptions.

\section{Reproduction}
\label{app:repro}

The artifact at \url{https://github.com/plan-lab-szu/biobind} contains the contracts, circuits,
evaluation scripts and the archived outputs behind every reported number.
Its README gives the toolchain versions, build instructions, the run
identifier of each result and the scope boundaries. Historical outputs are
retained; local proving keys are benchmark-only and are not included.

\section{Extended Trust Comparison}
\label{app:comparison}

\subsection{Re-authentication Trust}

World ID's WIP-201/202 draft~\cite{wip201,wip202} places the matching engine
inside a cloud TEE and uses a zero-knowledge proof to bind its signed result
to a credential and challenge. The cloud host and its operator are trusted for
availability; isolation and attestation protect the data, while the engine
provider is trusted for matching correctness. This comparison concerns that
draft, not the phone-local Face Auth described in the official
FAQ~\cite{worldfaq2026}.
\sysname{} places the trusted component on the user side: the $\AFE$ signs a
commitment to its own capture, and the match itself is proved in zero
knowledge and checked on chain. Capture integrity has a similar structure in
both designs, since World ID trusts app and device attestation and \sysname{}
trusts the $\AFE$; we count this as a shared assumption, not a difference.
The difference is where raw features are processed and who can check the
comparison. Table~\ref{tab:world} contrasts them.

\begin{table}[t]
\centering
\caption{Trust and privacy at re-authentication in World ID's draft design and in \sysname{}. At enrollment both designs process raw features in an attended, certified device (the Orb, the $\ES$).}
\label{tab:world}
\footnotesize
\begin{tabular}{@{}>{\raggedright\arraybackslash}p{2.1cm}>{\raggedright\arraybackslash}p{2.9cm}>{\raggedright\arraybackslash}p{2.9cm}@{}}
\toprule
 & World ID WIP-201/202 (draft) & \sysname{} \\
\midrule
Trusted component & Cloud enclave running a closed engine & User-side $\AFE$ \\
Raw face goes to & The enclave & Nowhere off the user side \\
Who compares & The enclave engine & The prover on the holder's host; the gateway checks the proof \\
Re-authentication root of trust & Enclave attestation and the cloud provider & $\AFE$ certificate, CRS and contract \\
Capture integrity & App and device attestation & $\AFE$ certificate \\
\bottomrule
\end{tabular}
\end{table}

\subsection{World ID-Inspired MPC Enrollment Interface}
\label{app:world-mpc}

World ID illustrates which enrollment conditions survive a backend change. The published iris MPC
protocol~\cite{bloemen2024ampc} evaluates a duplicate predicate over shared
templates under semi-honest, honest-majority assumptions; its second
construction protects both codes and masks. That construction supplies
the computational component of a proposed interface mapping, separate
from the re-authentication comparison above. This design analysis neither implements an MPC extension nor audits
World ID's current service.

The mapping realizes I1--I4 with secret shares. World ID's whitepaper
describes storing the fragments after a successful uniqueness
check~\cite{worldwhitepaper}. Reusing the checked shares can satisfy
query--insertion consistency without separately supplied insertion shares;
capture provenance and authorization of the credential and state transition
remain distinct conditions. In our proposed mapping, a certified enrollment
station derives the shares and credential commitment from one captured
template. Its signed manifest binds their canonical digests to the session,
matcher policy and queried gallery version. Each recipient authenticates
the station, checks the delivered shares against the manifest and uses the
checked shares for insertion. Share hashes authenticate bytes, but do not
prove correspondence with the committed template; this rests on trusted
station derivation, as in the BGV instance. Removing this derivation trust
requires verifiable sharing and a proof linking the shared template and
credential commitment.

Table~\ref{tab:world-mpc-map} summarizes the preserved conditions and
their proposed realization. The parties must agree on one logical gallery
history, although their local shares and state digests differ. A successful
finalization must atomically couple the authenticated state transition to
issuance, with the required shares durably prepared before commitment;
aborted sessions must not enter the searchable gallery. Distributed commit
and recovery are additional implementation requirements. MPC evaluation alone does not prevent two correct unique verdicts on one
snapshot from authorizing duplicate credentials.

\begin{table}[t]
\centering

\caption{Proposed World ID-inspired MPC interface mapping. These are
requirements for composition, not claims about deployed World ID checks.}
\label{tab:world-mpc-map}
\footnotesize
\setlength{\tabcolsep}{3pt}
\begin{tabular}{@{}>{\raggedright\arraybackslash}p{0.4cm}>{\raggedright\arraybackslash}p{2.6cm}>{\raggedright\arraybackslash}p{4.9cm}@{}}
\toprule
 & Preserved condition & MPC realization and trust \\
\midrule
I1 & One capture, query, insertion and issued record & Authenticate station shares and reuse them for insertion; station guarantees consistency with the capture and credential commitment. \\
I2 & Authenticated slot and state transition & Assign the slot from current state; authenticate each party's prepared update and finalize one logical transition with issuance. \\
I3 & Verdict valid at finalization & Bind the queried version; reject if an intervening insertion changed the append-only gallery. \\
I4 & Sound unique verdict & Correct MPC on the bound inputs and snapshot, plus authenticated result delivery; semi-honest parties must follow the protocol. \\
\bottomrule
\end{tabular}
\end{table}

The main change in trust concerns I4. BioZKFHE supplies verifiable
encrypted evaluation against a malicious compute node, subject to its
committee assumptions. In the cited MPC model, even corrupted computing
parties follow the protocol; an honest majority alone does not justify
correctness under active deviations. A signed or threshold-signed verdict
authenticates its origin, not its computation. The proposed interface
therefore needs authenticated output delivery under the passive-adversary
assumption, or a separately established malicious-secure MPC protocol or
verifiable computation mechanism if active deviations are allowed.
Capture provenance and enrollment-attempt limits remain separate trust
requirements in either case.

Lemma~\ref{lem:interface} supplies the common state argument,
while the concrete BGV theorem remains instance-specific. An MPC instantiation needs
its own input-consistency and verdict-soundness arguments for adaptive
histories and its chosen corruption model, replacing the BGV-specific
premises. Using iris templates also requires suitable capture and commitment
relations, a specified matcher and authentication circuit, and new biometric
error estimates. Our face-circuit costs, BGV capacity results and timings
do not measure this alternative.

\section{\texorpdfstring{Extended Literature Context}{Extended Literature Context}}
\label{app:related-literature}

This appendix preserves the extended source comparisons behind main-paper
Table~\ref{tab:related}. The cited systems have different goals and trust
models; their descriptions do not assert properties of unexamined deployments.

\subsection{Biometric Verification, Identification and Template Protection}

A 1:1 verification compares a probe with one enrolled template and accepts
when a distance or similarity crosses a threshold. A 1:N identification, or
deduplication, compares a probe with every entry in a gallery. Their error
behavior differs sharply: a per-comparison false match rate that is
acceptable for verification compounds over $N$ comparisons, so deduplication
needs much lower per-comparison error as the gallery grows. Template
protection standards such as ISO/IEC 24745~\cite{iso24745} ask for
irreversibility, unlinkability and renewability of stored references. This
paper does not propose a template-protection scheme; it uses commitments and
homomorphic ciphertexts so that raw features never leave the user side after
enrollment, and it
asks what must hold for the stored objects to come from one capture.

\subsection{Private 1:N Deduplication}

Private deduplication can use distributed computation on secret shares
or verifiable evaluation on ciphertexts. World ID's published iris MPC
protocol uses the former under semi-honest, honest-majority assumptions;
BioZKFHE uses the latter against a malicious compute
node~\cite{bloemen2024ampc,xin2026biozkfhe}. Neither representation alone
specifies how a query, gallery insertion and credential are bound to one
capture or how concurrent requests are finalized. We study these interface
conditions for both routes in main-paper Section~\ref{sec:gaps}.

Our concrete instance uses BioZKFHE's packed gallery encoding, publicly
verifiable block proofs and threshold
committee. Its honest-client assumption and snapshot-bound result integrity
leave malicious enrollment and gallery-update composition to the surrounding
protocol. We add and evaluate those checks, and
Appendix~\ref{app:world-mpc} maps the same conditions to MPC.

Wang et al. give a private token-based aid distribution system that
assumes recipients register only once~\cite{wang2023nyadid}. Janus addresses
that registration problem using MPC,
somewhat homomorphic encryption or a trusted execution environment, and
requires successful registrations to enter the database promptly. Its
deduplication component returns a binary result; the registration station's
decision to act on that result is handled by the surrounding
process~\cite{edalatnejad2024janus}. xDup instead studies private
deduplication across humanitarian organizations using fuzzy private set
intersection~\cite{rausch2026xdup}. These settings reinforce the need to
specify who may turn a private verdict into an authorized state change.

\subsection{Zero-Knowledge Biometric Authentication}

Several schemes verify a 1:1 match without revealing templates.
BAZKP~\cite{mao2023bazkp} combines a proof of squared Euclidean distance
under Pedersen commitments with a range proof, as described in the
article's ``Our scheme'' section. zkSABER~\cite[Sec.~III, pp.~3--5]{chinen2025zksaber} combines
quantized cosine matching, anonymous Merkle membership and nullifier-based
replay prevention in one circuit. BioZero~\cite{lin2025biozero} provides identity-bound, publicly verifiable
authentication over committed templates. It leaves uniqueness, issuance
and other lifecycle operations to the surrounding identity system; its
accepted witness does not independently establish a fresh sensor capture.
Kothari et al.~\cite{kothari2026fingerprint} recompute a cancelable
fingerprint match in Groth16, bind a verifier nonce into the proof and verify
on chain for 255{,}275 gas, independent of template size. They state the
limits that matter here: the probe is not constrained by the circuit, live
acquisition is outside the protocol, and the scheme provides 1:1
verification only. All of these show that whole-vector matching under
commitments with constant-size on-chain verification is established; none
addresses uniqueness or capture provenance.

\subsection{Identity Credentials and Deployed Proof-of-Personhood}

The broader credential literature separates trust in identity
issuance from later proof and use. Krul et al. systematize trust across
self-sovereign identity components~\cite{krul2024sok}; CanDID combines
decentralized credentials and Sybil resistance through a committee and
existing accounts~\cite{maram2021candid}. A formal personhood framework
defines uniqueness and unlinkability as distinct
properties~\cite{choudhuri2026pop}. \sysname{} focuses on the additional
biometric registration boundary: which protected query, gallery update and
credential arose from the same capture.

World ID~\cite{worldwhitepaper} provides a system-level reference combining
Orb-based acquisition, private iris uniqueness checking and credential use.
Its official Face Auth description places that check on the user's
phone~\cite{worldfaq2026}. Its draft re-authentication
design (WIP-201~\cite{wip201}) runs a closed face-matching engine inside a
cloud Nitro enclave, which sees raw biometrics and signs the result; WIP-202~\cite{wip202}
binds that signed result to the user's credential and the relying party's
challenge in a zero-knowledge proof. These are draft beta specifications, checked on October 31, 2026, and
are distinct from the phone-local Face Auth path. WIP-202 already binds the
matched image to the credential and challenge; our comparison concerns
trust allocation and registration-state guarantees, not an absence of
binding in World ID. WIP-201 discusses alternatives to its enclave design; the device-side
trusted capture assumed here is a different trust boundary.
Aadhaar and MOSIP certify capture devices; we do not analyze these
national systems.

\subsection{Anonymous Credentials and Membership Proofs}

The unlinkable profile of main-paper Section~\ref{sec:unlink} follows the
Semaphore pattern~\cite{semaphore}: a holder proves membership of a leaf in a
Merkle tree of valid credentials and publishes a scope-specific nullifier,
which gives one stable pseudonym per application without revealing which leaf
was used. We use Poseidon~\cite{grassi2021poseidon} for in-circuit hashing.
zk-creds shows how general-purpose zero-knowledge proofs can derive
anonymous claims from existing identity infrastructure~\cite{rosenberg2023zkcreds};
our unlinkable profile instead proves membership together with a
capture-bound biometric match.

Main-paper Table~\ref{tab:related} summarizes the compared operations and their input
assumptions. BioZero supplies an authentication relation, BioZKFHE supplies
a verifiable deduplication backend, and World ID supplies a broader identity
workflow with device and service trust. \sysname{} specifies and implements
the enrollment-state and capture bindings needed by its on-chain gateway.
The comparison does not establish missing properties in unexamined
deployments.

\section{Implementation Details}
\label{app:implementation}

\emph{Contracts.} The enrollment source exists in two versions with the same
checks: a reference version that keeps session fields in storage, and a
compact version that stores digests and takes records as calldata, as
discussed in main-paper Section~\ref{sec:construction}. A finalization adapter turns a final record
into an issuance receipt. The public-profile gateway implements typed
canonical policies, verifier pinning, live-state checks and the lifecycle
operations. The gateway of the unlinkable profile deploys two incremental Poseidon
Merkle trees with epoch-tagged root histories. For two-input Poseidon we
generate an inline-assembly library from the constants and optimized round
structure of circomlibjs; it agrees with circomlibjs on 69 inputs, including
$0$, $1$, $r-1$, where $r$ is the order of the circuit field, and 64 random inputs, and costs 21{,}560 gas per hash
against 32{,}183 for the generated external contract. circomlibjs cannot
generate a seven-input Poseidon contract within the EVM stack limit, so the
device authority computes device leaves off chain and publishes their fields
in events.

\emph{Circuits.} The public-profile circuit proves main-paper Eq.~\eqref{eq:dist} over
two commitments for $L=128$ or $512$. The circuit of the unlinkable profile adds the
credential tree, the device tree, EdDSA verification on BabyJubJub and the
nullifier. We recommend $L=512$, 12-bit coordinates ($\Gamma=2047$), commitment
format v2 and two depth-20 trees.

\emph{Devices.} An enrollment bridge plays the $\ES$: it receives a capture
package signed with P-256 and protected by AEAD, derives transparent
stand-ins for the two BGV ciphertexts and the commitment from the same vector, seals a pending record
before releasing anything, and signs declarations and links. A software
mock $\AFE$ follows the same wire format with an EdDSA key. Both use
software keys; the hardware capture module is deferred.

\emph{Committee.} The committee is simulated. It evaluates the verdict on a
plaintext model of the packed gallery over the session snapshot and recomputes merged
blocks over transparent stand-in ciphertexts; block proofs, the check that every block is proved, and
threshold decryption are not executed. A separate C++ experiment executes
real BGV encryption, loading, addition, serialization and query evaluation,
as discussed in main-paper Section~\ref{sec:bgv}; it is not integrated into this bridge or committee.

\emph{Biometrics.} Faces are aligned with MTCNN~\cite{zhang2016mtcnn} and
embedded with an InceptionResnetV1 model pretrained on
VGGFace2~\cite{cao2018vggface2} (facenet-pytorch 2.6.0, 512 dimensions, $L_2$
normalized), running on CPU with torch 2.2.2. Quantization multiplies by
$\Gamma$ and rounds half away from zero.

\emph{Toolchain and setup.} Contracts compile with solc 0.8.36 (200
optimizer runs, Shanghai) and run on Ganache 7.9.2; scripts run on Node.js
24.19.0. Circuits use Circom 2.2.3 with O2 and snarkjs 0.7.6, with
circomlib 2.0.5 and circomlibjs 0.1.7. Every Groth16 key comes from a local,
single-party Powers-of-Tau and phase-2 setup made for benchmarking. These keys
are not deployable, and the released artifacts will not include deployable
keys.

\subsection{BGV Confirmation and Serialization}
\label{app:bgv-confirmation}

The migration
fields are part of the finalization confirmation that BioZKFHE's committee
already sends, so the rule adds no confirmation round. Each member must
obtain the ciphertexts and perform $L$ additions and three block hashes;
serialization, validation and transfer also contribute to the cost. The
published BioZKFHE contract records each confirmation as a transaction from
the member's address, with no roster check and no structure for the result
digest. Our source therefore verifies $t$ signatures from a fixed roster over
a confirmation that covers the chain and contract, $\sid$, the session context, both
ciphertext digests, the verdict, the verified output digest and the migration
fields; a roster that changes over time would also need a committee epoch in
the message.

Two deployment assumptions support the rule. First, data availability:
members must obtain the full bytes of $C_M$ and $\Cen$, since digests alone
cannot be checked; the backend's storage and transport must guarantee this.
Second, canonical serialization: all parties hash the same bytes only if
they fix the SEAL version, the parameters, the ciphertext representation, the
number of rows and the length encoding, in addition to the uncompressed
mode.

\subsection{Digest-Only Source}
\label{app:digest-source}

During a session the source stores only the hash of the session record; the
full record is emitted in events, passed back as calldata in later calls and
checked against the stored hash. At finalization the source stores one
digest of the final record and emits the record. The adapter decodes the
record supplied with an issuance request, re-encodes it and compares it with
the source digest before running its field checks, so a non-canonical
encoding with trailing padding cannot pose as another record. Losing the
record affects availability only, since it can be rebuilt from the event
log.

\subsection{Commitment Format}
\label{app:commitment-format}

A commitment packs 12-bit coordinates, offset to non-negative values, into
field elements and hashes them with a Poseidon chain under a role- and
format-separated header. Format v1 absorbs two chunks per Poseidon call;
format v2 absorbs fifteen per $\mathsf{Poseidon}(16)$ call, which saves about
2{,}175 constraints per commitment with unchanged assumptions. The two
formats are domain-separated by a format version field.

\section{Evaluation Controls and Trace Details}
\label{app:evaluation-details}

\subsection{Attack Controls and End-to-End Measurements}

Table~\ref{tab:controls} summarizes the controls. Each attack is followed by
the honest version of the same operation, which succeeds. Fourteen honest
enrollments across a block boundary leave the on-chain last-block commitment
equal to the committee's recomputation, and the ten-file test suite passes all 45 tests.

\begin{table}[t]
\centering
\caption{Negative controls by component.}
\label{tab:controls}
\footnotesize
\setlength{\tabcolsep}{3pt}
\begin{tabular}{@{}>{\raggedright\arraybackslash}p{1.3cm}>{\raggedright\arraybackslash}p{4.9cm}>{\scriptsize}l@{}}
\toprule
Component & Attacks rejected & Run \\
\midrule
Source (both versions) & swapped $\Cen$ or missing declaration (A1); wrong $N$ (A2); slot mismatch; same-person race (A3$'$) and two-person race (A3); merges with $t{-}1$ corrupt, off-roster, duplicated or transplanted signatures; append with $u_{\mathsf{new}}\ne H(\Cen)$; adapter on a duplicate session & R-ENRSRC2 \\
Adapter & record differing from the source digest in any field & R-CALLDATA \\
Unlinkable gateway & swapped $\idc$; second issuance from one record; wrong beneficiary; reused one-time nullifier; proof on a root older than a revocation; revoked holder; revoked device; scope mismatch; tampered nullifier & R-PRIVGW2 \\
Unlinkable circuit & leaf not in tree; wrong identity secret; foreign scope; device not in tree; signature on another probe; distance over threshold; unopenable probe & R-PRIV \\
End to end & swapped $\idc$; host-chosen $\Cenr$; impostor in front of the $\AFE$ & R-PRIVE2EV2TAU \\
\bottomrule
\end{tabular}
\end{table}

The end-to-end run uses real LFW features in the recommended configuration
and passes eight of eight checks (R-PRIVE2EV2TAU-20261028). Two people
enroll through the bridge, the compact source, the simulated committee, the
adapter and the gateway of the unlinkable profile. Both authenticate with other images of
themselves and receive one stable pseudonym each; an impostor's capture
against the first person's credential fails at witness generation, with
squared distance 10{,}717{,}782 against a threshold of 5{,}253{,}245. With the
calibrated deduplication threshold, the first person's re-enrollment with
another image (inner product 2{,}580{,}826, below $\taudup=2{,}777{,}020$) was
\emph{not} detected, and that person obtained a second credential. This is the
$q_E\fnm$ term of main-paper Eq.~\eqref{eq:syb}: every protocol step executed correctly,
and the failure comes from the backend's accuracy at this operating point,
whose non-adversarial miss rate is about 17\%. Authentication costs
305{,}060--305{,}072 gas in that calibrated run; its proof times are
1.867--2.347\,s. These are three session timings, not the interleaved
benchmark median reported in main-paper Section~\ref{sec:signature-baseline}.

\subsection{Public-Profile Signature Baselines}
\label{app:public-baselines}

The on-chain setting does not itself require a zero-knowledge match: a
trusted matcher under the same capture and live-state checks can also
exclude E1--E3. We thus compare the cost of verifying opening and matching constraints
with that of additionally trusting the capture device for them. We reran the public-profile comparison on the current code
(R-BASE-20261030): bound ZK costs 336{,}239 gas, separate capture/match
signatures 123{,}102, and the merged signature B1$'$ 103{,}440. Each needs an
additional 108{,}121 gas challenge transaction. These are synthetic
128-dimensional public-profile runs. The honest matcher refuses non-matches;
a deliberately faulty authorized matcher can attest a non-match, violating
the baseline's trusted-matching premise.

\subsection{Enrollment Transaction Totals}

With $t=3$ and $n=5$, successful compact enrollment costs 228{,}746 gas
across three transactions, excluding credential issuance and BioZKFHE's block proofs.
Duplicate rejection costs 185{,}540 gas and a stale abort costs 188{,}249
gas over the session. Main-paper Table~\ref{tab:gas} reports the
per-operation costs; failed sessions incur creation and binding costs as
well as finalization.

Composed with BioZKFHE's published per-block costs, these totals give the
per-enrollment gas listed in Table~\ref{tab:compose} for the profiles that
BioZKFHE reports. Main-paper Figure~\ref{fig:cost} presents the
breakdown of the first three rows.

\begin{table}[t]
\centering
\caption{On-chain gas per enrollment, BioZKFHE alone and composed with \sysname{} (public-profile issuance, $t=3$, merge case). $\ell$: packed layers per block; $M$: blocks.}
\label{tab:compose}
\footnotesize
\setlength{\tabcolsep}{3pt}
\begin{tabular}{@{}lrrrrr@{}}
\toprule
Profile & $N$ & $M$ & BioZKFHE & Composed & Issuance \\
\midrule
FaceNet, $\ell{=}1$ & $10^4$ & 2 & 499{,}151 & 836{,}393 & 52\% \\
FaceNet, $\ell{=}1$ & $10^5$ & 13 & 1{,}431{,}885 & 1{,}769{,}127 & 25\% \\
FaceNet, $\ell{=}1$ & $10^6$ & 123 & 10{,}759{,}225 & 11{,}096{,}467 & 4\% \\
FaceNet, $\ell{=}2$ & $10^6$ & 62 & 6{,}160{,}105 & 6{,}497{,}347 & 7\% \\
FaceNet, $\ell{=}3$ & $10^6$ & 41 & 4{,}185{,}244 & 4{,}522{,}486 & 10\% \\
MobileFaceNet, $\ell{=}1$ & $10^6$ & 123 & 10{,}568{,}206 & 10{,}905{,}448 & 4\% \\
\bottomrule
\end{tabular}
\end{table}

\subsection{Deduplication Thresholds}
\label{app:dedup-thresholds}

Table~\ref{tab:dedup} lists the thresholds chosen on the calibration half
and the error rates measured on the identity-disjoint test half, which
main-paper Figure~\ref{fig:dedup} plots.

\begin{table*}[t]
\centering
\caption{Deduplication thresholds chosen on the calibration half, evaluated on the identity-disjoint test half (12-bit inner product). FD: false duplicate; MD: missed duplicate.}
\label{tab:dedup}
\footnotesize
\setlength{\tabcolsep}{4pt}
\begin{tabular}{@{}lrrrrrr@{}}
\toprule
Threshold & $\taudup$ (cosine) & Test FMR (95\% upper) & Test FNMR & FD, $N{=}1{,}645$ & MD, $N{=}1{,}645$ & Extrap.\ FD, $10^4/10^5/10^6$ \\
\midrule
$\Gamma^2/2$ & 2{,}095{,}104 (0.500) & $2.3\times10^{-3}$ & 2.6\% & 83.3\% & 0.5\% & $\approx$100\% / 100\% / 100\% \\
FMR $\le10^{-3}$ & 2{,}322{,}295 (0.554) & $8.3\times10^{-4}$ & 5.5\% & 53.7\% & 3.9\% & $\approx$100\% / 100\% / 100\% \\
FMR $\le10^{-4}$ & 2{,}777{,}020 (0.663) & $7.3\times10^{-5}$ ($7.9\times10^{-5}$) & 18.4\% & 6.5\% & 17.1\% & 52\% / 99.9\% / 100\% \\
FMR $\le10^{-5}$ & 3{,}162{,}706 (0.755) & $3.6\times10^{-6}$ ($5.3\times10^{-6}$) & 43.8\% & 0.4\% & 40.0\% & 3.6\% / 30\% / 97\% \\
FMR $\le10^{-6}$ & 3{,}406{,}345 (0.813) & $6.7\times10^{-7}$ ($1.6\times10^{-6}$) & 67.3\% & 0.1\% & 60.8\% & 0.7\% / 6.5\% / 49\% \\
\bottomrule
\end{tabular}
\end{table*}

\subsection{Evasive Re-enrollment on Bosphorus}
\label{app:bosphorus}

An enrollee who tries to obtain a second credential can change appearance
before re-enrolling. To measure how often such a change evades deduplication
(R-BOSPHORUS-20261109), we use the Bosphorus database~\cite{savran2008bosphorus},
which holds 4{,}666 face scans of 105 people, each with a colour image. Its
poses cover neutral faces, six emotional expressions, lower- and upper-face
action units and their combinations, yaw, pitch and diagonal rotations, and
occlusion of the eyes, mouth, glasses or hair. We use the colour images only.
Seven subjects whose images the database forbids to publish are left out
before any image is read, which leaves 98 subjects and 4{,}342 images once
images marked invalid are dropped.

The first neutral image of each subject is the enrolled entry, so the gallery
holds 98 people, and every other image of that subject is a re-enrollment
attempt. Faces are detected, embedded and scored with the face model and the
12-bit inner product of the LFW experiments of
Section~\ref{sec:eval}, and we reuse the
thresholds that the LFW calibration chose for per-pair false match rates of
$10^{-3}$ to $10^{-6}$ (Table~\ref{tab:dedup}) without retuning them. As in
the LFW experiment, an attempt is missed when its highest score over the
gallery is below the threshold. Scoring only against the person's own entry
gives the same rates at the three stricter thresholds and differs by at most
two percentage points at $10^{-3}$. A face detector failure leaves the station
without a capture to deduplicate; we count it conservatively as a miss.
Detection fails on 41 images (0.9\%): nine at $90^\circ$ yaw, eight among
lower-face action units, five each for eye occlusion and upper-face action
units, and 14 elsewhere, and none for the enrolled neutral images.

We group the conditions by whether an attended station could accept the
capture. The station accepts occlusion of the eyes, mouth, glasses or hair,
expressions, action units, yaw up to $30^\circ$ and slight pitch, and rejects
larger rotations, because an attended enrollment asks for a roughly frontal
face. Table~\ref{tab:bosphorus} lists the results. Over the accepted conditions, 1.8\% of 3{,}273 attempts are
missed at $10^{-4}$, 4.6\% at $10^{-5}$ and 10.8\% at $10^{-6}$. Occlusion is
the most effective single change, with hair occlusion missed at 6.2\% and
16.9\% for $10^{-4}$ and $10^{-5}$. For eye occlusion, five of the six misses at these two thresholds are failed detections, and the
sixth is one of the 93 detected images.
A rotation of $90^\circ$ is missed at 45.9\% and 83.2\%. Neutral repeats of
the same person are never missed.

The control rows show the limits of the calibration. When neutral images of
one person are compared with a gallery of the other 97 people, 31.0\% (95\%
interval 24.5--38.2\%) produce a false duplicate at the loosest threshold,
whereas independent comparisons at a per-pair false match rate of $10^{-3}$
would give about 9\%; at $10^{-4}$ the rate is 1.1\%, in line with LFW. The
subjects were captured with one scanner under similar conditions, which can
raise impostor scores above those of web photographs. We therefore rely on
the columns from $10^{-4}$ on, because at the loosest threshold a probe may be
flagged through a match to another person.

These rates describe 98 cooperative subjects who performed prescribed poses
in front of a 3D scanner. The images are the scanner's colour textures. The
intervals treat each attempt as independent although attempts of one subject
are correlated, and are therefore optimistic. An adversary who picks the most
effective disguise or repeats attempts may exceed them, which is why
\sysname{} limits $q_E$. Main-paper Theorem~\ref{thm:main} takes $\fnm$
under adversarial presentation, and the rates above are reference values for
controlled changes of appearance, which come closer to that presentation than
the LFW rates. The Bosphorus rates fall below the 17.1\% of
missed duplicates for second LFW images at $10^{-4}$ (Table~\ref{tab:dedup}),
because LFW photographs vary without constraint in lighting and pose,
whereas Bosphorus fixes the conditions of capture.

\begin{table*}[t]
\centering
\caption{Missed duplicates when 98 Bosphorus subjects re-enroll under changed appearance, in percent with 95\% Clopper--Pearson intervals (R-BOSPHORUS-20261109). Columns give the per-pair false match rate on which the LFW threshold was calibrated. An attempt without a detected face counts as missed, and $n$ includes these failures. The last row compares neutral images of one person against a gallery of the other 97 people and counts false duplicates.}
\label{tab:bosphorus}
\footnotesize
\setlength{\tabcolsep}{4pt}
\begin{tabular}{@{}>{\raggedright\arraybackslash}p{3.8cm}rrrrrr@{}}
\toprule
Condition & $n$ & Failed & $10^{-3}$ & $10^{-4}$ & $10^{-5}$ & $10^{-6}$ \\
\midrule
\multicolumn{7}{@{}l}{\textit{Conditions an attended station can accept}} \\
Eyes occluded & 98 & 5 & 6.1\,[2.3, 12.8] & 6.1\,[2.3, 12.8] & 6.1\,[2.3, 12.8] & 8.2\,[3.6, 15.4] \\
Mouth occluded & 98 & 0 & 0.0\,[0.0, 3.7] & 1.0\,[0.0, 5.5] & 7.1\,[2.9, 14.2] & 24.5\,[16.4, 34.2] \\
Glasses & 97 & 1 & 1.0\,[0.0, 5.6] & 4.1\,[1.1, 10.2] & 8.2\,[3.6, 15.6] & 17.5\,[10.5, 26.6] \\
Hair occluded & 65 & 2 & 3.1\,[0.4, 10.7] & 6.2\,[1.7, 15.0] & 16.9\,[8.8, 28.3] & 32.3\,[21.2, 45.1] \\
Six emotional expressions & 421 & 3 & 1.4\,[0.5, 3.1] & 1.9\,[0.8, 3.7] & 8.6\,[6.1, 11.6] & 15.9\,[12.6, 19.8] \\
Lower-face action units & 1{,}443 & 8 & 0.9\,[0.5, 1.5] & 1.7\,[1.1, 2.5] & 4.5\,[3.5, 5.7] & 11.7\,[10.1, 13.5] \\
Upper-face action units & 402 & 5 & 1.2\,[0.4, 2.9] & 1.7\,[0.7, 3.5] & 3.2\,[1.7, 5.5] & 6.2\,[4.1, 9.0] \\
Combined action units & 159 & 0 & 0.0\,[0.0, 2.3] & 0.0\,[0.0, 2.3] & 0.0\,[0.0, 2.3] & 3.1\,[1.0, 7.2] \\
Yaw up to $30^\circ$ & 294 & 1 & 0.3\,[0.0, 1.9] & 0.3\,[0.0, 1.9] & 0.7\,[0.1, 2.4] & 4.4\,[2.4, 7.4] \\
Slight pitch & 196 & 1 & 1.5\,[0.3, 4.4] & 1.5\,[0.3, 4.4] & 1.5\,[0.3, 4.4] & 2.0\,[0.6, 5.1] \\
All of the above & 3{,}273 & 26 & 1.1\,[0.8, 1.6] & 1.8\,[1.4, 2.3] & 4.6\,[3.9, 5.4] & 10.8\,[9.7, 11.9] \\
\addlinespace[2pt]
\multicolumn{7}{@{}l}{\textit{Conditions an attended station rejects}} \\
Yaw $45^\circ$ & 196 & 1 & 0.5\,[0.0, 2.8] & 1.0\,[0.1, 3.6] & 9.7\,[5.9, 14.7] & 25.0\,[19.1, 31.7] \\
Yaw $90^\circ$ & 196 & 9 & 13.8\,[9.3, 19.4] & 45.9\,[38.8, 53.2] & 83.2\,[77.2, 88.1] & 96.9\,[93.5, 98.9] \\
Full pitch & 195 & 4 & 3.1\,[1.1, 6.6] & 3.1\,[1.1, 6.6] & 5.1\,[2.5, 9.2] & 9.7\,[6.0, 14.8] \\
Diagonal turn & 197 & 1 & 0.5\,[0.0, 2.8] & 1.5\,[0.3, 4.4] & 8.1\,[4.7, 12.8] & 23.4\,[17.6, 29.9] \\
\addlinespace[2pt]
\multicolumn{7}{@{}l}{\textit{Controls}} \\
Same person, neutral repeat & 187 & 0 & 0.0\,[0.0, 1.9] & 0.0\,[0.0, 1.9] & 0.0\,[0.0, 1.9] & 0.0\,[0.0, 1.9] \\
False duplicates, 97-person gallery & 187 & 0 & 31.0\,[24.5, 38.2] & 1.1\,[0.1, 3.8] & 0.0\,[0.0, 1.9] & 0.0\,[0.0, 1.9] \\
\bottomrule
\end{tabular}
\end{table*}

\subsection{Public Observer Trace}
\label{app:observer-trace}

R-PRIVOBS-20261030 decodes actual calldata, logs and transaction metadata for
five successful authentications in the unlinkable profile: two captures of A in one
application, B in that application, and A and B in a second application.
The collector receives no fixture identity labels, openings or keys.
Same-scope nullifiers link A's first two sessions despite fresh sender
addresses. Across scopes the nullifiers differ, but all beneficiaries share
an observable funding source. That source alone does not identify a holder.
All probe commitments differ. Issuance calldata exposes both $\Cenr$ and
$\idc$, although neither is a direct authentication input. These six field
and equality-pattern checks pass; they are not a statistical privacy test
or a proof of anonymity. Appendix~\ref{app:privacy} states the conditional
protocol-level claim and the metadata it cannot hide.

The public-profile prototype's policy migration also links sessions through
the holder address, credential lineage and supersede event. The unlinkable
profile removes these authentication handles but retains the observable
issuance and metadata described above. The holder's host receives the probe opening
in the present prover configuration; hiding that opening would require an
encrypted channel to an attested prover.

\subsection{Circuit Optimization and Timing Protocol}
\label{app:circuit-details}

Table~\ref{tab:circuit} reports constraint counts and proving times
(R-PRIV512-20261027, R-PRIVV2-20261028, R-PROVETIME-20261028). Range checks on
1{,}024 coordinates dominate format v1, followed by the two Poseidon
commitment chains, the credential tree and EdDSA. Narrowing the distance
width leaves no room, since it is already the minimum that holds the sum, and
a shallower device tree would cap a policy at 128 devices although each
person carries one. Format v2 brings the 12-bit, 512-dimensional circuit below
$2^{15}$ constraints, which halves the Groth16 evaluation domain. We measure
v1 and v2 alternately on the same input, five rounds after one warm-up:
proving drops by 26\% and the total by 15\%, although witness generation is
slower because $\mathsf{Poseidon}(16)$ is heavier in WebAssembly. With depth 20
for both trees the total is 1{,}801\,ms. Real LFW vectors pass for the genuine user and fail impostor witness
generation in all three configured v2 circuits.

\section{BGV Capacity, Measurements and Parameter Controls}
\label{app:bgv-details}

{
\subsection{Unified configuration and capacity validation}
\label{app:bgv-unified}

Unified512-2 uses $L=512$, $\Gamma=2047$, degree $d=8192$, two layers and
$T=8{,}473{,}312$. Its plaintext prime is $p=281474976694273$; the coefficient
primes, in order, are $1152921504606830593$, $1125899905744897$,
$1125899906629633$ and $1125899906826241$. The last is the special prime;
the first three give the 160-bit data modulus. All stages use identical
parameters and retain SEAL's tc128 check. We enforce $|v_i|\le2047$ and
$\|\vv\|_2^2\le4236655$, consistent with the rounded unit-vector bound
$D=(2047+\sqrt{512}/2)^2$. The actual prime satisfies both C1 inequalities,
$T>2D$ and $p>2D(1+T)$.

Three separate processes generate fresh SEAL keys from OS entropy and use
plaintext seeds 20261031, 20261032 and 20261033. Each independently encrypts
and adds all 512 rows for every registration, reaching 16{,}384 entries
without refreshing the accumulated block. Each history uses one fixed
synthetic query. Inputs are normalized and quantized synthetic vectors;
registrations 1, 8192, 8193 and 16384 use admissible signed axis vectors.
At each checkpoint we decrypt all accumulated rows, evaluate the encrypted
query and compare all 8192 slots and both packed layers with the plaintext
reference. A fresh encryption of the same gallery provides a diagnostic
control, never a replacement for the accumulated block.

Table~\ref{tab:bgv-capacity} reports all 16 checkpoints per history.
All five error counters---row, query, packed-score, fresh-query and
fresh-packed-score---are zero at all 48 checkpoints. The 128-registration
precheck also passes its eight checkpoints. These tests cover specified
histories and one query per key; repeated checks of the same positions are
not independent trials or a general failure-probability estimate. Keys and key
fingerprints are not exported; parameter IDs identify parameters.

\begin{table}[t]
\centering
\caption{Unified512-2 capacity checkpoints. All noise budgets are in
bits. Every row, query and packing check, including fresh-gallery controls,
has zero errors. The row column is the minimum over all rows and three runs.}
\label{tab:bgv-capacity}
\footnotesize
{
\begin{tabular}{@{}rrrrr@{}}
\toprule
Registrations & Query 1 & Query 2 & Query 3 & Row min. \\
\midrule
1 & 31 & 31 & 29 & 103 \\
2 & 30 & 30 & 28 & 102 \\
4 & 29 & 29 & 27 & 101 \\
8 & 28 & 28 & 26 & 100 \\
16 & 27 & 27 & 25 & 100 \\
32 & 26 & 26 & 24 & 99 \\
64 & 25 & 25 & 23 & 98 \\
128 & 24 & 24 & 22 & 97 \\
256 & 23 & 23 & 21 & 96 \\
512 & 22 & 22 & 20 & 95 \\
1{,}024 & 21 & 21 & 19 & 94 \\
2{,}048 & 20 & 20 & 18 & 93 \\
4{,}096 & 19 & 19 & 17 & 92 \\
8{,}192 & 18 & 18 & 16 & 91 \\
8{,}193 & 18 & 18 & 16 & 91 \\
16{,}384 & 17 & 17 & 15 & 90 \\
\bottomrule
\end{tabular}}
\end{table}

\subsection{Source-dependent noise bound}
\label{app:bgv-bound}

We give a conservative sufficient bound for this operation path, to be
read with the implementation assumptions below. In
$R=\mathbb{Z}[X]/(X^d+1)$, the coefficient norm obeys
$\|ab\|_\infty\le d\|a\|_\infty\|b\|_\infty$; it is not the SIMD slot norm.
Let $Q$ be the product of the three data primes and $r_{\mathsf{sp}}$ the special prime.
SEAL 4.1.2 uses centered-binomial noise with coefficient magnitude at most
21 and ternary secret and ephemeral polynomials. Before dropping $r_{\mathsf{sp}}$, an
asymmetric encryption of zero has phase $pE$ modulo $Qr_{\mathsf{sp}}$, where
\begin{equation}
\label{eq:bgv-pre-switch}
 E=-e_{\mathrm{pk}}u+e_0+e_1s,
 \qquad \|E\|_\infty\le 21(2d+1).
\end{equation}
For each ciphertext component coefficient $c$, the zero-encryption switch
uses $a=c\bmod r_{\mathsf{sp}}$, $k=-ar_{\mathsf{sp}}^{-1}\bmod p$ in $[0,p-1]$, and
$\delta=a+r_{\mathsf{sp}}k$. Thus $0\le\delta<r_{\mathsf{sp}}p$, $\delta\equiv0\pmod p$ and
$\delta\equiv c\pmod{r_{\mathsf{sp}}}$. The switched phase has an integer representative
$z=(pE-\delta_0-\delta_1s)/r_{\mathsf{sp}}$ modulo $Q$. Its numerator is divisible by both
$r_{\mathsf{sp}}$ and $p$, so $z\equiv0\pmod p$. Since $r_{\mathsf{sp}}>344085$,
$\|z\|_\infty<p(d+2)$. Adding the centered plaintext polynomial $m$ gives
\begin{equation}
\label{eq:bgv-fresh-bound}
 \|m+z\|_\infty<p(d+2.5)<B_{\mathsf{ph}},
 \qquad B_{\mathsf{ph}}=2^{48}(d+3).
\end{equation}
Each row sums at most $K_{\mathsf{reg}}=16384$ such phases. A query sums $L$ products of
an accumulated row and a fresh query row. A sufficient condition for its
phase to remain inside the centered decryption range is therefore
\begin{equation}
\label{eq:bgv-capacity-bound}
 2LK_{\mathsf{reg}}dB_{\mathsf{ph}}^2<Q.
\end{equation}
Using the actual primes gives $Q/(2LK_{\mathsf{reg}}dB_{\mathsf{ph}}^2)\approx1.998536$, so the strict
inequality holds. This conservative ratio is not SEAL's measured invariant
noise budget. Together with C1, the bound addresses the specified
honestly generated ciphertext path, assuming correct ring arithmetic and
decoding. It does not cover arbitrary malicious ciphertexts or other
homomorphic circuits.

The bound requires SEAL's asymmetric zero encryption with one special-prime
drop, centered plaintext lift and correction factor one, followed only by
row additions and $L$ products plus $L-1$ additions. There are no rotations,
relinearizations, later modulus switches or refreshes. We checked these
paths against the frozen SEAL sources and the actual build; Gaussian noise
is disabled. A separate small integer-ring model checks 400 algebraic
instances, distinct from the SEAL histories. The source-derived argument
and its phase-representative reasoning remain subject to independent
mathematical review; arithmetic checks and finite runs are not that review.

\subsection{Timing and reproducibility}

Main-paper Table~\ref{tab:bgv} gives the serial Unified512-2 medians
from one warmup and five measured rounds. Saving the enrollment takes
188.64\,ms uncompressed or 678.77\,ms with zstd. In-memory Keccak-256
medians are 1004.80, 946.48 and 813.75\,ms for the tail, enrollment and
merged frames, respectively. Uncompressed frames contain 201{,}388{,}584
bytes, including 4136 bytes of framing. We rechecked their SHA-256 and
Keccak-256 hashes; large buffers remain in the local evidence directory,
with hashes included in the portable manifests. This timing path appends
once to a freshly encrypted almost-full gallery. It is distinct from the
three parallel capacity histories, whose elapsed times are not single-request
latencies. Two-context byte equality is a same-host check.

The native implementation uses GCC 16.2.0, SEAL 4.1.2 and Zstandard 1.5.7
on Windows 11 with an Intel Core i9-9900T. Each capacity process encrypts
rows with four threads and separate encoders, encryptors and memory pools;
serial timings do not sum parallel work. SEAL's supported C++17-off build
selects its stock atomic locks after a Windows shared-mutex handle leak was
diagnosed; cryptographic parameters and randomness are unchanged. Earlier
resource-stalled and console-interrupted runs are retained separately and
contribute no progress to the completed histories. Source, executable,
library, configuration and raw-record hashes are frozen with run IDs
R-BGV-RUNTIME-R3-FINAL-20261031 and R-BGV-RUNTIME-R3-MEASURE-20261031.
The reproducibility guide locates the audit and evidence package.
\par}

\subsection{Original parameter sets and repair controls}

We additionally implement actual public-key BGV enrollment and merge
operations, separately from the JavaScript pipeline. The experiment uses
official Microsoft SEAL 4.1.2 and the five parameter sets that BioZKFHE's
artifact, at commit \texttt{76ffb9f}, uses for its experiment E3. We refer
to them as the original parameter sets. The artifact's bundled SEAL
fork contains unfinished changes that do not compile in our environment;
this is an independent parameter-matched experiment, not a reproduction of
that binary. Compilation uses GCC 16.2.0, Release/O3 and Zstandard 1.5.7 on
Windows 11 with an Intel Core i9-9900T. SEAL's 128-bit parameter check is
enabled. The raw data include actual primes and parameter identifiers.

For each profile, one warmup precedes five measured rounds. We encrypt a
synthetic unit-normalized, quantized enrollment at the final empty slot of
a freshly encrypted full tail block, encrypt a query, load and merge the
frames, and compare all 8,192 decoded slots and their packed layers against
plaintext inner products. A second context loads the same input bytes and
recomputes identical merge bytes in every round. This is a second computation
in the same process, not a cross-platform determinism result. Table~\ref{tab:bgv-original} reports the component medians for the
original parameter sets (R-BGV-20261030); main-paper Table~\ref{tab:bgv} instead reports
Unified512-2. Construction, encoding and
public-key encryption are included in enrollment time; key generation,
capture, signatures, network transfer and chain execution are excluded.

\begin{table}[t]
\centering
\caption{Single-merge measurements with the original parameter sets,
retained as parameter controls. F: FaceNet ($L=128$, $\Gamma=128$); M: MobileFaceNet ($L=512$,
$\Gamma=512$). None/Def denote uncompressed/zstd frame sizes.}
\label{tab:bgv-original}
\footnotesize
\setlength{\tabcolsep}{3pt}
{
\begin{tabular}{@{}lrrrrr@{}}
\toprule
Profile & None (MiB) & Def (MiB) & Enroll (s) & Add (ms) & Noise$^*$ \\
\midrule
F, $\ell=1$ & 48.01 & 25.09 & 0.611 & 9.21 & 8 \\
F, $\ell=2$ & 48.01 & 31.43 & 0.624 & 9.27 & 11 \\
F, $\ell=3$ & 48.01 & 37.61 & 0.648 & 9.70 & 0 \\
M, $\ell=1$ & 192.06 & 106.95 & 2.745 & 41.86 & 13 \\
M, $\ell=2$ & 192.06 & 140.74 & 2.663 & 38.92 & 0 \\
\bottomrule
\end{tabular}\\[2pt]
\raggedright $^*$Minimum query noise budget in bits over six rounds.
All single-merge results decoded correctly, including the zero-budget
cases; the cumulative failures described below concern separate runs.}
\end{table}

The canonical experiment frame contains a version tag, $d,\ell,L,T$, and
length-prefixed rows saved with \texttt{compr\_mode\_type::none}. The actual
sizes are 50,347,176 bytes for 128 rows and 201,388,584 bytes for 512 rows;
framing contributes 1,064 and 4,136 bytes, respectively. Ciphertexts have
three data-level RNS limbs, size two, NTT representation and correction
factor one. Downloading only the enrollment frame therefore costs about
45.8--183.2\,TiB over a million registrations, even with a local tail replica.
This replaces the earlier compressed-block proxy for an uncompressed wire
format. Key/parameter identities and representation rules would also need
to be authenticated in a deployment.

Loading both inputs takes median 70.5--328.2\,ms and saving the merged frame
47.7--212.5\,ms. The sum of separately measured median Keccak-256 times for
the two inputs and result is 0.56--2.66\,s. These are in-memory component
costs, not a complete transaction latency; additions alone understate the
committee's work. The full ciphertext buffers and their hashes are retained
locally; portable raw samples and manifests identify every run.

\emph{Accumulated noise is a separate constraint.} A freshly encrypted
almost-full block plus one enrollment does not represent a block formed
by thousands of encrypted insertions. We therefore start with one enrollment
and independently encrypt and add every subsequent enrollment across all
$L$ rows. At powers-of-two checkpoints we evaluate the full encrypted query
and compare every slot and layer. With fresh keys in each of three runs,
FaceNet $\ell=3$ first fails at checked counts 4, 16 and 8. MobileFaceNet $\ell=2$
first fails at checked counts 16, 16 and 32. The failed query has zero reported noise
budget and all 8,192 slots disagree with the plaintext result. Checkpoints
do not identify the earliest possible failure, and a zero budget alone is
not our failure criterion.

These original configurations serve as baseline controls for parameter
repair. The failures concern incremental insertion, which BioZKFHE's
whole-block experiments do not exercise, and do not contradict its
published single-block results; every single merge in
Table~\ref{tab:bgv-original} decodes correctly. A diagnostic rerun of FaceNet $\ell=3$ fails
at four registrations while every accumulated row still decrypts correctly;
a fresh encryption of the same gallery, evaluated with the same encrypted
query, succeeds. This localizes the observed failure to the query's noise
budget rather than an incorrect row merge.

We therefore add an incremental-registration candidate with coefficient
modulus limb sizes $(60,50,50,50)$: 210 key-level bits and 160 data-level
bits at degree 8192, retaining SEAL's tc128 check. Plaintext moduli, packing,
inputs and query computation are unchanged. One fresh-key run per profile
performs 64 actual registrations; all seven checkpoints at counts
1,2,4,8,16,32,64 pass row, packed-score and fresh-gallery control checks.
Query noise budgets at count 64 are 29 bits for FaceNet $\ell=3$ and 40 bits
for MobileFaceNet $\ell=2$. These are short repair regressions for the two original vector
profiles, not full-capacity validation of those profiles. The complete
histories above instead use Unified512-2, with a different vector domain,
plaintext prime and packing base. Aligning that profile with authentication
does not integrate the encrypted backend into the bridge or committee.

\section{Extended Discussion and Design Alternatives}
\label{app:extensions}

\emph{Modality at scale.} Main-paper Section~\ref{sec:eval} identifies an accuracy tradeoff for the tested
public face model on LFW and in scale extrapolations assuming independent
comparisons. This does not establish an impossibility for face recognition,
and it matches the choice of the largest deployment: World ID uses the iris for
uniqueness and the face only for 1:1 re-authentication. Controlled
enrollment capture, several images per person, quality gating and stronger
face models would all improve the numbers; we did not measure any of them and
give no figures. An iris or multimodal backend must instantiate the
same binding and state conditions with its own matcher and error model;
Appendix~\ref{app:world-mpc} gives an MPC interface mapping.

\emph{Capture hardware.} The $\AFE$ is a software mock in this paper.
The contribution concerns protocol composition under explicit trusted
interfaces; validating a physical capture module is outside this evaluation.
A
hardware module would need a protected sensor path, a protected key, a
persistent anti-rollback counter for the budget and an attested firmware
measurement. Our mock follows the same wire format, so it establishes the
interface and the ordering of reservation, capture and signature, but not
sensor provenance. In the unlinkable profile the attempt budget rests on the
module alone, since the on-chain counter would identify the device.

\emph{Removing trust in the enrollment station.} The $\ES$ is trusted to
derive both ciphertexts and the commitment from one vector. A proof that a
BGV ciphertext and a Poseidon commitment encrypt and commit the same
plaintext would remove that trust for I1, but proofs over megabyte ciphertexts are far beyond current circuit
sizes; this remains open. Such a proof would not replace the $\ES$: deduplication is only
meaningful if an entity outside the enrollee's control witnesses one live person
and one capture, and limits enrollment attempts, which cannot be counted
per person before deduplication.

\emph{Self-enrollment variant.} The $\ES$ role can also be played by the
holder's own certified capture module, so that enrollment too stays on the user
side. The price is Sybil resistance. Attempts can then only be counted per
device, a person with several modules can retry until one capture is a missed
duplicate, and presentation attacks are easier without supervision; at the
operating point of main-paper Section~\ref{sec:eval}, a non-adversarial miss rate of
about 17\% compounds over retries. One enrollment ciphertext consists of $L$
BGV ciphertexts and is structurally a full gallery block. Our
measurements with BioZKFHE's original parameter sets require 48--192\,MiB under uncompressed framing and
about 0.61--2.75\,s to construct, encode and encrypt a registration on the
measured desktop CPU. These are separate backend profiles, not measurements
on a microcontroller or on the recommended authentication parameters.
They favor an adequately resourced attended station, but cost alone
does not establish its trustworthiness. Self-enrollment remains an extension requiring separate capture and
attempt-control evidence.

\emph{Choice of 1:N backend.} Our main instance inherits the packed
gallery encoding and the block proofs of BioZKFHE~\cite{xin2026biozkfhe};
these supply gallery encoding and verifiable encrypted evaluation. The present construction adds the capture,
issuance and state bindings around that engine. Appendix~\ref{app:world-mpc}
identifies the corresponding conditions for a World ID-inspired MPC
alternative and the different assumptions needed for I4. Public
verifiability is one way to establish verdict soundness, rather than a
requirement that every admissible backend use the same proof system.

\emph{Committee bandwidth and homomorphic digests.} A digest that is linear
in the ciphertext would let anyone check a merge from two digests and remove
the committee from root migration. A linear map modulo $q$ is not collision
resistant on arbitrary inputs, because its kernel can be found by Gaussian
elimination, so the naive version fails. A Pedersen-style commitment over the
unreduced integer block remains possible, but whether it saves bandwidth
depends on how BioZKFHE's block proofs bind blocks; we have not evaluated it experimentally.

\emph{Tree cost.} Credential issuance and revocation in the unlinkable profile
cost about 0.86 and 1.04 million gas. Storing tree nodes would let revocation
compute one path instead of two, and an off-chain root update proved by a
SNARK would allow batching; both trade storage or prover work for gas.

\emph{Template refresh and deletion.} The gallery only grows. Re-enrollment
after revocation, same-model template refresh and deletion all require a new
backend operation and a revised uniqueness game, and we do not support them.

\subsection{Deployment and Evaluation Limitations}
\label{app:deployment-limitations}

The prototype has no capture hardware: the $\ES$ is a software bridge and
the $\AFE$ a software mock, so nothing here shows sensor provenance or a
tamper-resistant budget, and enrollment has no presentation-attack
detection. The committee is simulated, block proofs are not executed, and our gas
figures exclude them; the composition with BioZKFHE uses its published
measurements. We do not implement BioZKFHE's block-proof submission or
the check that every snapshot block is proved; a deployment must accept only openings checked against their
quadratic arithmetic programs, which the published cost-measurement contract does not enforce. The
stale-session check compares the gallery count only, which suffices because
the gallery only grows; supporting replacement or migration with an
unchanged count would require comparing a gallery version or root instead.
Data availability of the ciphertexts that the committee rehashes is assumed,
not implemented. The aligned BGV profile passes three full-capacity
histories, each with one fixed synthetic query. This finite input/key
coverage does not establish a general failure probability, and the
source-dependent noise argument still requires independent review.
All Groth16 keys come from local single-party benchmark
setups and are not deployable. The biometric evaluation uses one public face
model and one dataset, LFW, which cannot resolve low FARs, and we have no
access to the pose and age stress sets or to a large controlled dataset. The
biometric terms $\fnm$ and $\fmax$ have only non-adversarial reference
values. The proofs have not been independently reviewed. The
authentication and privacy arguments for the unlinkable profile are conditional on explicit extraction,
Merkle-membership and related-output hiding assumptions; their concrete
instantiations still require review. The gateway of the unlinkable profile lacks policy
revocation and model migration. Gas figures are local transaction
measurements; proving-time experiments use the repetitions reported with
each run, rather than a deployment-wide performance guarantee.

\bibliographystyle{IEEEtran}
\bibliography{references}

\end{document}